\documentclass[aps, prd, letterpaper, amsmath, amssymb, preprintnumbers, showpacs, superscriptaddress, twocolumn, floatfix, nofootinbib]{revtex4-1}
\usepackage{graphicx}
\usepackage{mathtools}
\usepackage[usenames]{color}
\usepackage{siunitx} 
\usepackage{cancel}
\usepackage{hyperref}
\hypersetup{colorlinks=true, linkcolor=blue, citecolor=blue, filecolor=blue, urlcolor=blue}

\newcommand{\al}{\ensuremath{\alpha} }
\newcommand{\be}{\ensuremath{\beta} }
\newcommand{\ga}{\ensuremath{\gamma} }
\newcommand{\Ga}{\ensuremath{\Gamma} }
\newcommand{\de}{\ensuremath{\delta} }
\newcommand{\De}{\ensuremath{\Delta} }
\newcommand{\eps}{\ensuremath{\epsilon} }
\newcommand{\la}{\ensuremath{\lambda} }
\newcommand{\La}{\ensuremath{\Lambda} }
\renewcommand{\si}{\ensuremath{\sigma} }
\newcommand{\Ibb}{\ensuremath{\mathbb I} }
\newcommand{\cO}{\ensuremath{\mathcal O} }
\newcommand{\Zbb}{\ensuremath{\mathbb Z} }
\newcommand{\psibar}{\ensuremath{\overline\psi} }
\newcommand{\chibar}{\ensuremath{\overline\chi} }
\newcommand{\muhat}{\ensuremath{\widehat\mu} }
\newcommand{\cf}{\textit{cf}.\ }
\newcommand{\X}{\ensuremath{\!\times\!} }
\newcommand{\gsim}{\ensuremath{\gtrsim} }
\newcommand{\lsim}{\ensuremath{\lesssim} }
\newcommand{\llra}{\ensuremath{\longleftrightarrow} }
\newcommand{\Sb}{\ensuremath{\cancel{S^4}} }
\newcommand{\Ozpp}{\ensuremath{\cO_{0^{++}}} }
\newcommand{\nn}{\nonumber}
\newcommand{\ReTr}{\ensuremath{\mbox{ReTr}} }
\newcommand{\Tr}[1]{\ensuremath{\mbox{Tr}\left[ #1 \right]} }
\newcommand{\vev}[1]{\ensuremath{\left\langle #1 \right\rangle} }
\newcommand{\ME}[3]{\ensuremath{\langle #1 \left| #2 \right| #3 \rangle} }
\newcommand{\pbp}{\ensuremath{\vev{\psibar\psi}} }
\newcommand{\eq}[1]{Eq.~(\ref{#1})}
\newcommand{\fig}[1]{Fig.~\ref{#1}}
\newcommand{\tab}[1]{Table~\ref{#1}}
\newcommand{\secref}[1]{Section~\ref{#1}}
\newcommand{\refcite}[1]{Ref.~\cite{#1}}

\begin{document}
\title{Nonperturbative investigations of SU(3) gauge theory with eight dynamical flavors}

\author{T.~Appelquist}
\affiliation{Department of Physics, Sloane Laboratory, Yale University, New Haven, Connecticut 06520, USA}
\author{R.~C.~Brower}
\affiliation{Department of Physics and Center for Computational Science, Boston University, Boston, Massachusetts 02215, USA}
\author{G.~T.~Fleming}
\affiliation{Department of Physics, Sloane Laboratory, Yale University, New Haven, Connecticut 06520, USA}
\author{A.~Gasbarro}
\affiliation{Department of Physics, Sloane Laboratory, Yale University, New Haven, Connecticut 06520, USA}
\affiliation{Physical and Life Sciences, Lawrence Livermore National Laboratory, Livermore, California 94550, USA}
\affiliation{Nuclear Science Division, Lawrence Berkeley National Laboratory, Berkeley, CA 94720, USA}
\author{A.~Hasenfratz}
\affiliation{Department of Physics, University of Colorado, Boulder, Colorado 80309, USA}
\author{X.~Y.~Jin}
\affiliation{Computational Science Division, Argonne National Laboratory, Argonne, Illinois 60439, USA}
\author{E.~T.~Neil}
\email{ethan.neil@colorado.edu}
\affiliation{Department of Physics, University of Colorado, Boulder, Colorado 80309, USA}
\affiliation{RIKEN BNL Research Center, Brookhaven National Laboratory, Upton, New York 11973, USA}
\author{J.~C.~Osborn}
\affiliation{Computational Science Division, Argonne National Laboratory, Argonne, Illinois 60439, USA}
\author{C.~Rebbi}
\affiliation{Department of Physics and Center for Computational Science, Boston University, Boston, Massachusetts 02215, USA}
\author{E.~Rinaldi}\email{erinaldi@bnl.gov}
\affiliation{RIKEN BNL Research Center, Brookhaven National Laboratory, Upton, New York 11973, USA}
\affiliation{Physical and Life Sciences, Lawrence Livermore National Laboratory, Livermore, California 94550, USA}
\affiliation{Nuclear Science Division, Lawrence Berkeley National Laboratory, Berkeley, CA 94720, USA}
\author{D.~Schaich}\email{schaich@itp.unibe.ch}
\affiliation{AEC Institute for Theoretical Physics, University of Bern, 3012 Bern, Switzerland}
\affiliation{Department of Physics, Syracuse University, Syracuse, New York 13244, USA}
\author{P.~Vranas}
\affiliation{Physical and Life Sciences, Lawrence Livermore National Laboratory, Livermore, California 94550, USA}
\affiliation{Nuclear Science Division, Lawrence Berkeley National Laboratory, Berkeley, CA 94720, USA}
\author{E.~Weinberg}
\affiliation{Department of Physics and Center for Computational Science, Boston University, Boston, Massachusetts 02215, USA}
\author{O.~Witzel}
\affiliation{Department of Physics, University of Colorado, Boulder, Colorado 80309, USA}
\collaboration{Lattice Strong Dynamics (LSD) Collaboration}
\noaffiliation

\date{24 July 2018}

\preprint{RBRC-1286; LLNL-JRNL-753511}

\begin{abstract}
  We present our lattice studies of SU(3) gauge theory with $N_f = 8$ degenerate fermions in the fundamental representation.
  Using nHYP-smeared staggered fermions we study finite-temperature transitions on lattice volumes as large as $L^3\X N_t = 48^3\X 24$, and the zero-temperature composite spectrum on lattice volumes up to $64^3\X 128$.
  The spectrum indirectly indicates spontaneous chiral symmetry breaking, but finite-temperature transitions with fixed $N_t \leq 24$ enter a strongly coupled lattice phase as the fermion mass decreases, which prevents a direct confirmation of spontaneous chiral symmetry breaking in the chiral limit.
  In addition to the connected spectrum we focus on the lightest flavor-singlet scalar particle.
  We find it to be degenerate with the pseudo-Goldstone states down to the lightest masses reached so far by non-perturbative lattice calculations.
  Using the same lattice approach, we study the behavior of the composite spectrum when the number of light fermions is changed from eight to four.
  A heavy flavor-singlet scalar in the 4-flavor theory affirms the contrast between QCD-like dynamics and the low-energy behavior of the 8-flavor theory.
\end{abstract}

\maketitle

\section{\label{sec:intro}Introduction} 
The discovery of a Higgs particle at the Large Hadron Collider (LHC)~\cite{Aad:2012tfa, Chatrchyan:2012ufa} was a major step towards the longstanding goal of determining the mechanism of electroweak symmetry breaking.
The properties of this particle, which are so far consistent with the Standard Model~\cite{Khachatryan:2016vau}, could also result from new strong dynamics at or above the TeV scale.
Lattice gauge theory is an indispensable tool to study the relevant strongly coupled systems, which will generally differ qualitatively from QCD in order to be phenomenologically viable.

In recent years lattice investigations have begun to explore novel near-conformal strong dynamics that emerge upon increasing the light fermion content of non-Abelian gauge theories (\cf the recent reviews~\cite{DeGrand:2015zxa, Nogradi:2016qek, Svetitsky:2017xqk} and references therein).
A particularly significant result of these efforts is increasing evidence~\cite{Aoki:2013pca, Appelquist:2013sia, Aoki:2013zsa, Aoki:2013hla, Aoki:2014oha, Athenodorou:2014eua, Fodor:2015vwa, Rinaldi:2015axa, Brower:2015owo, Fodor:2016pls, Hasenfratz:2016gut, DelDebbio:2015byq, Appelquist:2016viq, Aoki:2016wnc, Gasbarro:2017fmi, Athenodorou:2017dbf, Fodor:2017nlp} that such near-conformal dynamics might generically give rise to scalar ($0^{++}$) Higgs candidates substantially lighter than the analogous $f_0(500)$ meson of QCD~\cite{Patrignani:2016xqp}.

The flavor-singlet scalar meson (labelled interchangeably by \si or $0^{++}$ in the following) is of particular interest as a potential composite Higgs candidate.
The strength of current LHC experimental bounds on new particles favors a ``little hierarchy''~\cite{Azatov:2017gas} between the Higgs boson and the other states arising from this new sector, which can arise dynamically when the $0^{++}$ state is found to be light compared to the rest of the spectrum.
To assess the viability of models of dynamical electroweak symmetry breaking built on gauge theories with a particularly light $0^{++}$ meson, detailed knowledge of the appropriate low-energy effective theory in the presence of this light state is required.
Lattice calculations provide crucial input into discriminating candidate low-energy effective theories.

In this paper we continue to explore these issues in the context of SU(3) gauge theory with $N_f = 8$ light fermions in the fundamental representation.
Previous lattice studies of this system have identified several features quite distinct from QCD, which make it a particularly interesting representative of the broader class of near-conformal gauge theories.
These features include slow running of the gauge coupling (a small \be function)~\cite{Hasenfratz:2014rna, Fodor:2015baa}, a reduced electroweak $S$~parameter~\cite{Appelquist:2014zsa}, a slowly evolving mass anomalous dimension $\ga_m$~\cite{Cheng:2013eu}, and changes to the composite spectrum including a light flavor-singlet $0^{++}$ scalar~\cite{Aoki:2013xza, Schaich:2013eba, Aoki:2014oha, Appelquist:2014zsa, Rinaldi:2015axa, Appelquist:2016viq, Aoki:2016wnc, Gasbarro:2017fmi} and a heavier flavor-singlet $0^{-+}$ pseudoscalar~\cite{Aoki:2017fnr} (these states are referred to as the \si and $\eta'$ mesons in QCD).
Several lattice groups continue to investigate the 8-flavor theory in order to learn more about its low-energy dynamics and relate it to phenomenological model building.

\begin{figure*}[tbp]
  \includegraphics[width=0.7\linewidth]{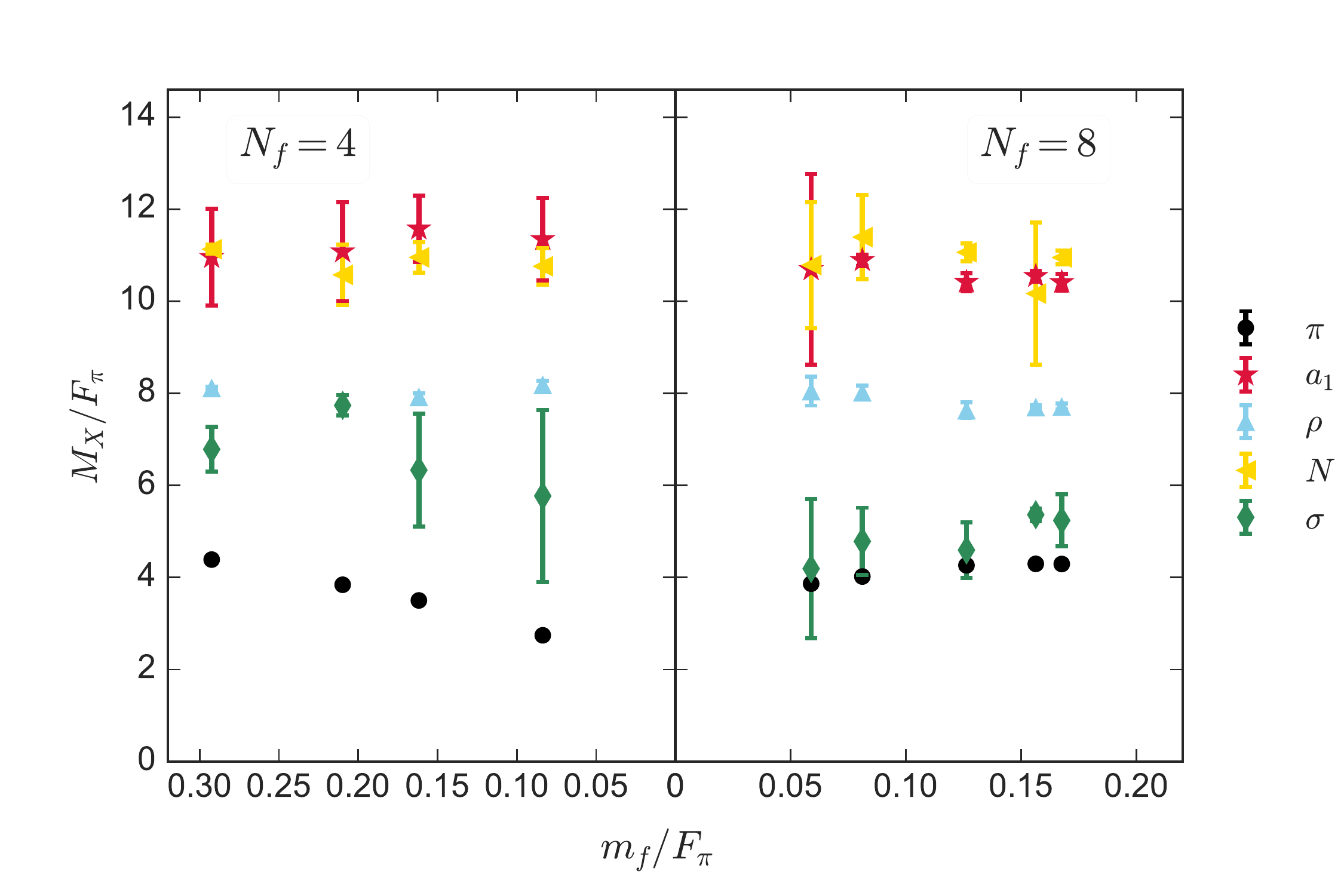}
  \caption{\label{fig:spectrum_4f8f}Comparison of our spectroscopy results for $N_f = 4$ (left) and $N_f = 8$ (right).  Hadron masses (vertical axis) and the fundamental fermion mass (horizontal axis) are both shown in units of the pion decay constant $F_{\pi}$; the chiral limit $m_f = 0$ is at the center of the plot for both theories.  The hadrons shown are the lightest $0^{++}$ meson ($\si$), $0^{-+}$ PNGB meson ($\pi$), $1^{--}$ vector meson ($\rho$), $1^{++}$ axial-vector meson ($a_1$), and the nucleon ($N$).  The major qualitative difference between the two values of $N_f$ is the degeneracy of the light scalar \si with the pions at $N_f = 8$.}
\end{figure*}

These investigations employ a wide variety of methods, including the computation of the running coupling and its discrete \be function~\cite{Appelquist:2007hu, Appelquist:2009ty, Hasenfratz:2014rna, Fodor:2015baa}, exploration of the phase diagram through calculations at finite temperature~\cite{Deuzeman:2008sc, Miura:2012zqa, Jin:2010vm, Schaich:2012fr, Hasenfratz:2013uha, Lombardo:2014mda, Schaich:2015psa}, analysis of hadron masses and decay constants~\cite{Fodor:2009wk, Cheng:2011ic, Aoki:2013xza, Aoki:2014oha, Schaich:2013eba, Appelquist:2014zsa, Rinaldi:2015axa, Aoki:2016fxd, Appelquist:2016viq, Aoki:2016bfp, Aoki:2016wnc, Gasbarro:2017fmi, Aoki:2017fnr}, study of the eigenmodes of the Dirac operator~\cite{Fodor:2009wk, Cheng:2011ic, Hasenfratz:2012fp, Cheng:2013eu, Aoki:2016fxd}, and more~\cite{Hasenfratz:2010fi, Hasenfratz:2011xn, Petropoulos:2012mg, Ishikawa:2013tua, Ishikawa:2015iwa, daSilva:2015vna, Aoki:2015aqa, Noaki:2015xpx, Aoki:2016yrm}.
While most of these studies obtain results consistent with spontaneous chiral symmetry breaking in the massless limit for $N_f = 8$~\cite{Appelquist:2007hu, Appelquist:2009ty, Hasenfratz:2014rna, Fodor:2015baa, Deuzeman:2008sc, Miura:2012zqa, Jin:2010vm, Schaich:2012fr, Hasenfratz:2013uha, Lombardo:2014mda, Schaich:2015psa, Fodor:2009wk, Aoki:2013xza, Aoki:2014oha, Schaich:2013eba, Appelquist:2014zsa, Appelquist:2016viq, Aoki:2016bfp, Aoki:2016wnc, Gasbarro:2017fmi, Cheng:2013eu}, this has not yet been established definitively and some recent works favor the existence of a conformal infrared fixed point (IRFP)~\cite{Ishikawa:2015iwa, daSilva:2015vna, Noaki:2015xpx}.
For example, although all lattice studies of the 8-flavor discrete \be function obtain monotonic results, with no non-trivial IR fixed point where $\be(g_{\star}^2) = 0$, it remains possible that an IRFP could exist at some stronger coupling that these works were not able to access.

This possibility can be tested through complementary studies of phase transitions at finite temperature $T = 1 / (aN_t)$, where `$a$' is the lattice spacing and $N_t$ is the temporal extent of the lattice.
In a chirally broken system such as QCD, the bare (pseudo-)critical couplings $g_{cr}(N_t)$ of these transitions must move to the asymptotically free UV fixed point $g_{cr} \to 0$ as the UV cutoff $a^{-1} \to \infty$ and $N_t \to \infty$ holding $T(N_t) = T_{cr}$ fixed.
In an IR-conformal system, in contrast, the finite-temperature transitions must accumulate at a finite bare coupling as $N_t \to \infty$, so that $T_{cr}$ is independent of $N_t$, and remain separated from the weak-coupling conformal phase by a bulk transition.

Unlike running coupling studies, finite-temperature lattice calculations use non-zero bare fermion mass $am$ to give mass $aM$ to the pseudo-Nambu--Goldstone bosons (PNGBs) which appear in the chirally broken phase.
If the Compton wavelength of the PNGBs $\sim 1 / (aM)$ is not small relative to the spatial extent of the lattice, significant finite-volume effects will occur.
Results must be extrapolated to the $am \to 0$ chiral limit to ensure that the chiral symmetry breaking is truly spontaneous.
Although previous works observed QCD-like scaling of $g_{cr}$ for $N_f = 8$ with sufficiently large masses $am \gsim 0.01$~\cite{Deuzeman:2008sc, Miura:2012zqa, Schaich:2012fr, Hasenfratz:2013uha}, this did not persist at smaller $am \leq 0.005$, where the finite-temperature transitions merged with a bulk transition into a lattice phase.

In \secref{sec:phase} we revisit this finite-temperature analysis, employing larger $N_t$ than those previous works.
Although these larger lattices allow us to consider smaller masses down to $am = 0.0025$, we find that the finite-temperature transitions still run into a strongly coupled lattice phase before reaching the chiral limit.
That is, we are not able to directly confirm spontaneous chiral symmetry breaking.
The details of the lattice phase depend on our lattice action, which we also review in \secref{sec:phase}.
We use improved nHYP-smeared staggered fermions, which conveniently represent $N_f = 8$ continuum flavors as two (unrooted) lattice fields.
Staggered fermions (with or without various forms of improvement) are also used by almost all of the other studies summarized above, with the exceptions of Refs.~\cite{Ishikawa:2013tua, Ishikawa:2015iwa} (Wilson fermions) and Refs.~\cite{Appelquist:2014zsa, Noaki:2015xpx} (domain wall fermions).

Beginning in \secref{sec:ensembles} we turn to the main topic of this paper, large-scale studies of the 8-flavor composite spectrum at zero temperature.
\secref{sec:ensembles} describes the zero-temperature ensembles that we have generated for this work, which reach the lightest masses considered to date.
We use the Wilson flow to set the scale of these ensembles, and observe the Wilson flow scale to be more sensitive to the fermion mass than would be expected for lattice QCD.
In \secref{sec:spectrum} we review the details of our staggered spectrum analyses, separately considering the flavor-singlet scalar \si that involves contributions from fermion-line-disconnected diagrams.

Our results for the 8-flavor spectrum are discussed in \secref{sec:results}.
Their most significant feature is the presence of a remarkably light flavor-singlet scalar particle ($\si$), which remains degenerate with the PNGBs ($\pi$) down to the lightest masses reached so far by lattice calculations.
This is a significant contrast with QCD, which we strengthen by carrying out a similar 4-flavor spectrum calculation.
Using the same lattice action and analysis procedure, our QCD-like 4-flavor results do not produce a light scalar, as expected.
The comparison is reported in \fig{fig:spectrum_4f8f} where the masses of composite states are normalized by the $\pi$ decay constant ($F_{\pi}$).

However, other aspects of our 8-flavor spectrum results are qualitatively similar to QCD.
At the most basic level, the ratio of the vector mass to the pseudoscalar mass steadily increases as we approach the chiral limit, providing indirect indication that the theory exhibits spontaneous chiral symmetry breaking.
We also find the ratio of the vector mass to the pseudoscalar decay constant to be comparable to its QCD value, $M_{\rho} / F_{\pi} \approx 8$, and rather constant as we decrease the masses.
In the context of models of dynamical electroweak symmetry breaking, this suggests (via the Kawarabayashi--Suzuki--Riazuddin--Fayyazuddin (KSRF) relations~\cite{Kawarabayashi:1966kd, Riazuddin:1966sw}) a multi-TeV-scale vector resonance with a large decay width $\Ga / M \simeq 0.2$, comparable to that of the QCD $\rho$ meson.
This is broader than the typical width assumed in past LHC searches for such states~\cite{Azatov:2017gas}; dedicated searches for broad resonances, although challenging, are well-motivated by the lattice results.

We summarize our conclusions and prospects for further progress in \secref{sec:conclusion}.
In particular, we focus on the issue of the appropriate low-energy effective field theory (EFT) to describe the 8-flavor spectrum we observe.
A consequence of the light flavor-singlet scalar is that we cannot expect to carry out chiral extrapolations by fitting our data to chiral perturbation theory ($\chi$PT), which assumes that the PNGBs are much lighter than all other particles.
Finally, in the appendices we provide additional information about auto-correlations and topological charge evolution, more technical details about fitting correlation functions for the flavor-singlet scalar, and studies of finite-volume and discretization effects.

\section{\label{sec:phase}Lattice action and finite-temperature phase diagram} 
Our numerical calculations use improved nHYP-smeared staggered fermions~\cite{Hasenfratz:2001hp, Hasenfratz:2007rf} with smearing parameters $\al = (0.5, 0.5, 0.4)$, and a gauge action that includes both fundamental and adjoint plaquette terms with couplings $\be_F$ and $\be_A$, respectively, related by $\be_A / \be_F = -0.25$~\cite{Cheng:2011ic}.
This lattice action was used in several previous studies of the 8-flavor system, including explorations of the phase diagram~\cite{Cheng:2011ic, Schaich:2012fr, Hasenfratz:2013uha}, the composite spectrum~\cite{Schaich:2013eba}, the discrete \be function~\cite{Hasenfratz:2014rna} and the scale-dependent mass anomalous dimension $\ga_m(\mu)$~\cite{Cheng:2013eu}.
Using the same lattice action for all of these complementary investigations makes it easier to compare their results and thereby gain more comprehensive insight into the dynamics of $N_f = 8$.

The first work using this action observed a strongly coupled ``$\Sb$'' lattice phase in which the single-site shift symmetry ($S^4$) of the staggered action is spontaneously broken~\cite{Cheng:2011ic}.
In the massless limit, a first-order bulk (zero-temperature) transition around $\be_F \approx 4.6$ separates the \Sb phase from the weak-coupling phase where the continuum limit is defined.
At even stronger couplings there is a second bulk transition into a chirally broken lattice phase.
A similar phase structure has been seen by other many-flavor lattice investigations using different improved staggered actions~\cite{Deuzeman:2012ee, Fodor:2012et}.\footnote{Investigations using unimproved staggered fermions with either improved or unimproved gauge actions see a simpler bulk phase structure with only a single, chirally broken strong-coupling phase~\cite{Brown:1992fz, Appelquist:2009ty, Jin:2012dw}.}
However, the characteristics of these strong-coupling phases are not universal and depend on the details of the lattice action.
Although in this section we scan the lattice phase diagram, including the transition into the \Sb phase, our zero-temperature calculations reported in the rest of the paper will consider a coupling $\be_F = 4.8$ safely on the weak-coupling side of this bulk transition.

The presence of the \Sb phase prevents lattice investigations from reaching arbitrarily strong couplings.
For example, \refcite{Hasenfratz:2014rna} was only able to determine the continuum-extrapolated discrete \be function for renormalized couplings up to $g_c^2 \lsim 14$ (in finite-volume Wilson flow renormalization schemes introduced by \refcite{Fodor:2012td}).
As summarized in \secref{sec:intro}, although this \be function is monotonic throughout the accessible range of couplings, this does not guarantee that the 8-flavor theory exhibits spontaneous chiral symmetry breaking.
It remains possible that, at stronger couplings, the \be function might reach an extremum and then return to $\be(g_{\star}^2) = 0$ at some large $g_{\star}^2 \gsim 15$.
(Indeed, this happens in four-loop perturbation theory in the $\overline{\textrm{MS}}$ scheme, which predicts $g_{\star}^2 \approx 19.5$~\cite{Ryttov:2010iz, Pica:2010xq}, but perturbation theory seems unlikely to be reliable at such strong couplings.)

In the remainder of this section we present a complementary search for spontaneous chiral symmetry breaking in the 8-flavor system, by studying its finite-temperature phase diagram.
Initial results from this work appeared in \refcite{Schaich:2015psa}.
As described in \secref{sec:intro}, in order to establish spontaneous chiral symmetry breaking, the finite-temperature transitions we observe at non-zero bare fermion mass $am$ must persist in the chiral limit.
Previous finite-temperature studies with $am \geq 0.005$ and $N_t \leq 20$ instead found that these transitions merged with the bulk transition into the \Sb phase~\cite{Schaich:2012fr, Hasenfratz:2013uha}.
Here we move to larger $40^3\X 20$ and $48^3\X 24$ lattice volumes, which allows the exploration of smaller masses, $0.0025 \leq am \leq 0.01$.

For each combination of lattice volume and mass $am$, we generate ensembles of gauge configurations with $\be_F$ ranging from the strong-coupling \Sb phase to the deconfined phase at weak coupling.
To generate these ensembles we use either QHMC/FUEL~\cite{Osborn:2014kda} or a modified version of the MILC code,\footnote{\texttt{http://www.physics.utah.edu/$\sim$detar/milc/} \\ \texttt{https://github.com/daschaich/KS\_nHYP\_FA/}} in both cases employing the hybrid Monte Carlo (HMC) algorithm with a second-order Omelyan integrator~\cite{Takaishi:2005tz} accelerated by additional heavy pseudofermion fields~\cite{Hasenbusch:2002ai} and multiple time scales~\cite{Urbach:2005ji}.
(The only exception is the $N_f = 4$ zero-temperature ensemble with $am_f = 0.003$, which used an approximate force-gradient integrator with $\theta = 0.109$~\cite{Omelyan:2002fg, Yin:2011np}.)
We monitor several observables to identify both bulk and finite-temperature transitions, including $\pbp$, the Polyakov loop, \Sb order parameters introduced in \refcite{Cheng:2011ic}, and the massless Dirac eigenvalue spectrum $\rho(\la)$.

\begin{figure}[tbp]
  \includegraphics[width=\linewidth]{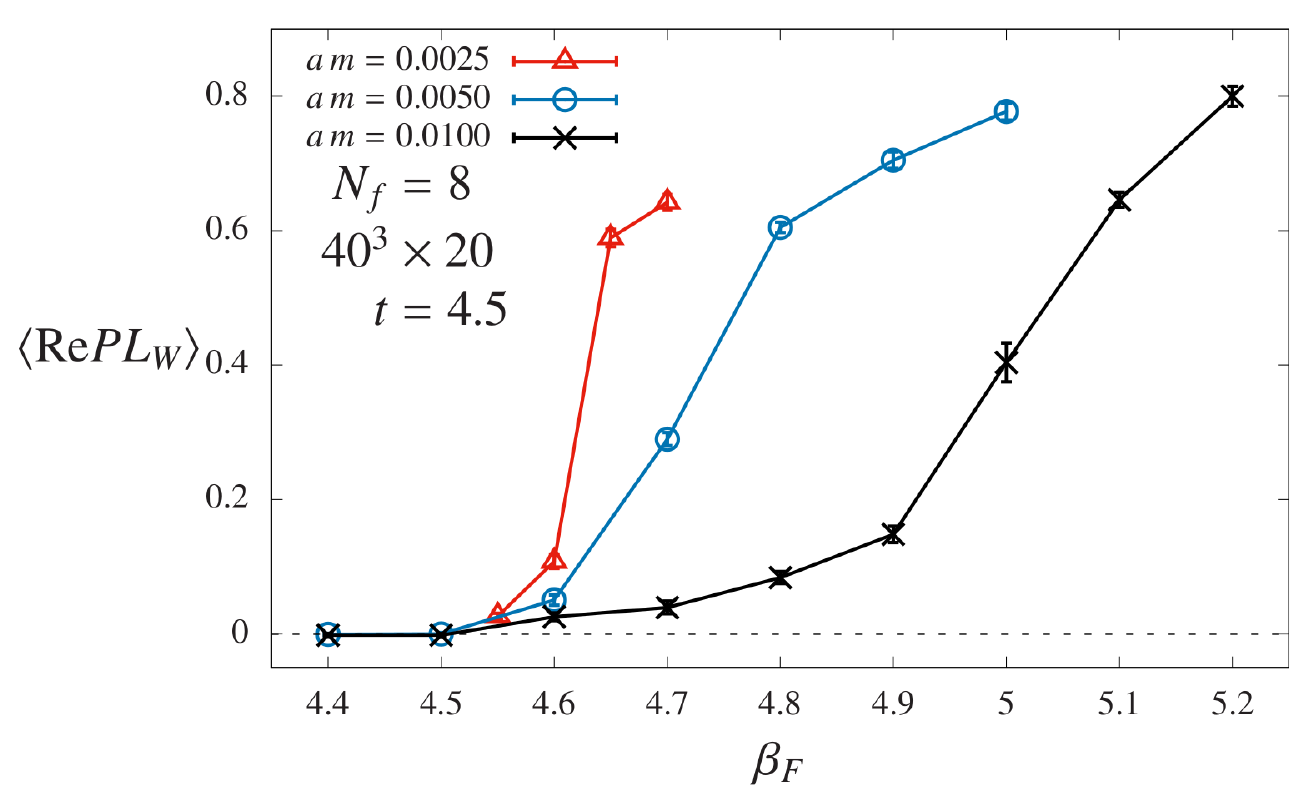}
  \caption{\label{fig:Wpoly} The real part of the Polyakov loop computed at Wilson flow time $t = 4.5$ (corresponding to $\sqrt{8t} / N_t = 0.3$) for all of our $40^3\X 20$ ensembles vs.\ the bare lattice coupling $\be_F$.  As the fermion mass decreases from $am = 0.01$ the transitions in $PL_W$ sharpen and move to stronger coupling.  When $am = 0.0025$ the transition merges with the zero-temperature bulk transition into the \Sb phase at $\be_F \approx 4.625$.}
\end{figure}

\begin{figure}[tbp]
  \includegraphics[width=\linewidth]{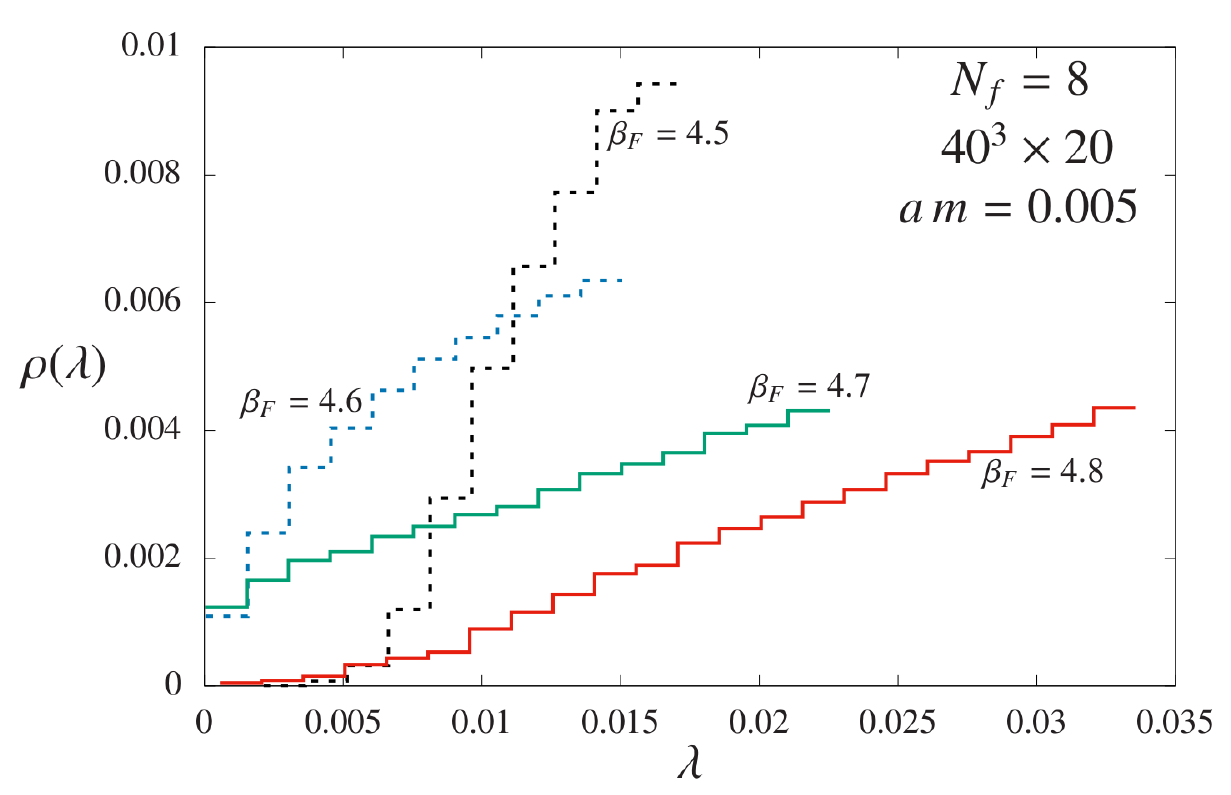}
  \caption{\label{fig:rho} The spectral densities $\rho(\la)$ as histograms of the 200 smallest eigenvalues \la of the massless Dirac operator, for four of the six $40^3\X 20$ ensembles with $am = 0.005$.  The gap at weak coupling $\be_F = 4.8$ indicates a chirally symmetric system, while $\rho(0) > 0$ at $\be_F = 4.7$ implies chiral symmetry breaking.  The dotted results from systems in the \Sb phase ($\be_F = 4.6$ and 4.5) show the onset of a soft edge, $\rho(\lambda) \propto \sqrt{\la - \la_0}$ with $\la_0 > 0$.}
\end{figure}

Of these observables, the most useful are the Polyakov loop and $\rho(\la)$, for which representative results are shown in Figs.~\ref{fig:Wpoly} and \ref{fig:rho}, respectively.
To improve the Polyakov loop signal we compute it after applying the Wilson flow, a continuous transformation that smooths lattice gauge fields to systematically remove short-distance lattice cutoff effects~\cite{Narayanan:2006rf}.
For sufficiently large flow time $t$ the Wilson-flowed Polyakov loop $PL_W$ shows a clear contrast between confined systems with vanishing or small $PL_W \ll 1$ and deconfined systems with large $PL_W \sim 1$, even for large $N_t \geq 20$ which tend to produce noisy results for the unimproved Polyakov loop.
This observable is a modern variant of the RG-blocked Polyakov loop investigated in previous studies, which showed that the improvement in the signal does not affect the location of the transition~\cite{Schaich:2012fr, Hasenfratz:2013uha}.

In \fig{fig:Wpoly} we consider the real part of the Wilson-flowed Polyakov loop computed at flow time $t = 4.5$ for $40^3\X 20$ lattices with $am = 0.01$, 0.005 and 0.0025 vs.\ the bare lattice coupling $\be_F$.
This flow time was chosen to produce $\sqrt{8t} / N_t = 0.3$, which we keep fixed by considering $t = 6.48$ for $N_t = 24$.
As the fermion mass decreases the finite-temperature transitions in \fig{fig:Wpoly} steadily sharpen and move to stronger coupling.
At $am = 0.0025$ the finite-temperature transition merges with the bulk transition into the \Sb phase, implying that even larger volumes are required to establish whether chiral symmetry breaking occurs spontaneously in the massless limit.
Our $N_t = 24$ results behave similarly, as we discuss further below.

First, we review how we determine the bulk transition into the \Sb phase.
\refcite{Cheng:2011ic} identified two order parameters of the single-site staggered shift symmetry, which take the form of differences between local observables on neighboring lattice sites,
\begin{align}
  \De P_{\mu} & = \vev{\ReTr \square_{n, \mu} - \ReTr \square_{n + \muhat, \mu}}_{n_{\mu}\, {\rm even}} \label{eq:plaq_diff} \\
  \De L_{\mu} & = \langle\alpha_{\mu}(n - \muhat) \chibar(n - \muhat) U_{\mu}(n - \muhat) \chi(n)       \label{eq:link_diff} \\
              & \qquad\qquad\quad - \alpha_{\mu}(n)\chibar(n) U_{\mu}(n) \chi(n + \muhat)\rangle_{n_{\mu}\, {\rm even}}.    \nn
\end{align}
Here $\ReTr \square_{n, \mu}$ is the average real trace of the six plaquettes that include the gauge link $U_{\mu}(n)$ connecting sites $n$ and $n + \muhat$, $\chi(n)$ is the staggered fermion field, and $\displaystyle \alpha_{\mu}(n) = (-1)^{\sum_{\nu < \mu} n_\nu}$ is the usual staggered phase factor.
Finally, the expectation value $\vev{\cdots}_{n_{\mu}\, {\rm even}}$ is taken only over sites whose $\mu$ component is even.

In addition, \refcite{Cheng:2011ic} found that the eigenvalue spectrum of the massless Dirac operator exhibits an unusual `soft edge'~\cite{Bowick:1991ky, Akemann:1996vr, Damgaard:2000cx} in the \Sb phase, $\rho(\lambda) \propto \sqrt{\la - \la_0}$ with $\la_0 > 0$.
This allows the spectral density to distinguish between all three phases of interest, as illustrated in \fig{fig:rho} for $40^3\X 20$ ensembles with $am = 0.005$.
At weak couplings, including $\be_F = 4.8$ in the figure, the system is deconfined and chirally symmetric, with $\rho(0) = 0$ and a gap below the smallest eigenvalue.
The gap becomes larger for weaker couplings omitted from the plot.
At intermediate couplings such as $\be_F = 4.7$ we observe the expected chiral symmetry breaking, with $\rho(0) \neq 0$ and a small slope $\frac{d\rho}{d\la}$.
Finally, once we enter the \Sb phase at stronger couplings $\be_F \lsim 4.6$ the spectral density shows the onset of a soft edge with $\rho(\lambda) \sim \sqrt{\la}$.
This is most clear at $\be_F = 4.5$, where a gap has reopened and $\rho(0) = 0$ indicates the restoration of chiral symmetry.

The four curves shown in \fig{fig:rho} therefore indicate that the $N_t = 20$ chiral transition with $am = 0.005$ is between $4.7 < \be_F^{(c)} < 4.8$, consistent with the Wilson-flowed Polyakov loop in \fig{fig:Wpoly}.
In addition, we can see that the transition into the \Sb phase is between $4.6 < \be_F^{(c)} < 4.7$.
While the $\be_F = 4.6$ spectral density does not show a clear gap and may possess a non-zero $\rho(0)$ in the infinite-volume limit, its dependence on \la is consistent with the square-root behavior of the soft edge, in contrast to the curves with $\be_F \geq 4.7$.
Since the \Sb order parameters discussed above indicate that our $\be_F = 4.6$ ensemble is in the \Sb phase, it seems most likely that the lack of a clear gap is related to the non-zero sea mass $am = 0.005$.

\begin{figure}[tbp]
  \includegraphics[width=\linewidth]{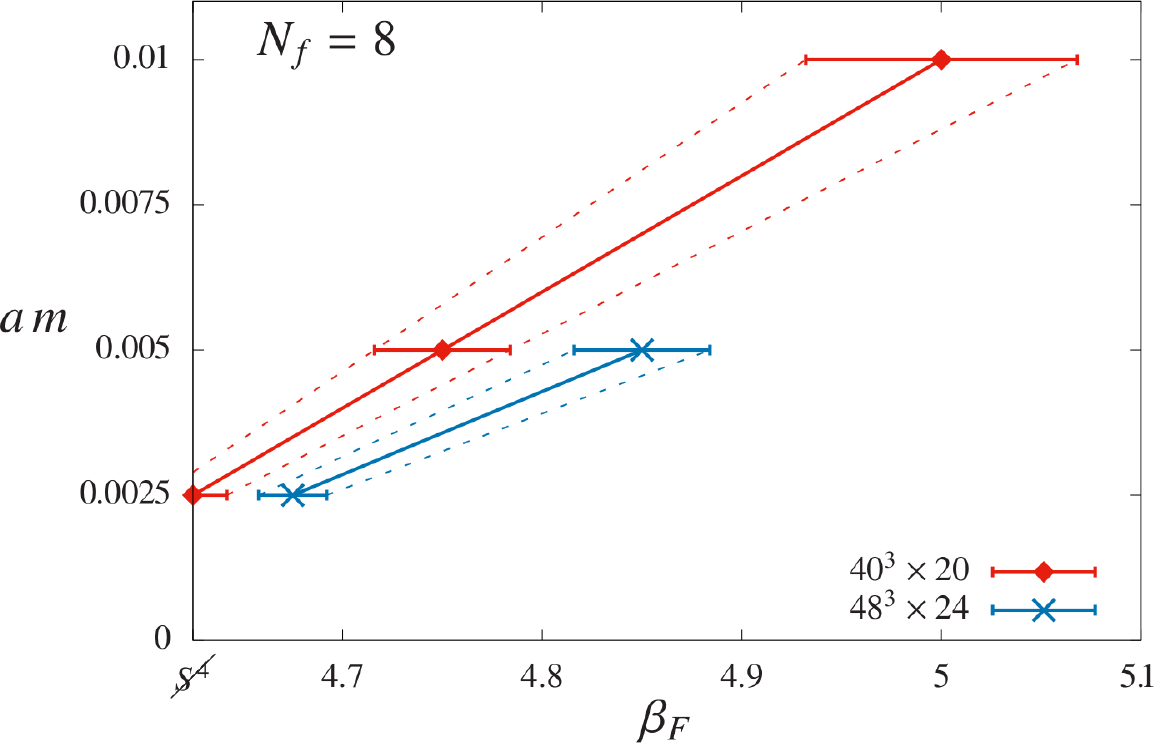}
  \caption{\label{fig:phase_diag} Finite-temperature transitions from lattices with temporal extents $N_t = 20$ and 24, with lines connecting points to guide the eye.  The region above these lines is confined and chirally broken, while the region below is deconfined and chirally symmetric.  The left edge of the plot indicates the bulk transition into the \Sb lattice phase.  The finite-temperature transitions merge with this bulk transition at $am > 0$, preventing a direct confirmation of spontaneous chiral symmetry breaking.}
\end{figure}

Based on the observables described above, we identify the $N_t = 20$ and 24 finite-temperature transitions shown in \fig{fig:phase_diag}.
For $N_t = 20$ and $am = 0.0025$ we find the system at $\be_F = 4.65$ to be in the weak-coupling phase while that at $\be_F = 4.6$ is in the \Sb phase.
Although a finer scan of intermediate values $4.6 < \be_F < 4.65$ might still reveal a narrow chirally broken phase, we conclude that the $N_t = 20$ finite-temperature transitions have effectively merged with the bulk transition by $am \approx 0.0025$.
For the larger $N_t = 24$ we see that the finite-temperature transitions at a fixed mass move to weaker couplings.
While this is the expected behavior for a chirally broken system, it is not sufficient to conclusively establish that the 8-flavor theory exhibits spontaneous chiral symmetry breaking in the massless limit.
Even though the $N_t = 24$ transitions manage to reach slightly smaller $am$ before running into the \Sb phase, it is clear that these transitions will also merge with the bulk transition at a non-zero mass.


Although our new $40^3\X 20$ and $48^3\X 24$ finite-temperature investigations do not suffice to establish spontaneous chiral symmetry breaking, it is still significant that the $am \geq 0.0025$ finite-temperature transitions move to smaller masses as the temporal extent of the lattice increases.
%
In principle it might be possible to construct an alternate lattice action that would allow us to reach stronger couplings before encountering a lattice phase.
Then we would be able to obtain comparable results from smaller lattices.
However, an attempt to do this by adding a second nHYP smearing step was not successful~\cite{Hasenfratz:2014rna}.
We do not currently plan to generate the larger-volume $64^3\X 32$ and $72^3\X 36$ lattice ensembles that would be needed to pursue a more definitive demonstration of spontaneous chiral symmetry breaking using our current action.
While such lattice volumes are within the reach of existing algorithmic and computing technology (\cf the zero-temperature $64^3\X 128$ investigations presented below), they would consume significant resources that we prefer to invest in more promising directions discussed in \secref{sec:conclusion}.

\section{\label{sec:ensembles}Zero-temperature lattice ensembles and scale setting} 
We now focus on our main $N_f = 8$ investigations, which determine the composite spectrum of the theory at zero temperature.
On the basis of the phase diagram discussed above, we carry out our computations at a relatively strong coupling $\be_F = 4.8$ that is still safely on the weak-coupling side of the \Sb phase.
This relatively strong coupling, in combination with large lattice volumes up to $64^3\X 128$, allows us to consider the lightest masses yet to be reached by lattice studies of the 8-flavor theory.
\tab{tab:sim_pars} summarizes the ensembles of gauge configurations that we have generated using the same software and algorithm as described in the previous section.
These include four 4-flavor ensembles, two of which (with lattice volume $24^3\X 48$) are matched in alternate ways to the 8-flavor $24^3\X 48$ ensemble with the largest $am_f = 0.00889$: one (with $\be_F = 6.6$) matches the $\pi$ and $\rho$ masses in lattice units, while the other (with $\be_F = 6.4$) matches the $\pi$ mass and $\sqrt{8t_0}$ in lattice units.
Additional ensembles at matched bare parameters but with smaller lattice volumes, used to study finite-volume effects, are presented in Appendix~\ref{app:disc_FV}.

\begin{table*}[tbp]
  \centering
  \renewcommand\arraystretch{1.2}  
  \addtolength{\tabcolsep}{3 pt}   
  \begin{tabular}{cccS[table-format=1.5]c|S[table-format=5]cS[table-format=4,table-space-text-post=$^{\star}$]S[table-format=3]|S[table-format=1.8]}
    \hline
    $N_f$ & $\be_F$ & $L^3\X N_t$  & $am_f$  & $\tau$ & {MDTU} & Sep. & {\# Est.}      & {Bins} & $\sqrt{8t_0} / a$ \\ 
    \hline
    8     & 4.8     & $64^3\X 128$ & 0.00125 & 0.5    &  4314  &  6   &  720$^{\dag}$  &  72    &                   \\ 
          &         &              &         &        &  4494  &  6   &  750           &  75    &                   \\ 
          &         &              &         &        &  5094  &  6   &  850           &  85    & 4.7345(14)        \\ 
    \hline
    8     & 4.8     & $48^3\X 96$  & 0.00222 & 1.0    & 10640  & 16   &  666           & 111    &                   \\ 
          &         &              &         &        &  8912  & 16   &  558           &  93    & 4.5742(21)        \\ 
    \hline
    8     & 4.8     & $32^3\X 64$  & 0.005   & 2.0    &  5100  & 20   &  256           & 128    &                   \\ 
          &         &              &         &        &  5100  & 20   &  256           & 128    &                   \\ 
          &         &              &         &        &  6380  & 20   &  320           & 160    & 4.1371(33)        \\ 
    \hline
    8     & 4.8     & $32^3\X 64$  & 0.0075  & 1.0    & 24590  & 10   & 2460           & 492    & 3.7820(19)        \\ 
    \hline
    8     & 4.8     & $24^3\X 48$  & 0.00889 & 1.0    & 24500  & 20   & 1226           & 613    & 3.6152(32)        \\ 
    \hline
    4     & 6.4     & $24^3\X 48$  & 0.0125  & 1.0    & 24620  & 20   & 1232           & 616    & 3.7214(13)        \\ 
    \hline
    4     & 6.6     & $48^3\X 96$  & 0.003   & 2.0    & 20680  & 40   &  518$^{\S}$    & 518    & 4.9779(10)        \\ 
    \hline
    4     & 6.6     & $32^3\X 64$  & 0.007   & 2.0    & 15840  & 20   &  792$^{\star}$ & 198    & 4.8478(20)        \\ 
    \hline
    4     & 6.6     & $24^3\X 48$  & 0.015   & 1.0    & 24380  & 20   & 1220           & 610    & 4.6390(22)        \\ 
    \hline
  \end{tabular}
  \caption{\label{tab:sim_pars}For each of our main ensembles we record the number of flavors, bare coupling $\be_F \simeq 12 / g_0^2$, lattice volume, fermion mass $am_f$, and trajectory length $\tau$ in molecular dynamics time units (MDTU).  Working with the stated number of thermalized MDTU we compute estimates of the observables after every `Sep.' MDTU.  The resulting number of estimates are divided into the given number of bins for jackknife and bootstrap analyses, including those producing the Wilson flow scale $\sqrt{8t_0}$ discussed in the text.  Ensembles with $am_f \leq 0.005$ consist of multiple streams, and we combine all bins from every stream in our analyses.  Three of our flavor-singlet scalar spectrum analyses differ slightly: $^{\dag}$For the $64^3\X 128$ ensemble we have only 410 estimates (41 bins) from the first stream along with the full sets of data from the second and third streams.  $^{\S}$For the $N_f = 4$ $48^3\X 96$ ensemble we have 1034 estimates in 517 bins.  $^{\star}$For the $N_f = 4$ $32^3\X 64$ ensemble we have only 198 estimates, one per bin.}
\end{table*}

In addition to recording the lattice volume, fermion mass and available statistics, \tab{tab:sim_pars} also reports values for the reference scale $\sqrt{8t_0}$ introduced in \refcite{Luscher:2010iy}.
This reference scale is defined through the Wilson flow (discussed in the previous section), by requiring that $\left\{t^2\vev{E(t)}\right\}_{t = t_0} = 0.3$, where the energy density
\begin{equation}
  \label{eq:Wflow}
  E(L, t) = -\frac{1}{2} \frac{\mbox{ReTr}\left[F_{\mu\nu}(t) F^{\mu\nu}(t)\right]}{1 + \de(L, t)}
\end{equation}
is evaluated after flow time $t$ using the standard clover-leaf construction of $F_{\mu\nu}$.
The factor of $\de(L, t)$ is the tree-level finite-volume perturbative correction introduced by \refcite{Fodor:2014cpa}, which reduces discretization artifacts in the energy density.
The specific form of this correction depends on the gauge action, on the (Wilson) action used in the gradient flow transformation, and on the (clover) operator used to define $E(t)$.
At tree level our fundamental--adjoint plaquette gauge action is equivalent to the Wilson gauge action, so in the terminology of \refcite{Fodor:2014cpa} we use the WWC scheme.

\begin{figure}[tbp]
  \includegraphics[width=\linewidth]{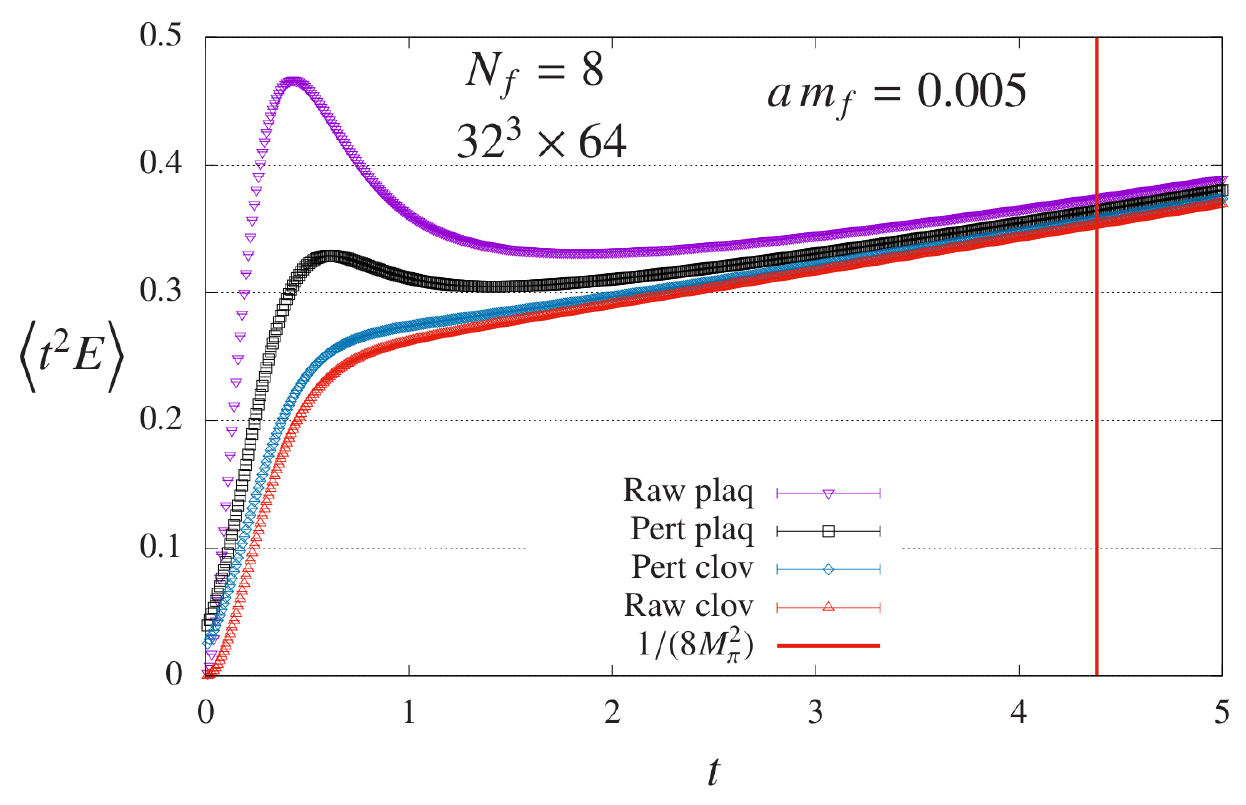}
  \caption{\label{fig:Wflow_pert}Four determinations of $t^2\vev{E(t)}$ (using clover- and plaquette-based definitions of the energy density, with and without perturbative improvement) illustrate discretization artifacts for the $N_f = 8$ $32^3\X 64$ ensemble with $am_f = 0.005$.  These artifacts are most significant for small $t \lsim 1$, and are ameliorated by the corresponding tree-level finite-volume perturbative corrections, which reduce the larger plaquette-based results while increasing the smaller clover-based results.  For future EFT analyses we want $\sqrt{8t_0} < 1 / M_{\pi}$, corresponding to $t$ smaller than the vertical red line at $t = 1 / (8M_{\pi}^2)$.}
\end{figure}

In \fig{fig:Wflow_pert} we confirm that this perturbative correction does indeed reduce discretization artifacts, by comparing our clover-based results against the Wilson plaquette operator $t^2 E_{\text{plaq}} = 12t^2 (3 - \Box)$ where $\Box$ is the trace of the plaquette normalized to 3.
We consider the $N_f = 8$ $32^3\X 64$ ensemble with $m = 0.005$; the other ensembles show similar behavior.
The two lattice definitions of the energy density differ by discretization artifacts, which are most significant for small $t \lsim 1$ where $E_{\text{plaq}}$ can be more than twice as large as the clover-based result.
Appropriately, the perturbative correction reduces the plaquette-based results while increasing the clover-based results, reducing the overall discretization artifacts.
To simplify comparisons with other groups' results we do not include the $t$-shift improvement of \refcite{Cheng:2014jba} in our analyses.

\begin{figure}[tbp]
  \includegraphics[width=\linewidth]{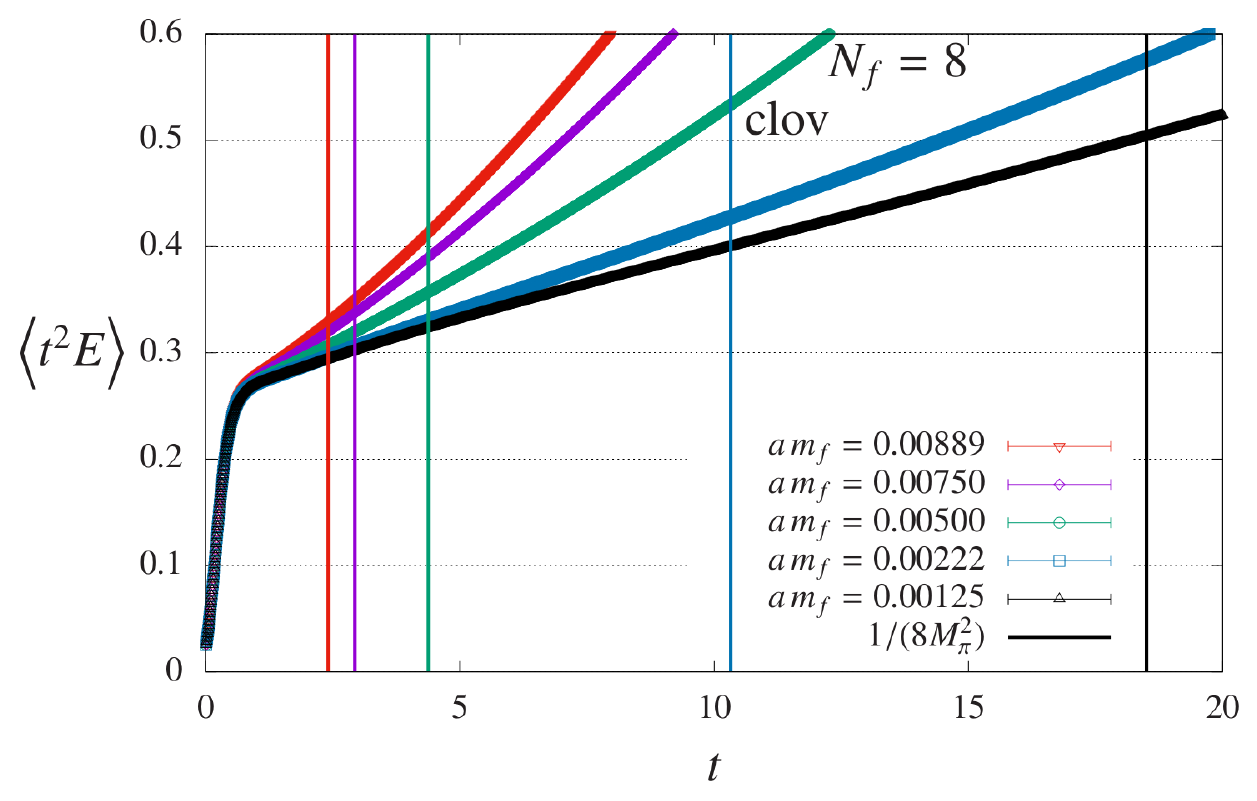}
  \caption{\label{fig:Wflow_tSqE}Perturbatively improved clover-based $t^2\vev{E(t)}$ (\protect\eq{eq:Wflow}) for all 8-flavor ensembles in \protect\tab{tab:sim_pars}, with fermion masses $0.00889 \geq am \geq 0.00125$ from top to bottom.  The vertical lines show the corresponding $t = 1 / (8M_{\pi}^2)$.}
\end{figure}

In addition to quantifying discretization artifacts, \fig{fig:Wflow_pert} also indicates that these artifacts are more severe for $E_{\text{plaq}}$ than for the clover discretization of the energy density.
For sufficiently large $t$ the discretization artifacts are removed by the Wilson flow and $t^2\vev{E(t)}$ becomes approximately linear.
The perturbatively improved clover-based curve is roughly linear already for $t \gsim 1$, where significant non-linearities are still visible in the plaquette-based results.
This motivates our choice of the perturbatively improved clover discretization in \eq{eq:Wflow}.
Figure~\ref{fig:Wflow_tSqE} shows the corresponding results for all 8-flavor ensembles in \tab{tab:sim_pars}.
We note that the condition $\left\{t^2\vev{E(t)}\right\}_{t = t_0} = 0.3$ produces $\sqrt{8t_0} < 1 / M_{\pi}$ for all these ensembles, with the separation growing as the fermion mass decreases.
Investigation of the behavior of the scale $\sqrt{8t_0}$ under different EFT descriptions of this theory, as has been carried out for chiral perturbation theory~\cite{Bar:2013ora}, may allow discrimination between different possible EFTs.

\begin{figure}[tbp]
  \includegraphics[width=\linewidth]{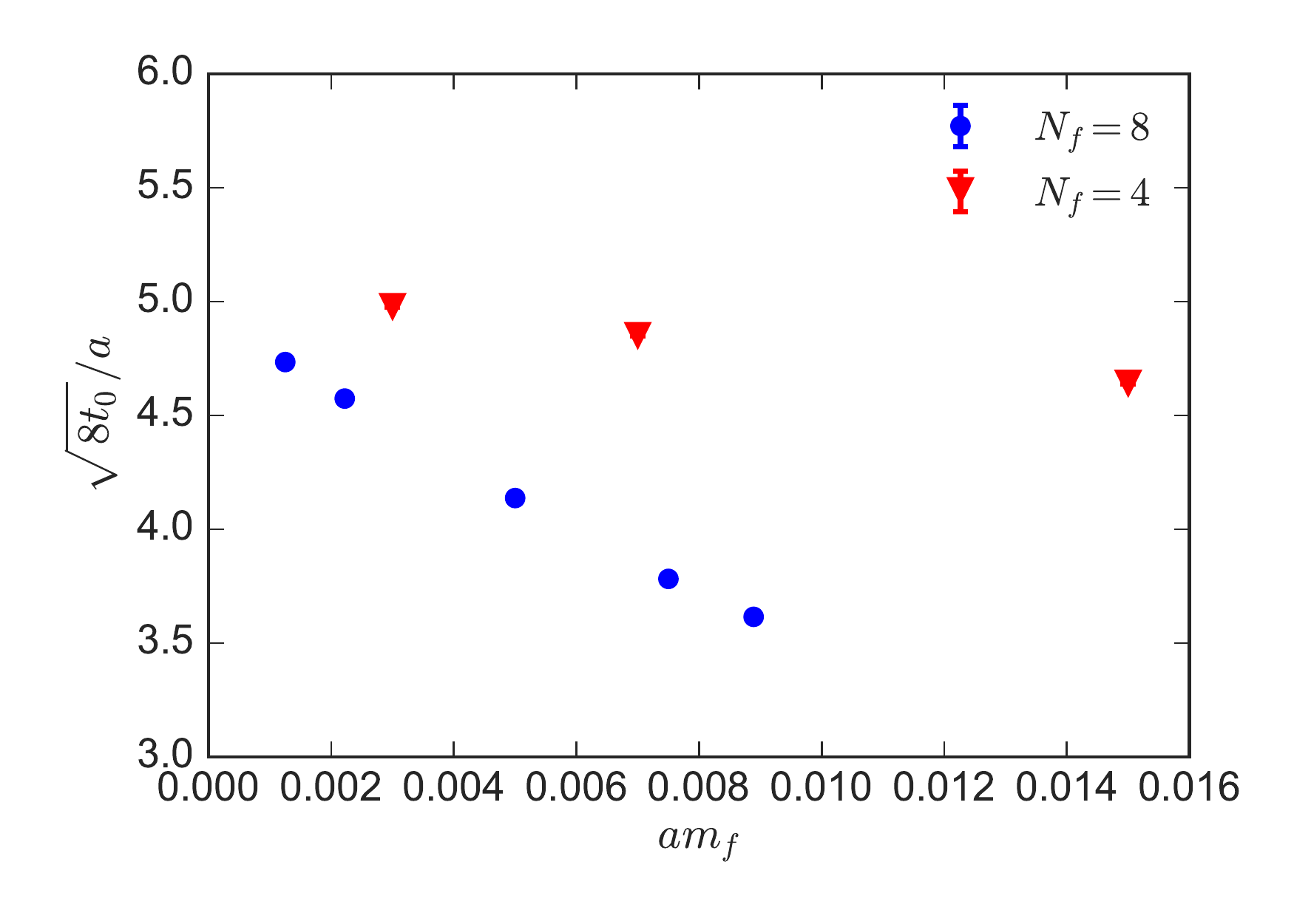}
  \caption{\label{fig:Wflow_scale}The Wilson flow scale $\sqrt{8t_0}$ vs.\ the fermion mass $am_f$ for our 8-flavor (blue circles) and 4-flavor (red triangles) ensembles listed in \protect\tab{tab:sim_pars}.  For $N_f = 4$ we show only the $\be_F = 6.6$ ensembles.  (The $\be_F = 6.4$ ensemble approximately matches the $\sqrt{8t_0}$ of the 8-flavor ensemble with $am_f = 0.00889$.)  The scale shows significantly more sensitivity to the fermion mass in the 8-flavor theory compared to lattice QCD or the 4-flavor theory.}
\end{figure}

Finally, in \fig{fig:Wflow_scale} we plot our results in \tab{tab:sim_pars} for the 4- and 8-flavor Wilson flow scale $\sqrt{8t_0}$ vs.\ the fermion mass $am_f$.
The figure shows that the $N_f = 8$ scale increases rapidly as $am_f$ decreases, with significantly more sensitivity to the fermion mass than is seen for $N_f = 4$ or for lattice QCD (\cf Fig.~4 of \refcite{DeGrand:2017gbi}).

\section{\label{sec:spectrum}Spectrum analysis details} 
In this section we discuss our analysis of hadronic two-point correlation functions to extract estimates for hadron masses and decay constants of the 8-flavor theory.
In the continuum theory the global symmetries after spontaneous symmetry breaking are $\mathrm{SU}(8)_V \times \mathrm{U}(1)_B$.
On the lattice, the staggered discretization further reduces the flavor symmetry to $\mathrm{U}(2) \times \mathrm{SW}_4 \subset \mathrm{SU}(8)_V \times \mathrm{U}(1)_B$~\cite{Kilcup:1986dg}.
The continuum theory has 63 massless Nambu--Goldstone bosons (NGBs) in the chiral limit whereas the staggered theory has 4 massless NGBs, corresponding to the unbroken U(2) subgroup, plus 59 light PNGBs which become massless in the continuum limit.
We use a typical QCD naming scheme for the flavor-nonsinglet mesons, including the NGB pseudoscalar ($\pi$), the vector ($\rho$) and the axial-vector ($a_1$) mesons, to indicate the $J^{PC}$ quantum numbers of the mesons.
In the continuum limit, these mesons will transform as the adjoint representation of SU(8)$_V$ but at finite lattice spacing they will transform under representations of the $\mathrm{U}(2) \times \mathrm{SW}_4$ staggered flavor group.
For three states we compute both the mass ($M_{\pi_5}$, $M_{\rho_s}$, $M_{a_{1,5}}$) and the decay constants ($F_{\pi}$, $F_{\rho}$, $F_{a_1}$).
We use the decay constant normalization corresponding to the QCD value $F_{\pi} = 92.2(1)$~MeV (\cf section 5.1.1 of \refcite{Aoki:2016frl}).
We also compute the masses of other flavor-nonsinglet mesons, several baryons including the lightest nucleon ($M_N$), and the flavor-singlet scalar meson ($M_{\si}$).
A complete list of computed masses is given in \tab{tab:spec_list8}.
The flavor-singlet \si state is the most challenging state to study because it mixes with the vacuum and receives contributions from fermion-line-disconnected diagrams.

\begin{table*}[tbp]
  \centering
  \renewcommand\arraystretch{1.2}  
  \addtolength{\tabcolsep}{3 pt}   
  \begin{tabular}{cc|cc|cl}
    \hline
    Direct State   & Spin--Taste         & Oscillating State & Spin--Taste         & Source   & Comments                        \\
    \hline
    $\pi_5$        & $\ga_5 \xi_5$       & ---               & ---                 & C,P,W    & NGB, for $F_{\pi}$              \\
    $\pi_{45}$     & $\ga_{45} \xi_{45}$ & $a_{0, s}$        & 1                   & W        &                                 \\
    $\pi_{i5}$     & $\ga_5 \xi_{i5}$    & ---               & ---                 & W        &                                 \\
    $\pi_{ij}$     & $\ga_{45} \xi_{ij}$ & $a_{0, i}$        & $\xi_i$             & W        &                                 \\
    \hline
    $\rho_s$       & $\ga_i$             & $b_{1, 45}$       & $\ga_{jk} \xi_{45}$ & C,W      & For $F_{\rho}$                  \\
    $\rho_4$       & $\ga_{i4} \xi_4$    & $a_{1, 5}$        & $\ga_{i5} \xi_5$    & C,W      & For $F_{a_1}$                   \\
    $\rho_{i5}$    & $\ga_{j4} \xi_{i5}$ & $a_{1, i4}$       & $\ga_{j5} \xi_{i5}$ & W        &                                 \\
    $\rho_{ij}$    & $\ga_k \xi_{ij}$    & $b_{1, k}$        & $\ga_{ij} \xi_k$    & W        &                                 \\
    \hline
    $N$            & ---                 & \La               & ---                 & W        & Corresponds to $\mathbf{8}$ rep \\
    \hline
    \si [$0^{++}$] & 1                   & $\eta_{45}$       & $\ga_{45}\xi_{45}$  & $0^{++}$ & Flavor-singlet meson            \\
    \hline
  \end{tabular}
  \caption{\label{tab:spec_list8}List of the states we analyze on our 8-flavor lattice ensembles.  Each row indicates a type of hadronic correlator and the columns indicate the spin and flavor (taste) quantum numbers of the states in the direct (non-oscillating) and oscillating channels, \cf~\eq{eq:corr_model}.  The source labels are ``$W$'' for wall sources, ``$C$'' for conserved currents, and ``$P$'' for the point pseudoscalar operator; the $0^{++}$ analysis uses a combination of regular and stochastic sources as described in the text.  In the last column we also identify the states we used to extract the decay constants.}
\end{table*}

The use of staggered fermions requires some discussion of the staggered flavor (or taste) quantum number.
A complete analysis of the staggered meson correlators constructed from fermion bilinear interpolating operators requires considering operators distributed, at a minimum, over the $2^4$ unit hypercube.
In this paper we only consider the operators listed in \tab{tab:spec_list8}, all of which are at most one-link separated.
Below we will discuss the spin and taste structures of our interpolating operators in more detail.

We split the discussion of our spectrum analyses into three parts.
In the next subsection we consider the computation and analysis of the connected spectrum, including decay constants.
We then separately discuss the additional disconnected observables needed for the computation of the flavor-singlet scalar \si correlator, and the corresponding analysis.
Finally, we explain our procedure for numerical analysis of the correlators including determination of systematic errors associated with the choice of fit range $[t_{\text{min}}, t_{\text{max}}]$, which is common to all of the spectroscopy.
Although we have also analyzed gluonic operators that couple to the scalar channel, these produce much noisier data from which we could not extract meaningful results, so we will not discuss them further.

\subsection{Connected spectrum} 
To extract the mass of the $\pi$ (both PNGB and taste-split), $\rho$, $a_1$ and $N$, we use Coulomb gauge-fixed wall sources~\cite{Gupta:1990mr}.
To extract the decay constants $F_{\pi}$, $F_{\rho}$ and $F_{a_1}$, we additionally use the conserved current interpolating operators $A_4$, $V_i$ and $A_i$, respectively, where $i$ indexes spatial directions.
Special to the PNGB $\pi_5$, we additionally use the point pseudoscalar operator, $P$, and a random Gaussian wall source thereof.
In all cases we use point operators at the sink.
We compute all two-point functions using $N_t / 8$ sources distributed evenly in Euclidean time, averaging over those data to obtain a single estimate in each channel from each analyzed configuration.

As summarized in \tab{tab:sim_pars}, we bin all estimates of observables to reduce the effects of auto-correlations, setting the bin size to be of the same order as the auto-correlation time of the chiral condensate $\pbp$, which we determine with \texttt{UWerr}~\cite{Wolff:2003sm}.
In Appendix~A we provide more information about auto-correlations.

In general, a meson two-point function where both the source and sink interpolating operators have the same quantum numbers is parameterized as
\begin{widetext}
\begin{align}
  \label{eq:corr_model}
  C_X(t) = & \Tr{\vev{\sum_{\vec x} \chibar(\vec x + \vec \de', t) \tau^a \Ga(\vec \de') \chi(\vec x, t) \chibar(\vec \de, 0) \tau^a \Ga(\vec \de) \chi(\vec 0, 0)}} \cr
         = & \sum_{i = 0}^{N_{\text{exc}}} \left[A_{X, i} \left(e^{-E_{X, i} t} \pm e^{-E_{X, i} (N_t - t)}\right) + (-1)^t A_{X, i}' \left(e^{-E_{X, i}' t} \pm e^{-E_{X, i}' (N_t - t)}\right)\right],
\end{align}
\end{widetext}
where $\Ga(\vec \de)$ refers to the spin-taste structure of the interpolating operator in channel $X$, and $\tau^a$ is an appropriate generator of the flavor symmetry (since we are computing flavor-nonsinglet states).
In the staggered formulation a meson operator with quantum numbers \Ga will also couple to a ``parity partner'' state with quantum numbers $\Ga \ga_4 \ga_5 \xi_4 \xi_5$~\cite{Golterman:1985dz}.
A single interpolating operator will also couple to multiple states with the same quantum numbers.
We parameterize the direct-state energies and amplitudes as $E_{X, i}$ and $A_{X, i}$, respectively, with oscillating-state $E_{X, i}'$ and $A_{X, i}'$.
In both cases $0 \leq i \leq N_{\text{exc}}$ for $N_{\text{exc}}$ excited states in addition to the $i = 0$ ground states; our central analysis sets $N_{\text{exc}} = 1$ for all states.
With this in mind, $E_{X, 0}'$ refers to the ground-state energy of the parity partner state in a given channel.
This contribution is non-zero except for the Goldstone channel, where the would-be parity partner has the quantum numbers of the staggered vector charge.
The sign of the term $e^{-E_{X, i} (N_t - t)}$ is positive for all correlators except $\vev{A_4(0) P(t)}$, where it is negative due to odd time-reversal symmetry.
Since meson correlators are symmetric or anti-symmetric about the middle timeslice $N_t / 2$, the data are `folded' by averaging $C(t)$ with $C(N_t / 2 - t)$ or $-C(N_t / 2 - t)$, respectively.

Staggered baryons have a more complicated structure, and similarly more complicated interpretation of parity partners.
\refcite{Golterman:1984dn} discusses the parameterization of baryon two-point functions.
For baryon correlators we do not include the backwards-propagating $e^{E_{X,i}(N_t - t)}$ states and fit to $C(t)$ only for $t \leq 3N_t / 8$.
The signal-to-noise in our correlators is poorest for $t > 3N_t / 8$, so we do not expect this restriction to significantly reduce the precision of our results.

For the pseudoscalar, vector, and axial-vector channels, we carry out joint fits to two correlators, one containing the appropriate conserved-current source and the other using a wall source with the same quantum numbers.
This joint fit improves the precision of the ground-state energy and thus our determinations of the mass and decay constant in each channel.
The particular operators used for the determination of the decay constants are noted in \tab{tab:spec_list8}.

We define the continuum decay constants via
\begin{align}
  \ME{0}{A_4}{\pi}        & = \sqrt{2} F_{\pi} M_{\pi} \label{eq:Fpi} \\
  \ME{0}{V_i}{\rho^{(i)}} & = \sqrt{2} F_{\rho} M_{\rho} \eps^{(i)}   \\
  \ME{0}{A_i}{a_1^{(i)}}  & = \sqrt{2} F_{a_1} M_{a_1} \eps^{(i)} \label{eq:FA}
\end{align}
for the pseudoscalar, vector, and axial-vector channels, respectively.
Here $\eps_i$, with $i = 1, 2, 3$, are polarization vectors.
Because we use conserved staggered currents no renormalization factors appear (i.e., the $Z$-factors are exactly 1).

We can also define $F_{\pi}$ from the Goldstone pseudoscalar interpolating operator~\cite{Kilcup:1986dg}
\begin{equation}
  \ME{0}{P}{\pi} = \sqrt{2} F_{\pi} M_{\pi}^2 / m_f,
\end{equation}
where $m_f$ is the fermion mass.
Since staggered fermions preserve an exact chiral symmetry this fermion mass receives no additive renormalization.

Since each staggered species corresponds to multiple continuum Dirac fermions, there is a non-trivial conversion factor between the continuum matrix elements and the lattice matrix elements~\cite{Kilcup:1986dg}:
\begin{equation}
  \ME{0}{\cO}{s} \llra \frac{1}{\sqrt 4} \ME{0}{\cO_{lat}}{s}_{\text{lat}},
\end{equation}
where $1 / \sqrt{4}$ comes from the four continuum flavors per staggered fermion, $s$ refers to one of the meson states $\pi$, $\rho$, $a_1$, and \cO is the corresponding interpolating operator.

\subsection{\label{ssec:spectrum_disc}Disconnected spectrum} 
Unlike the connected spectrum, estimating the mass of the flavor-singlet $0^{++}$ (also denoted \si here) requires computing fermion-line-disconnected diagrams.
We only use one interpolating operator for the $0^{++}$ meson, the zero-momentum local operator
\begin{equation}
  \Ozpp(t) = \sum_{\vec x} \chibar(\vec x, t) \chi(\vec x, t),
\end{equation}
which has a spin--taste-singlet, as well as flavor-singlet, structure.
Because the operator is a flavor singlet, there are two ways to contract the staggered fermion fields, which differ by a minus sign because of anticommutation relations, corresponding to two different estimates of the staggered Dirac propagator $G_F$:
\begin{widetext}
  \begin{align}
    S(t) = & \vev{\Ozpp(0) \Ozpp(t)} = \vev{\sum_{\vec x} \overbracket{\chibar(\vec x, 0)\chi}(\vec x, 0) \sum_{\vec y}\overbracket{\chibar(\vec y, t) \chi}(\vec y, t) - \sum_{\vec x} \overbracket{\chibar(\vec x, 0) \underbracket{\chi(\vec x, 0) \sum_{\vec y}\chibar}(\vec y, t) \chi}(\vec y, t)} \cr
         = & \vev{\frac{N_f}{4} \sum_{\vec x} \Tr{G_F(\vec x, 0; \vec x, 0)} \sum_{\vec y} \Tr{G_F(\vec y, t; \vec y, t)} - \sum_{\vec x} \sum_{\vec y} \Tr{G_F(\vec x, 0; \vec y, t) G_F(\vec y, t; \vec x, 0)}}                                                                                        \\
         = & \sum_{\vec x, \vec y} \left\{\frac{N_f}{4} \vev{\Tr{G_F(\vec x, 0; \vec x, 0)} \Tr{G_F(\vec y, t; \vec y, t)}} - \vev{\Tr{G_F(\vec x, 0; \vec y, t) G_F(\vec y, t; \vec x, 0)}}\right\} \equiv \frac{N_f}{4} D(t) - C(t).                                                                   \nn
  \end{align}
\end{widetext}

The two types of contractions split the computation of the $0^{++}$ correlator into two pieces: a disconnected (double-trace) piece $D(t)$ and a connected (single-trace) piece $-C(t)$.
The factor of $N_f / 4$ is a relative loop counting factor between the two pieces, where the factor of four corresponds to the four continuum Dirac flavors encoded by each staggered lattice field.

Exactly computing the disconnected piece requires evaluating all diagonal elements of the inverse of the Dirac matrix.
This is computationally impractical.
Instead of explicitly computing the diagonal of the inverse, we use an improved stochastic trace estimator with dilution, which we now review step by step.

First, we consider a stochastic estimate for any element of the inverse of a complex matrix $S(x, y)$, where $x$ and $y$ index rows and columns, respectively.
We can construct a set of noise vectors, $\eta_i(x)$, satisfying
\begin{equation}
  \label{eq:stochestim}
  \lim_{N_i \to \infty} \frac{1}{N_i} \sum_{i = 1}^{N_i} \eta_i(x) \eta_i^{\dag}(y) = \Ibb(x, y),
\end{equation}
where $\Ibb(x, y)$ is the identity matrix.
Some common choices for $\eta_i(x)$ are U(1), $\Zbb_2$, $\Zbb_4$, or an appropriately scaled normal distribution.
In this paper we always use either U(1) or $\Zbb_4$ noise.
We use $N_i = 2$ or 3 noise sources for $L = 64$ and $N_i = 6$ for all $L < 64$.
For each $\eta_i(x)$ we define the vector
\begin{equation}
  \phi_i(x) = S^{-1}(x, y) \eta_i(y).
\end{equation}
A stochastic estimate of the inverse matrix $S^{-1}(x, y)$ is given by
\begin{widetext}
\begin{align}
  S^{-1}(x, y) & = \lim_{N_i \to \infty} \frac{1}{N_i} \sum_{i = 1}^{N_i} \phi_i(x) \eta_i^{\dag}(y) = \lim_{N_i \to \infty} \frac{1}{N_i} \sum_{i = 1}^{N_i} S^{-1}(x, z) \eta_i(z) \eta_i^{\dag}(y) \cr
               & = S^{-1}(x, z) \left[\lim_{N_i \to \infty} \frac{1}{N_i} \sum_{i = 1}^{N_i} \eta_i(z) \eta_i^{\dag}(y)\right] = S^{-1}(x, z) \Ibb(z, y).
\end{align}
For finite $N_i$ the error of this approximation decreases $\propto 1 / \sqrt{N_i}$ as expected.

In our context $S(x, y)$ is the staggered Dirac matrix.
The inverse of the Dirac matrix is the Green's function $G_F(\vec x, t; \vec y, t')$, where we suppress color indices.
The trace that enters the disconnected (double-trace) piece of the \Ozpp correlator can be estimated by
\begin{align}
  \label{eq:tracestim}
  \lim_{N_i \to \infty} \frac{1}{N_i} \sum_{\vec x} \sum_{i = 1}^{N_i} \phi_i(\vec x, t) \eta_i^{\dag}(\vec x, t) & = \lim_{N_i \to \infty} \frac{1}{N_i} \sum_{\vec x} \sum_{i = 1}^{N_i} G_F(\vec x, t; \vec y, t') \eta_i(\vec y, t') \eta_i^{\dag}(\vec x, t)              \cr
                                                                                                                  & = \sum_{\vec x} G_F(\vec x, t; \vec y, t') \left[\lim_{N_i \to \infty} \frac{1}{N_i} \sum_{i = 1}^{N_i} \eta_i(\vec y, t') \eta_i^{\dag}(\vec x, t)\right] \cr
                                                                                                                  & = \sum_{\vec x} G_F(\vec x, t; \vec y, t') \Ibb(\vec y, t'; \vec x, t) = \sum_{\vec x} G_F(\vec x, t; \vec x, t) = \Ozpp(t).
\end{align}

When forming the double-trace correlator, we need to avoid taking the product of two estimates from the same source.
More explicitly, the double-trace correlator must be computed as
\begin{equation}
  \label{eq:doubletrace}
  D(t) = \frac{N_f}{4} \lim_{N_i \to \infty} \frac{1}{N_i (N_i - 1)} \sum_{i = 1}^{N_i} \sum_{j = 1, j \neq i}^{N_i} \left[\sum_{\vec x} \phi_i(\vec x, t) \eta_i^{\dag}(\vec x, t)\right] \left[\sum_{\vec y} \phi_j(\vec y, 0) \eta_j^{\dag}(\vec y, 0)\right],
\end{equation}
\end{widetext}
where we impose $j \neq i$ to avoid quadratic noise contributions.
This is modified, and to some extent relaxed, in the context of dilution.

Away from the infinite-source limit, stochastic sources can suffer from noisy couplings to other states.
In particular, stochastic estimates of single-trace correlation functions can be sensitive to states with lighter masses.
This is solved by dilution, where instead of computing the propagator from a full-volume noise source, we first partition the source into separate pieces and then compute propagators for each individual piece~\cite{Foley:2005ac}.
The ultimate effect of this partitioning is that some off-diagonal components of the sum in \eq{eq:stochestim} are exactly zero even with finite $N_i$.
In practice, appropriate use of dilution improves convergence: instead of the $\cO(1 / \sqrt{N_i})$ convergence of stochastic methods, dilution can give an effective $\cO(1 / N_i)$ convergence for an equal amount of computational work.

We take advantage of dilution in time, color, as well as even/odd in space to individually compute the single-trace pieces of the disconnected correlator.
We can then construct the disconnected correlator, $D(t)$, by computing the all-to-all correlator of these single-trace operators.
Dilution in time allows us to relax the constraint $j \neq i$ in \eq{eq:doubletrace} for $t \neq 0$: the contributions from $t \neq 0$ come from different noise sources because the diluted $\eta$ sources are non-zero on different timeslices.
The issue at $t = 0$ can be avoided by imposing $j \neq i$ uniquely for that separation.
The construction of the disconnected correlator can be efficiently done in Fourier space via the convolution theorem.
Because the $0^{++}$ channel has the same quantum numbers as the vacuum, the disconnected piece suffers from a vacuum contamination that must be removed, as we discuss below.

Instead of using \eq{eq:tracestim} for $\Ozpp(t)$ we employ the improved stochastic estimator
\begin{equation}
  \label{eq:chiralward}
  G_F(x_{\mu}; x_{\mu}) = m \sum_{z_{\mu}} |G_F(x_{\mu}; z_{\mu})|^2,
\end{equation}
which follows from the Ward identity for the staggered chiral symmetry~\cite{Kilcup:1986dg}.
With this definition, the stochastic estimate for $\Ozpp(t)$ can be written as
\begin{equation}
  \Ozpp(t) = m \lim_{N_i \to \infty} \frac{1}{N_i} \sum_{i = 1}^{N_i} \sum_{\vec x} \phi_i(\vec x, t) \phi_i^{\dag}(\vec x, t).
\end{equation}
This is an improved estimator because it involves only positive-definite contributions (corresponding to the positive Goldstone propagator in \eq{eq:chiralward}).

The connected piece $-C(t)$ is much simpler to compute.
While point sources could be used for this computation, in this work we reuse the stochastic propagators discussed above, simply contracting them in a different way as described by \refcite{Foley:2005ac}.
Whereas the full correlator $S(t)$ couples to the $0^{++}$ meson as its lightest single-particle state, the positive-definite correlator $-C(t)$ by itself instead couples to the $a_0$ meson.

Neglecting excited states and scattering states, we parameterize $S(t)$ and $-C(t)$ by
\begin{align}
   S(t) & = a_{\text{vac}} + a_d \cosh\left[M_{\si}(N_t / 2 - t)\right] \label{eq:Scorr} \\
        & \qquad + b_d (-1)^t \cosh\left[M_{\eta_{45}}(N_t / 2 - t)\right]               \cr
  -C(t) & = a_c \cosh\left[M_{a_0}(N_t / 2 - t)\right]                                   \\
        & \qquad + b_c (-1)^t \cosh\left[M_{\pi_{45}}(N_t / 2 - t)\right]. \nn
\end{align}
The staggered oscillating partner of the $0^{++}$ meson is a pseudoscalar which is a singlet under the exact U(2) flavor subgroup but a non-singlet under the SW$_4$ staggered flavor group, so we label it $\eta_{45}$ (\cf \tab{tab:spec_list8}).

On the other hand, $D(t)$ is not a well-defined correlator in the sense that it does not uniquely couple to one set of quantum numbers.
Numerically, we can still parameterize it as a linear combination of $S(t)$ and $-C(t)$,
\begin{align}
  D(t) & = \frac{4}{N_f} \left\{S(t) - (-C(t)) \right\}                                    \cr
       & = a_{\text{vac}} + a_d \cosh[M_{\si} t'] - a_c \cosh[M_{a_0} t'] \label{eq:Dcorr} \\
       & \quad + (-1)^t \left\{b_d \cosh[M_{\eta_{45}} t'] - b_c \cosh[M_{\pi_{45}} t']\right\}, \nn
\end{align}
with $t' \equiv N_t / 2 - t$.

\begin{figure}[tbp]
  \includegraphics[width=\linewidth]{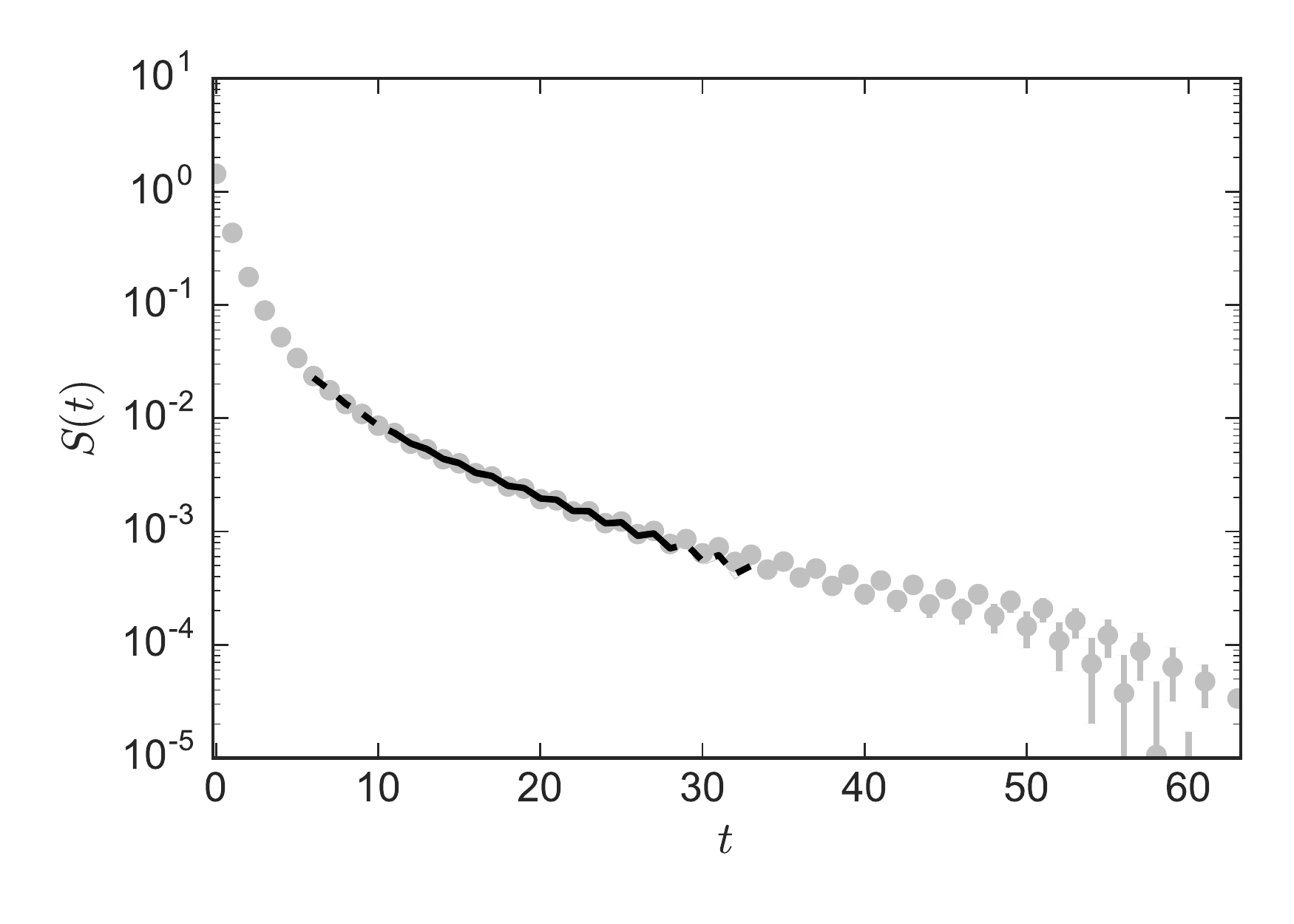} \\
  \includegraphics[width=\linewidth]{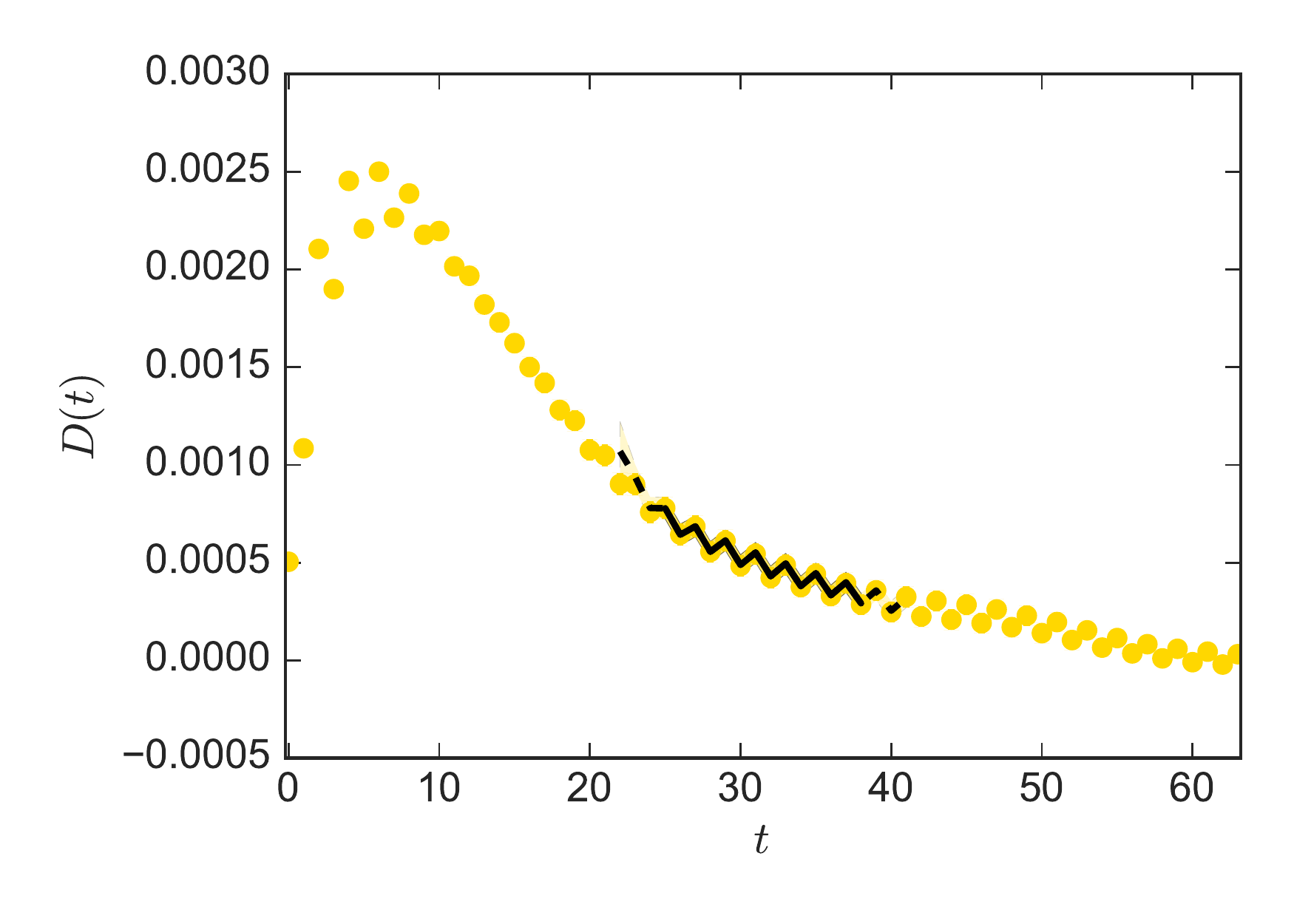}
  \caption{\label{fig:scalar_corrs}Comparison of $D(t)$ and $S(t)$ on a representative ensemble with $am_f = 0.00125$.  The negative contribution of excited states at small $t$ in $D(t)$ is manifest.  The solid line shows the best-fit prediction from individual fits to each correlator over the range of $t$ values used in the fit, while the dashed line shows the best-fit prediction extended beyond the data used in the fit.  The colored band shows the 1\si error on the fit prediction.}
\end{figure}

For $M_{0^{++}} < M_{a_0}$, we can extract the mass of the $0^{++}$ from $D(t)$ alone for asymptotically large $t$~\cite{Aoki:2014oha}.
This correlator appears to have smaller excited-state contamination than $S(t)$, leading to longer plateaus in effective-mass plots.
However, consistent with the construction of $D(t)$ as a difference between correlators, the excited states are not positive definite, and cause the correlator to curve down instead of up at small $t$, as shown in \fig{fig:scalar_corrs}.

In practice, instead of fitting $S(t)$ or $D(t)$ directly to the functional forms shown above, we fit to the finite difference $S(t + 1) - S(t)$.
This has the advantage that the vacuum contribution $a_{\text{vac}}$ is removed entirely before we attempt any nonlinear fits.
Even with the finite difference, it is difficult to extract the ground state from the individual correlators $S(t)$ and $-C(t)$: both channels suffer from strong excited-state contamination, especially because, in the infinite stochastic source limit, we are effectively using the point, or local, source and sink operators.
Instead, we carry out a joint fit to $S(t)$ and $D(t)$ simultaneously.
The opposite sign of the excited-state contamination in $D(t)$ helps to counterbalance the significant positive contamination in $S(t)$, leading to a more robust joint estimate of $M_{0^{++}}$.

In Appendix~B we give further numerical details of our extraction of $M_{0^{++}}$, including comparisons between our main joint-fit-based analysis and individual fits to the $D(t)$ and $S(t)$ finite-difference correlators.

\subsection{\label{ssec:spectrum_syst}Fit selection and fit-range systematic error} 
To determine the masses and decay constants, we carry out fully correlated fits over a fit range $[t_{\text{min}}, t_{\text{max}}]$ for all of our correlators, subject to the models described above.
Those restrictions allow for more reliable determination of the correlator covariance matrix with finite statistics, and also suppress contributions from higher excited states at very small $t$.
We make use of the \texttt{lsqfit} Python package~\cite{Lepage:lsqfit} for our fits.
All results presented here include one excited state in each of the oscillating and non-oscillating channels, i.e., we include four states in total to describe any correlator with an oscillating parity-partner contribution.
We include priors on the energies of all states for numerical stability only: the prior values are $aE = 1(5)$ for ground states, and $\log(\De E) = -1(4)$ for the splitting between each excited state and its respective ground state.

For many lattice studies, it is common practice to fix $[t_{\text{min}}, t_{\text{max}}]$ based on some empirical analysis of the correlators.
We select our central fits based on an empirical fit quality criterion~\cite{Bitar:1990cb},
\begin{equation}
  \label{eq:criterion}
  C \equiv \frac{p \times N_{\text{dof}}}{\sum_n (\si_n / \mu_n)^2}.
\end{equation}
Here $p$ is the unconstrained $p$-value of the fit, $N_{\text{dof}}$ is the number of degrees of freedom in the fit, and $\si_n / \mu_n$ is the best-fit uncertainty on the $n$th fit parameter normalized by its mean value.
We include only the ground-state parameters in the sum in the denominator.
The central value and statistical uncertainty for each quantity is taken from the best fit, determined to be the fit with the maximum value of $C$.

The choice of a specific fit window in principle introduces a systematic effect into the analysis if the resulting masses or decay constants have some residual dependence on $[t_{\text{min}}, t_{\text{max}}]$.
To account for this possibility, we assign an additional fit-range systematic error~\cite{Ayyar:2017qdf}.
Over all fits considered for a particular correlator, we compare the central result (mass or decay constant) as obtained for all fits passing a minimal fit quality cut, in our case $p \geq 0.1$, and passing the other cuts described in the next paragraph.
We further discard all fits which agree with the central result at 1$\si$, as well as fits which are consistent with zero at the 3\si level, to avoid unnecessarily inflating our errors due to outlier fits with large uncertainties.
The fit-range systematic is then taken to be half of the maximum difference between central values of fits passing all cuts.

To carry out the analysis described above, we consider all possible combinations of $[t_{\text{min}}, t_{\text{max}}]$ subject to the following constraints.
First, we require $t_{\text{max}} - t_{\text{min}} < 2\sqrt{N_s}$, where $N_s$ is the number of independent Monte Carlo samples; this cut is placed to avoid including fits where the data covariance matrix is ill-determined by our statistics.
We also require that $N_{\text{dof}} > 0$, which imposes a minimum separation $t_{\text{max}} - t_{\text{min}}$.
To exclude data with poor signal-to-noise which could inflate the overall error, we require $t_{\text{max}} \leq 3N_t / 8$, and to avoid a regime where the ground state is poorly determined we require $t_{\text{max}} \geq N_t / 4$.
Finally, we fix $t_{\text{min}} \geq 4$, as the data at smaller $t$ suffer from substantial excited-state contamination.

\begin{figure}[tbp]
  \includegraphics[width=\linewidth]{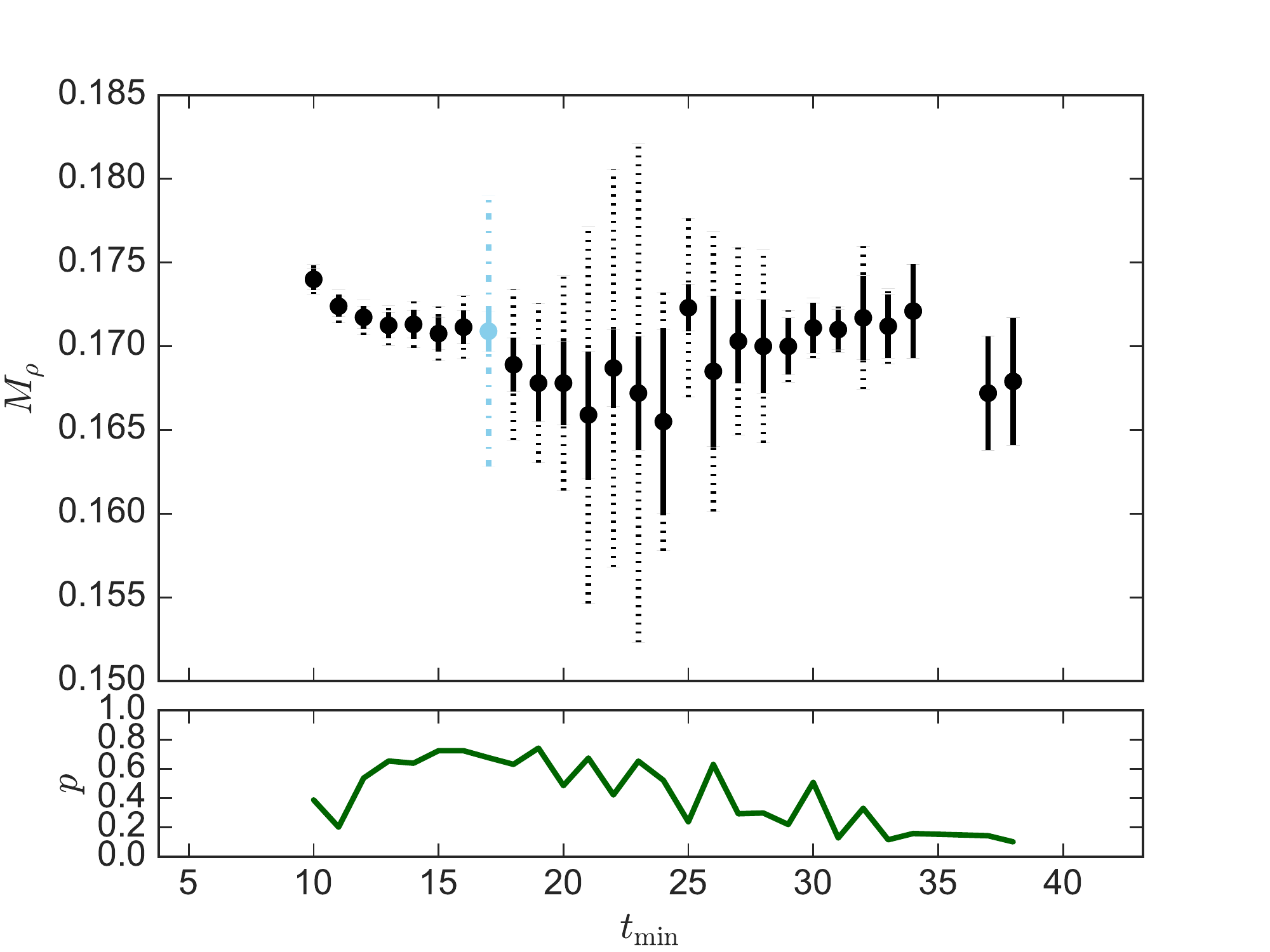}
  \caption{\label{fig:fitrange}Fit-range dependence of the vector meson mass on a representative ensemble with $am_f = 0.00222$, showing results at each $t_{\text{min}}$ satisfying the condition $p \geq 0.1$.  Dashed error bars on each point show the systematic variation of the central result with $t_{\text{max}}$ at a given $t_{\text{min}}$.  The blue point at $t = 17$ shows the best-fit point as determined by the fit quality criterion~(\ref{eq:criterion}); the dash-dotted error bar on this point shows the overall fit-range systematic error, determined as described in the text.}
\end{figure}

A representative plot showing the determination of this fit-range systematic error is shown in \fig{fig:fitrange}.
For each fixed $t_{\text{min}}$, the point with solid error bars represents the best fit, while the dashed error bars, if any, represent the fit-range systematic determined by varying $t_{\text{max}}$ at that particular value of $t_{\text{min}}$.
The colorized point shows the best fit as determined from the fit quality criterion~(\ref{eq:criterion}), and the dashed line on that point represents the total fit-range systematic from consideration of all $[t_{\text{min}}, t_{\text{max}}]$.

\section{\label{sec:results}Spectrum results and discussion} 
\begin{table*}[tbp]
  \centering
  \renewcommand\arraystretch{1.2}  
  \addtolength{\tabcolsep}{3 pt}   
  \begin{tabular}{ccS[table-format=1.5]|S[table-format=1.9]S[table-format=1.7]S[table-format=1.8]S[table-format=1.7]S[table-format=1.7]}
    \hline
    $N_f$ & $L^3\X N_t$  & $am_f$  & $M_{\pi}\sqrt{8t_0}$ & $M_{\si}\sqrt{8t_0}$ & $M_{\rho}\sqrt{8t_0}$ & $M_{a_1}\sqrt{8t_0}$ & $M_N \sqrt{8t_0}$ \\
    \hline
    8     & $64^3\X 128$ & 0.00125 & 0.3885(30)           & 0.42(15)             & 0.809(31)             & 1.07(21)             & 1.08(14)          \\
    8     & $48^3\X 96$  & 0.00222 & 0.5036(55)           & 0.599(91)            & 1.005(17)             & 1.364(13)            & 1.43(11)          \\
    8     & $32^3\X 64$  & 0.005   & 0.6988(99)           & 0.753(99)            & 1.251(26)             & 1.707(30)            & 1.813(29)         \\
    8     & $32^3\X 64$  & 0.0075  & 0.7798(25)           & 0.973(25)            & 1.3979(29)            & 1.914(20)            & 1.85(28)          \\
    8     & $24^3\X 48$  & 0.00889 & 0.824(16)            & 1.01(11)             & 1.4797(76)            & 1.997(32)            & 2.101(24)         \\
    \hline
    4     & $24^3\X 48$  & 0.0125  & 0.8518(31)           & 1.716(48)            & \textemdash           & 2.46(24)             & 2.34(15)          \\
    4     & $48^3\X 96$  & 0.003   & 0.48965(55)          & 1.03(33)             & 1.461(14)             & 2.03(16)             & 1.921(70)         \\
    4     & $32^3\X 64$  & 0.007   & 0.7346(33)           & 1.33(26)             & 1.662(16)             & 2.43(15)             & 2.298(68)         \\
    4     & $24^3\X 48$  & 0.015   & 1.044(11)            & 1.61(12)             & 1.9315(29)            & 2.61(25)             & 2.649(22)         \\
    \hline
  \end{tabular}
  \caption{\label{tab:spec_results}Results for masses from each of our ensembles.  All uncertainties include both statistical and fit-range systematic error.}
\end{table*}
\begin{table*}[tbp]
  \centering
  \renewcommand\arraystretch{1.2}  
  \addtolength{\tabcolsep}{3 pt}   
  \begin{tabular}{ccS[table-format=1.5]|S[table-format=1.9]S[table-format=1.8]S[table-format=1.8]S[table-format=1.8]S[table-format=1.7]}
    \hline
    $N_f$ & $L^3\X N_t$  & $am_f$  & $F_{\pi}\sqrt{8t_0}$ & $F_{\rho}\sqrt{8t_0}$ & $F_{a_1}\sqrt{8t_0}$ & $F_{\rho} / F_{\pi}$ & $g_{\rho \pi \pi}$ \\
    \hline
    8     & $64^3\X 128$ & 0.00125 & 0.10052(66)          & 0.143(14)             & 0.111(43)            & 1.43(14)             & 5.69(22)           \\
    8     & $48^3\X 96$  & 0.00222 & 0.12524(78)          & 0.170(25)             & 0.178(50)            & 1.36(20)             & 5.67(10)           \\
    8     & $32^3\X 64$  & 0.005   & 0.1639(11)           & 0.206(27)             & 0.199(17)            & 1.25(17)             & 5.40(12)           \\
    8     & $32^3\X 64$  & 0.0075  & 0.18150(72)          & 0.2383(49)            & 0.237(35)            & 1.313(28)            & 5.446(24)          \\
    8     & $24^3\X 48$  & 0.00889 & 0.1918(13)           & 0.195(54)             & 0.297(23)            & 1.02(28)             & 5.455(46)          \\
    \hline
    4     & $24^3\X 48$  & 0.0125  & 0.22168(71)          & 0.3213(17)            & 0.3029(71)           & 1.4492(91)           & \textemdash        \\
    4     & $48^3\X 96$  & 0.003   & 0.17847(95)          & 0.277(11)             & 0.277(15)            & 1.553(62)            & 5.788(63)          \\
    4     & $32^3\X 64$  & 0.007   & 0.2100(11)           & 0.3229(82)            & \textemdash          & 1.538(40)            & 5.602(58)          \\
    4     & $24^3\X 48$  & 0.015   & 0.23793(84)          & 0.343(13)             & 0.328(43)            & 1.441(57)            & 5.740(22)          \\
    \hline
  \end{tabular}
  \caption{\label{tab:f_results}Results for decay constants from each of our ensembles, as well as the quantity $F_{\rho} / F_{\pi}$ which is predicted to be equal to $\sqrt{2}$ by the KSRF relations and $g_{\rho \pi \pi}$ as inferred from KSRF.  All uncertainties include both statistical and fit-range systematic error.}
\end{table*}

\begin{figure}[tbp]
  \includegraphics[width=\linewidth]{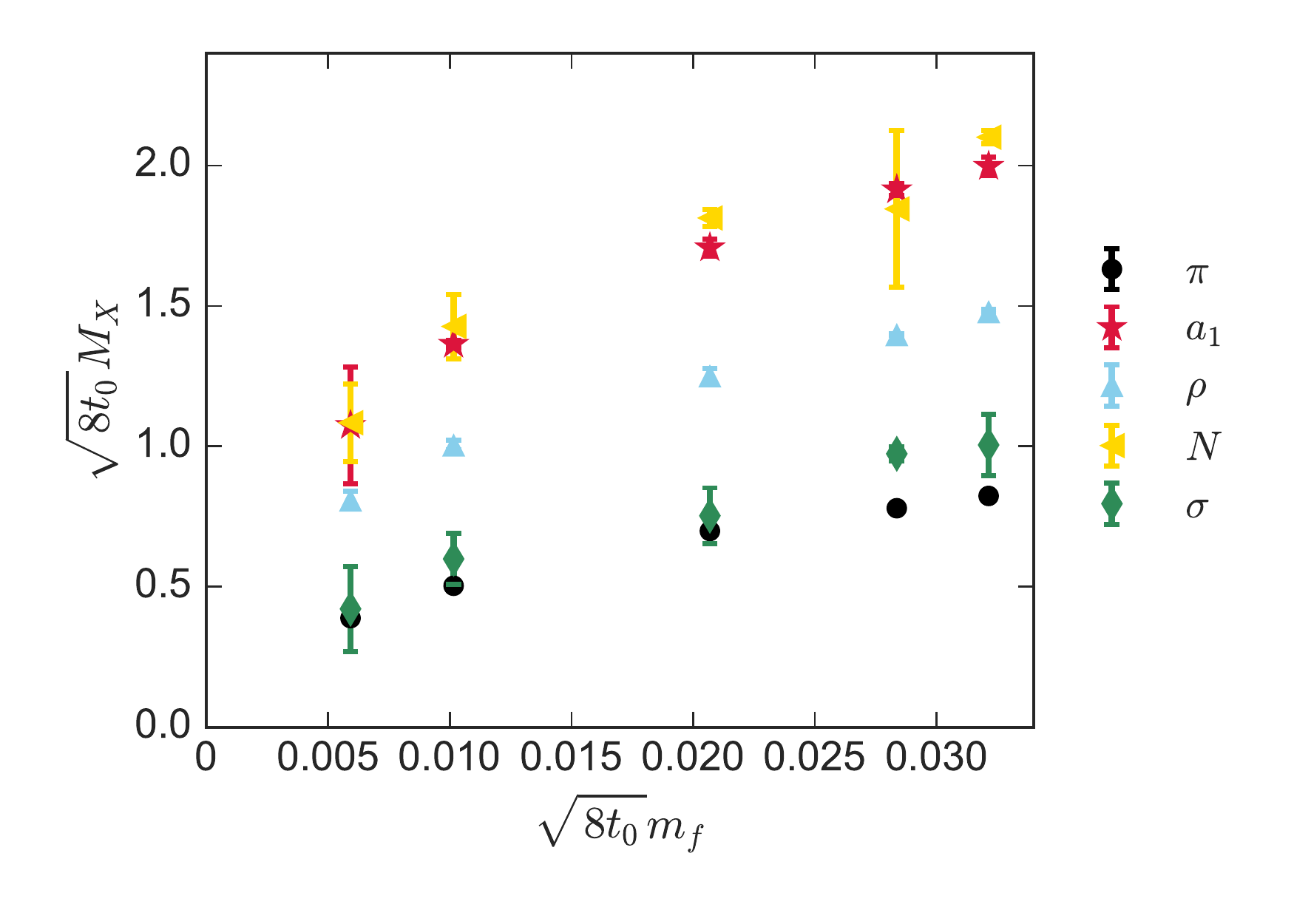}
  \caption{\label{fig:spectrum}The spectrum of the 8-flavor theory, in units of the Wilson flow scale $\sqrt{8t_0}$.  A clear separation exists on all ensembles between the light $\pi$ and \si states and the $\rho$ vector meson and other heavier states.}
\end{figure}

\begin{figure}[tbp]
  \includegraphics[width=\linewidth]{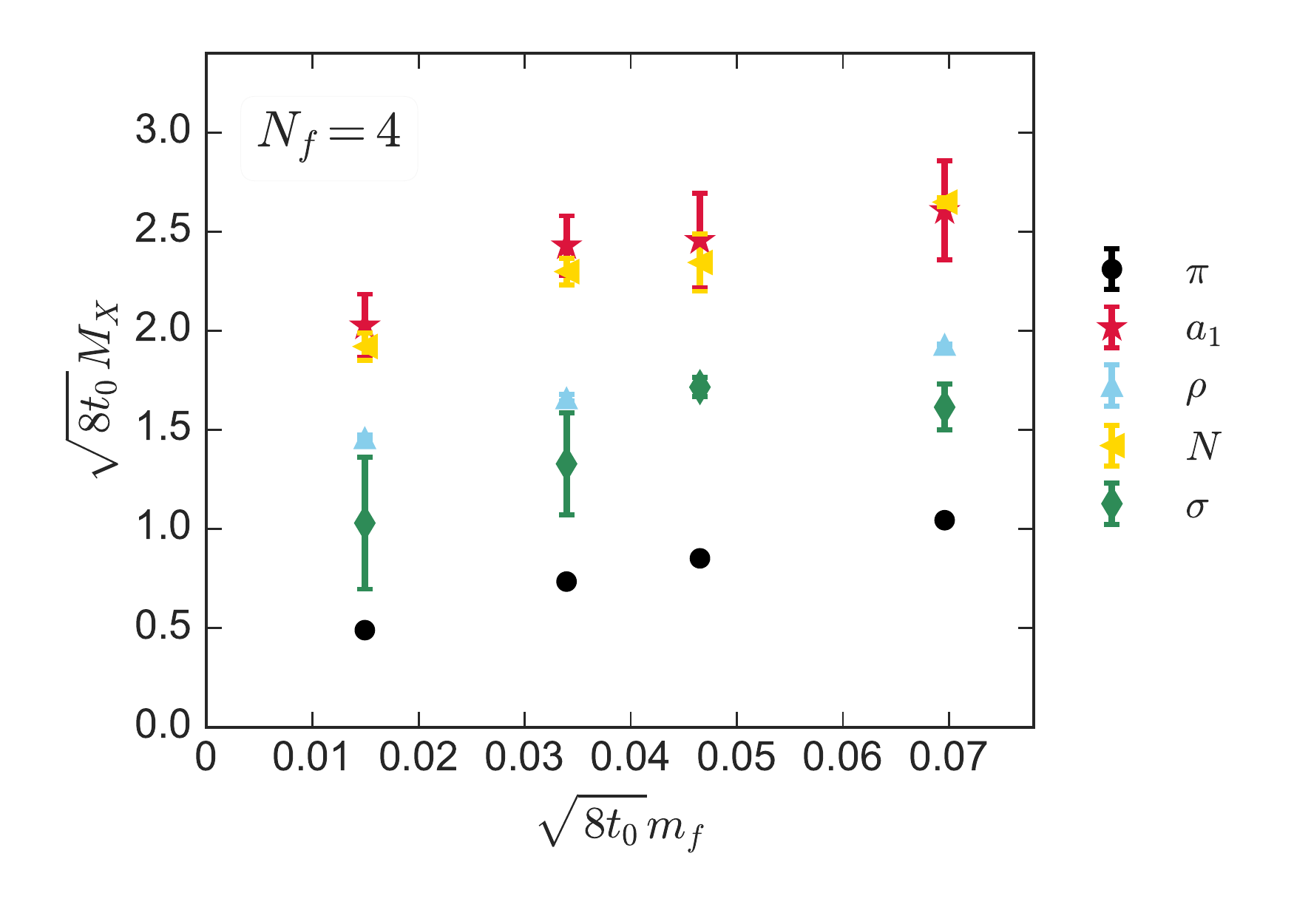}
  \caption{\label{fig:spectrum_4f}The spectrum of the 4-flavor theory, in units of the Wilson flow scale $\sqrt{8t_0}$.  This spectrum is qualitatively similar to what we calculate for the 8-flavor theory, with the exception of the \si state, which here is significantly heavier than the $\pi$ and nearly degenerate with the $\rho$.}
\end{figure}

Our results for the spectrum of light bound states in the 8-flavor and 4-flavor theories are collected in Tables~\ref{tab:spec_results} and \ref{tab:f_results}, and plotted in Figs.~\ref{fig:spectrum_4f8f}, \ref{fig:spectrum} and \ref{fig:spectrum_4f}.
The uncertainties shown for the results include both statistical and fit-range systematic errors, combined in quadrature.
No systematic uncertainty for finite volume or lattice discretization is included.
We believe that the former effect is negligible for our results, but discretization effects are difficult to estimate since we work at a single value of the bare coupling $\be_F$.
Appendix~\ref{app:disc_FV} contains a more detailed study of possible finite-volume and discretization effects.

The most striking feature apparent in the $N_f = 8$ spectrum is the relative lightness of the flavor-singlet scalar state, $\si$.
On most ensembles it is degenerate with the pion within our uncertainty; in all cases, in the range of fermion masses we study both \si and $\pi$ are much lighter than the next heaviest hadron, which is the $\rho$ meson.
This result stands in contrast with QCD, where phenomenological estimates place the $f_0$ flavor-singlet scalar between 400--550~MeV~\cite{Patrignani:2016xqp}, above the threshold for two-pion decay.
Lattice QCD calculations with heavier-than-physical pion masses show that the $f_0$ state remains heavy relative to the $\pi$ states, eventually becoming heavier than $M_{\rho}$~\cite{Kunihiro:2003yj, Howarth:2015caa, Briceno:2016mjc}.

Our study of the 4-flavor theory provides a direct test for systematic effects in our analysis of the \si scalar that might give an artificially small value for its mass.
The spectrum results for $N_f = 4$ shown in \fig{fig:spectrum_4f} indicate that the \si mass in this case is qualitatively heavier, closer to the $\rho$ meson than to the pions.
We surmise that the appearance of a light \si scalar meson is not an artifact of our analysis procedure, and the additional fermion species in the $N_f = 8$ theory seem to be important for \si to be light.

\begin{figure}[tbp]
  \includegraphics[width=\linewidth]{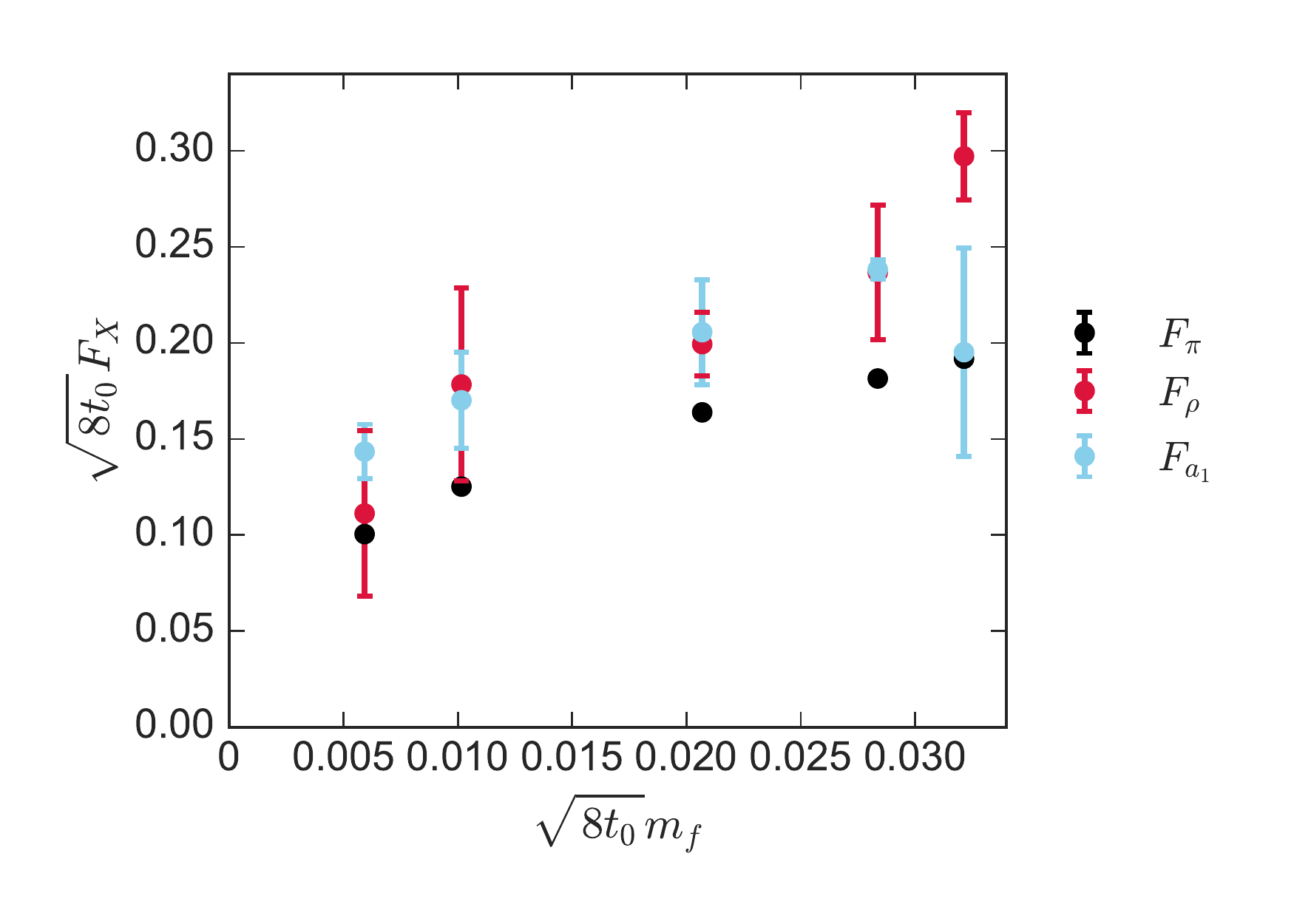}
  \caption{\label{fig:decay}Decay constants of the 8-flavor theory, defined in Eqs.~(\ref{eq:Fpi})--(\ref{eq:FA}).  Both the vector and axial-vector decay constants are fairly imprecise, suffering from significant fit-range systematic effects in our analysis.}
\end{figure}

Decay constants for the 8-flavor theory are plotted in \fig{fig:decay}.
The decay constants $F_{\rho}$ and $F_{a_1}$ suffer from significant fit-range systematic errors in our analysis.
We can observe qualitatively that they tend to be larger than $F_{\pi}$ and roughly degenerate with each other.

Here we do not attempt to extrapolate our results to the chiral limit $am_f \to 0$, as this requires the choice of a specific chiral EFT to model the data, and the presence of the light \si state indicates that a careful study of possible EFT descriptions is needed.
We defer such a study to future work, and here present some ratios of quantities at finite $am_f$ that may provide useful insights.

\begin{figure}[tbp]
  \includegraphics[width=\linewidth]{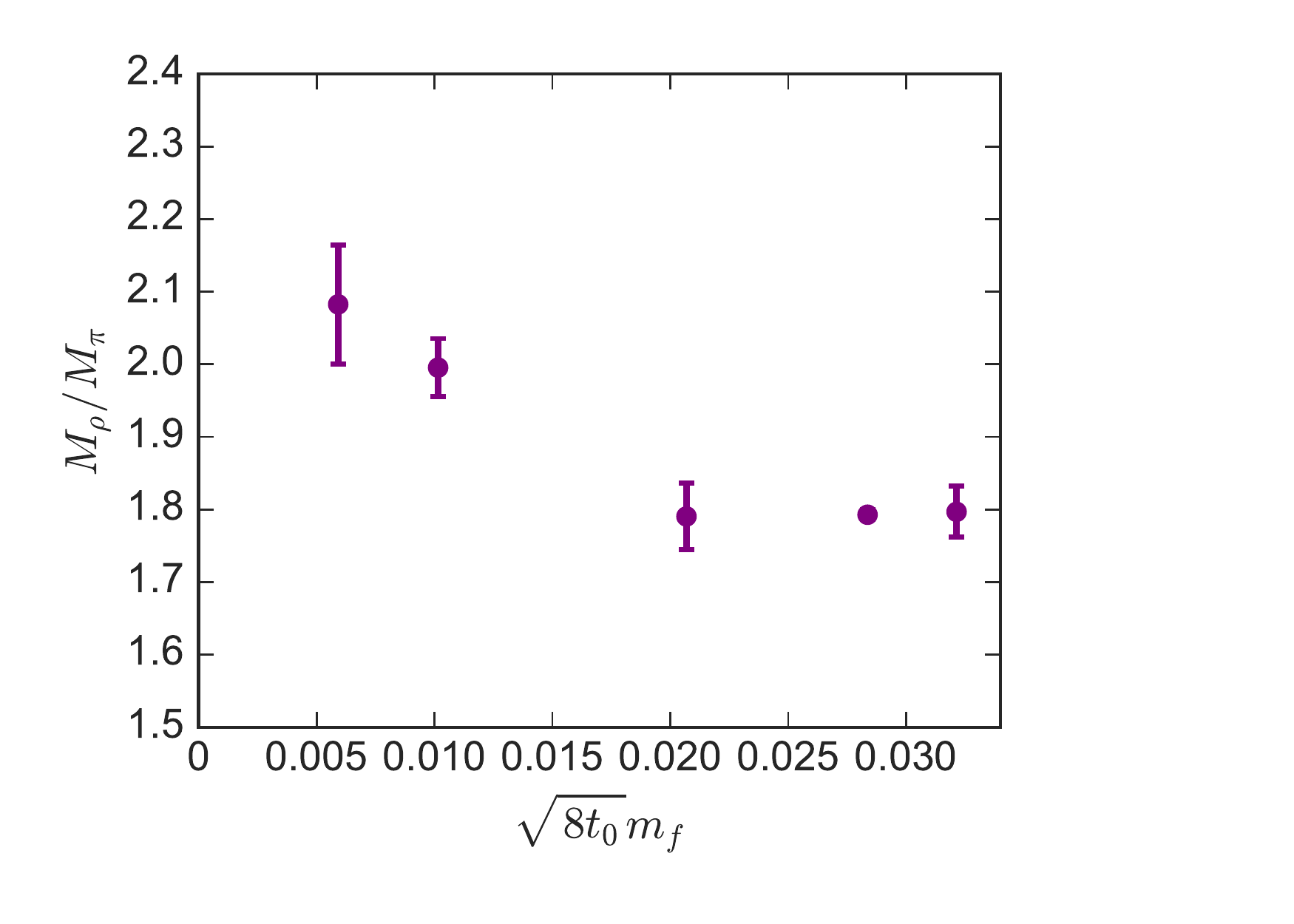}
  \caption{\label{fig:mV_over_mP}The ratio of masses $M_{\rho} / M_{\pi}$; this quantity is expected to be constant for an infrared-conformal theory exhibiting hyperscaling, while for a theory with spontaneous chiral symmetry breaking it should diverge in the $am_f \to 0$ limit as $M_{\pi} \to 0$ and $M_{\rho}$ approaches a finite value.}
\end{figure}

We first present the ratio of meson masses $M_{\rho} / M_{\pi}$ in \fig{fig:mV_over_mP}.
This ratio should diverge in the $am_f \to 0$ limit for a theory with spontaneous chiral symmetry breaking, as the pion mass vanishes while the $\rho$ mass remains finite.
On the other hand, in either an infrared-conformal theory whose masses exhibit hyperscaling~\cite{DelDebbio:2010ze} or in a theory with very heavy fermion masses which exceed the binding energy, this ratio should remain roughly constant.
Our results in the fermion mass range studied show growth of this ratio with decreasing $am_f$, but no clear indication of divergence.
This effect has been argued~\cite{daSilva:2015vna} to be due to finite lattice volume, which we discuss in Appendix~\ref{app:disc_FV}.

\begin{figure}[tbp]
  \includegraphics[width=\linewidth]{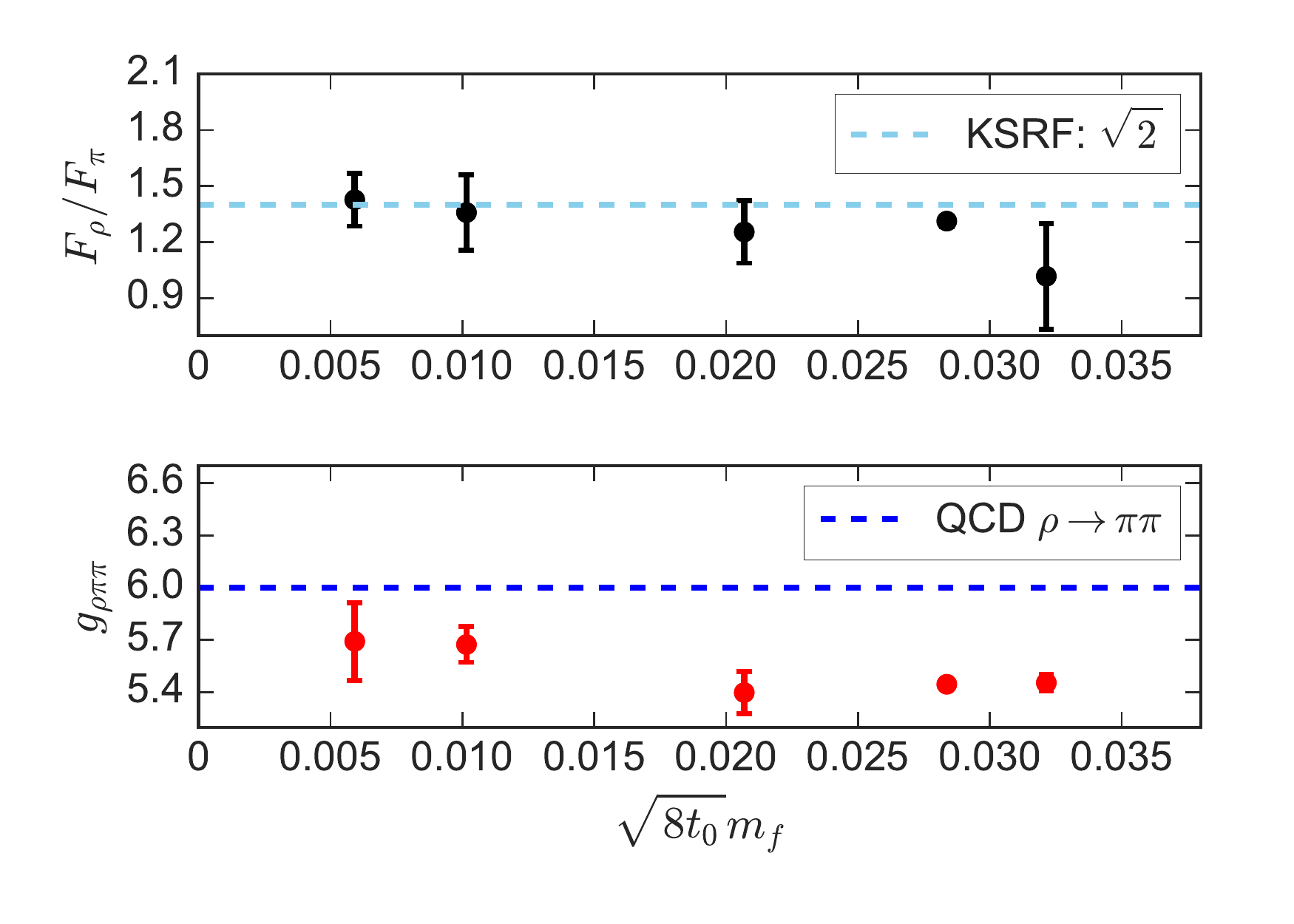}
  \caption{\label{fig:ksrf}Upper panel: $F_{\rho} / F_{\pi}$, predicted to be equal to $\sqrt{2}$ by the KSRF relations.  Lower panel: assuming the validity of the KSRF relations, the coupling $g_{\rho \pi \pi}$ extracted from $M_{\rho}$ and $F_{\pi}$ as described in the text.}
\end{figure}

The KSRF relations~\cite{Kawarabayashi:1966kd, Riazuddin:1966sw} are a pair of equations relating the vector and PNGB masses and decay constants, based on current algebra and vector-meson dominance.
In our conventions, they read
\begin{align}
  \label{eq:KSRF}
  F_{\rho} & = \sqrt{2} F_{\pi} &
  g_{\rho \pi \pi} & = \frac{M_{\rho}}{\sqrt{2} F_{\pi}}.
\end{align}
These relations, and their implications for the strong decay process $\rho \to \pi \pi$, were discussed previously in \refcite{Appelquist:2016viq} in more detail.
Figure~\ref{fig:ksrf} shows our numerical results for the KSRF relations.
The top panel shows $F_{\rho} / F_{\pi}$, which is again found to be relatively independent of $am_f$ and consistent with the KSRF prediction of $\sqrt{2}$.
The bottom panel assumes the validity of the second KSRF relation and estimates the coupling $g_{\rho \pi \pi}$, which is again relatively mass-independent and qualitatively similar to the value inferred from QCD experiment.
This implies a broad decay width $\Ga / M \sim 0.2$ for a vector resonance in any model of electroweak symmetry breaking built on this theory, given the assumptions discussed in \refcite{Appelquist:2016viq}.

\section{\label{sec:conclusion}Conclusions} 
The presence of a light, unflavored stable scalar meson \si in the spectrum of the SU(3) $N_f = 8$ theory which is approximately degenerate with the PNGBs $\pi$ when $M_{\pi} / M_{\rho} < 0.5$ is perhaps the most dramatic difference between this theory and QCD where $M_{\pi} / M_{\rho} \approx 0.2$ and the lightest $f_0$ meson is a broad, unstable resonance well above decay threshold~\cite{Patrignani:2016xqp}.
Our result remains consistent with studies of the same theory reported by the LatKMI collaboration~\cite{Aoki:2016wnc} at heavier fermion masses.
There is growing evidence from QCD that for heavier PNGB masses $M_{\pi} / M_{\rho} \gsim 0.46$ the $f_0$ may become stable with a mass just below the decay threshold~\cite{Wilson:2015dqa, Briceno:2016mjc, Briceno:2017qmb}, but this is still very different from the SU(3) gauge theory with larger number of flavors or other near-conformal theories with light scalars~\cite{Aoki:2013zsa, Aoki:2014oha, Athenodorou:2014eua, Fodor:2015vwa, Rinaldi:2015axa, Brower:2015owo, Hasenfratz:2016gut, DelDebbio:2015byq, Appelquist:2016viq, Aoki:2016wnc, Gasbarro:2017fmi, Athenodorou:2017dbf}.

However, there are other significant differences between $N_f = 8$ and QCD which have been noted in the past~\cite{Appelquist:2014zsa} and remain present in this work.
Of particular interest is the relatively steep slope of the pion decay constant $F_{\pi}$ vs.\ $am_f$ in the light-fermion-mass regime (\cf \fig{fig:decay}).
The leading-order prediction from chiral perturbation theory is that $F_{\pi}$ should be independent of the fermion mass, so that significant $am_f$ dependence indicates the presence of large higher-order contributions; past studies of SU(3) theories with $N_f > 2$ have noted the rapid growth of NLO contributions with $N_f$~\cite{Appelquist:2009ka, Neil:2010sc, Fleming:2013tra}.
The poor convergence of $\chi$PT signaled by the large size of these higher-order contributions may be related to the existence of a light scalar which has been omitted from the EFT.
It would be interesting to explore this possibility, and the influence of the light scalar's presence on other physical quantities of interest such as the chiral condensate.

On the other hand, we also find some qualitative similarities between the $N_f = 8$ theory and QCD.
In particular, the ratios of $\rho$ and $\pi$ masses and decay constants appearing in the KSRF relations (\fig{fig:ksrf}) show only mild mass dependence and indicate a vector meson with properties quite similar to that of the QCD $\rho$, including a large strong decay width.
This is an important qualitative feature which should be accounted for in dedicated LHC searches for heavy vector resonances associated with composite Higgs models built on theories such as this one.

Another set of differences between $N_f = 8$ and QCD has been long predicted under the name \textit{parity doubling}.
The expectation is that the $\rho$ and $a_1$ meson masses will become approximately degenerate as they are parity partners.
Similarly the $a_0$ meson is expected to become substantially lighter as a parity partner of the $\pi$~\cite{Kurachi:2006ej, Appelquist:2010xv}.
In the range of masses we study, we find no evidence for parity doubling between $\rho$ and $a_1$.
This is in contrast to previously reported results for $N_f = 6$ and 8~\cite{Appelquist:2014zsa}, where degeneracy between the $\rho$ and $a_1$ masses appeared to set in at light fermion masses.
We have not reported any results for the mass of the $a_0$ state, due to concerns about large systematic effects in our attempts to determine it; such a study is deferred to future work.

Ultimately, a deeper understanding of these results requires investigation of candidate low-energy EFTs appropriate for describing the observed spectrum with a light \si state.
Several proposals for such EFTs have been made~\cite{Soto:2011ap, Matsuzaki:2013eva, Golterman:2016lsd, Hansen:2016fri, Appelquist:2017wcg, Appelquist:2017vyy, Meurice:2017jry, Meurice:2017zng, Gasbarro:2017ccf, Golterman:2018mfm, DeFloor:2018xrp, Appelquist:2018tyt}; distinguishing between these proposals requires investigation of quantities beyond the spectrum of light states, both on the EFT and lattice sides.
Although not strictly an EFT, comparison to holographic predictions for the spectrum~\cite{Erdmenger:2014fxa} would be interesting to explore as well.
We have initiated new studies of PNGBs $2\to 2$ elastic scattering~\cite{Gasbarro:2017fmi} and form factors, with the goal of exploring which of these EFTs provide the best description of the 8-flavor theory at low energies.

\subsection*{Acknowledgments} 
We thank Kieran Holland for useful discussions.
We thank the Lawrence Livermore National Laboratory (LLNL) Multiprogrammatic and Institutional Computing program for Grand Challenge allocations and time on the LLNL BlueGene/Q supercomputer.
We also thank Argonne National Laboratory (ANL) for allocations through the ALCC and INCITE programs.
Computations for this work were carried out in part on facilities of the USQCD Collaboration, which are funded by the Office of Science of the U.S.~Department of Energy (DOE).
Additional numerical analyses were carried out on clusters at LLNL and on computers at the Massachusetts Green High Performance Computing Center, in part funded by the U.S.~National Science Foundation (NSF), and on computers allocated under the NSF XSEDE program to the project TG-PHY120002.
Part of this work was performed at the Aspen Center for Physics (R.C.B., G.T.F., A.H., C.R.\ and D.S.) supported by NSF grant 1066293, and at the Kavli Institute for Theoretical Physics (R.C.B., G.T.F., A.H., E.T.N., E.R., D.S.\ and O.W.) supported by NSF grant PHY-1748958.
D.S.\ was supported by DOE grants {DE-SC0008669} and {DE-SC0009998}.
A.H., O.W.\ and E.T.N.\ were supported by DOE grant {DE-SC0010005}; Brookhaven National Laboratory is supported by the DOE under contract {DE-SC0012704}.
R.C.B., C.R.\ and E.W.\ were supported by DOE grant {DE-SC0015845}.
In addition, R.C.B.\ and C.R.\ acknowledge the support of NSF grant OCI-0749300.
G.T.F.\ was supported by NSF grant PHY-1417402.
E.R.\ was supported by a RIKEN SPDR fellowship.
E.R.\ and P.V.\ acknowledge the support of the DOE under contract DE-AC52-07NA27344 (LLNL).
A.G.\ acknowledges support under contract number {DE-SC0014664} and thanks LLNL and Lawrence Berkeley National Laboratory for hospitality during the completion of this work.
ANL is supported by the DOE under contract DE-AC02-06CH11357.

\appendix
\section{\label{app:therm}Auto-correlations and topological charge evolution} 
\begin{figure}[bp]
  \includegraphics[width=\linewidth]{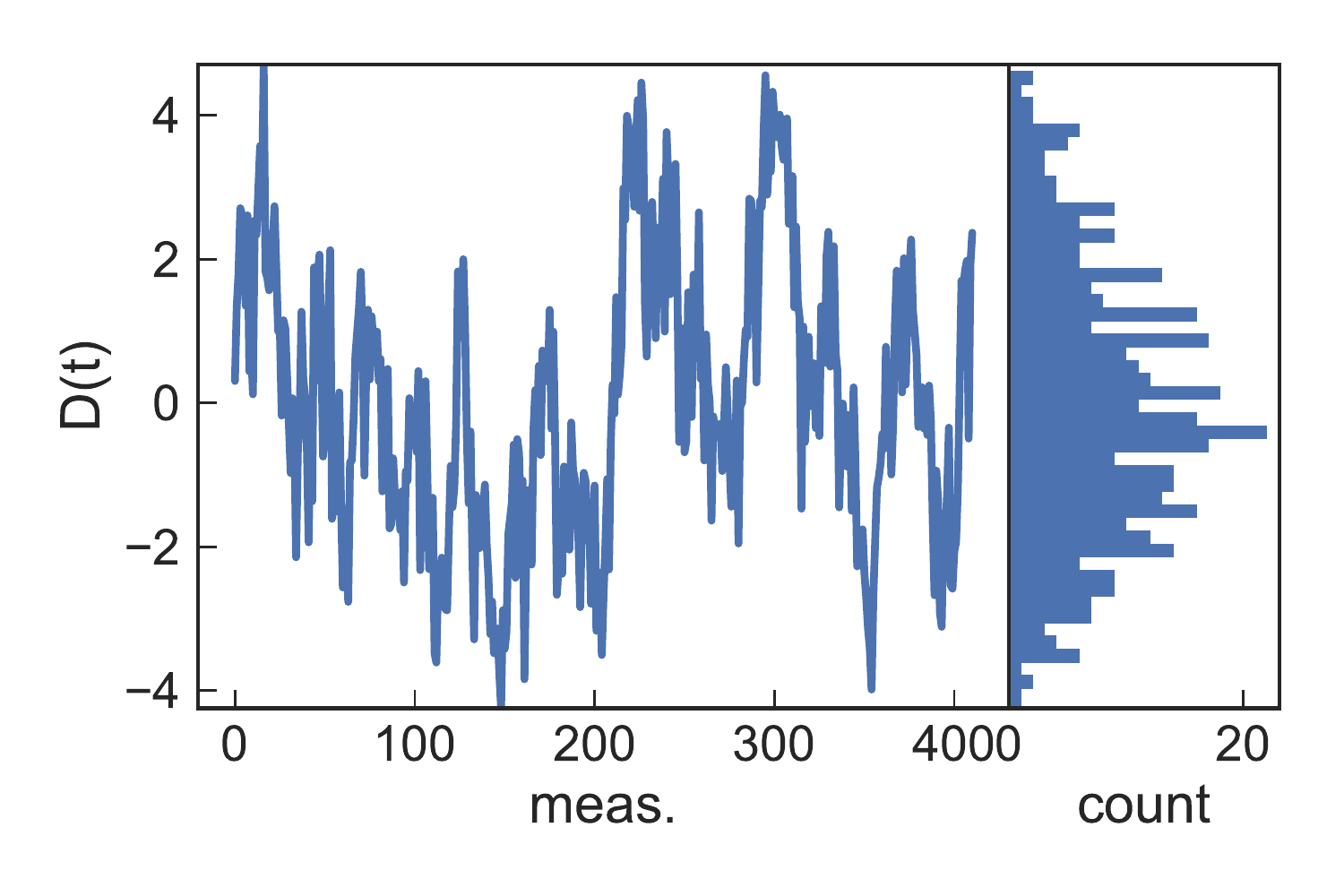} \\
  \includegraphics[width=\linewidth]{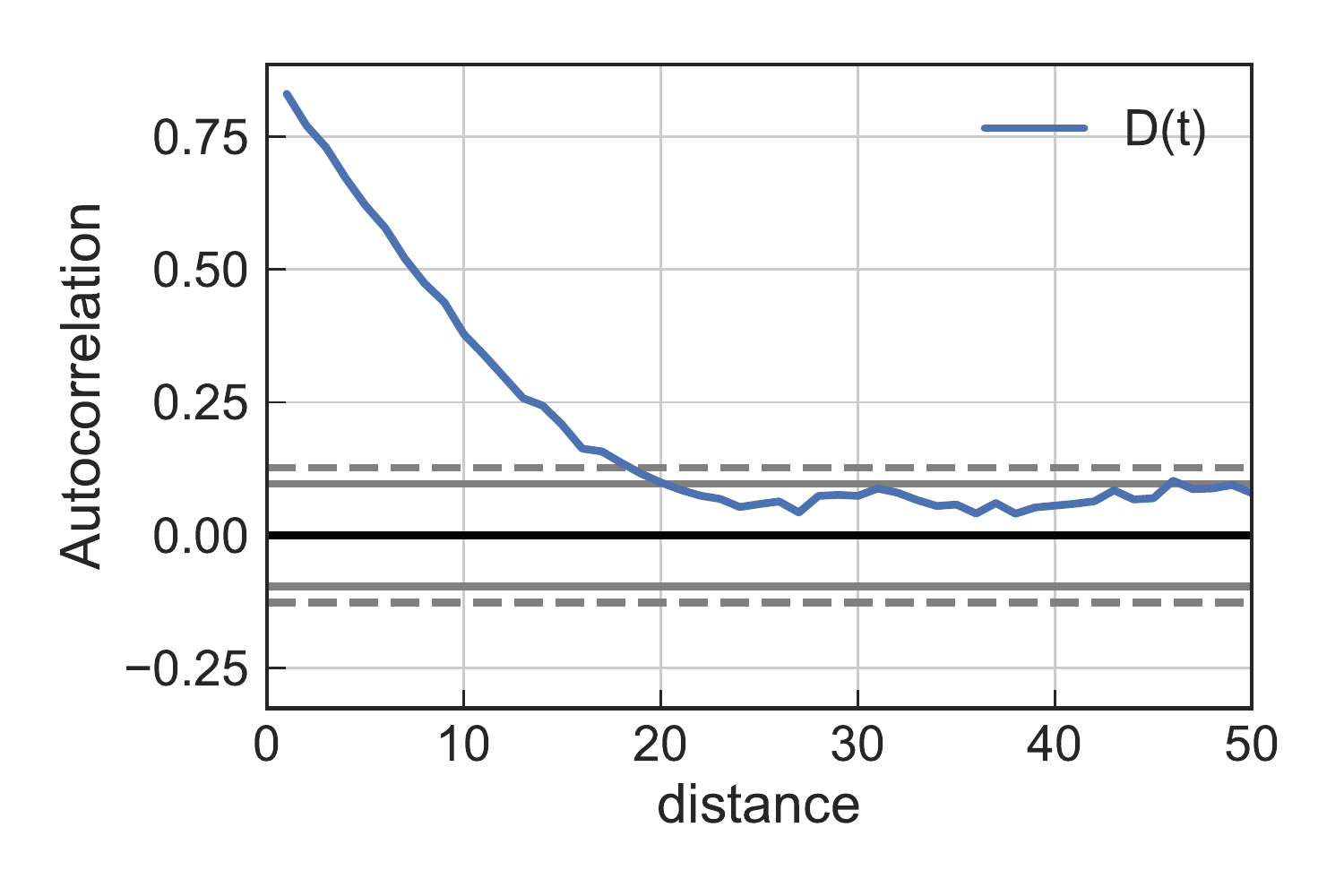}
  \caption{\label{fig:D_autocorr_ensA_s0}Autocorrelation function, history and histogram for the disconnected correlator $D(t)$ on the first stream of the $am_f = 0.00125$ $N_f = 8$ ensemble at a fixed representative time separation $t = 9$.  The dashed grey lines show compatibility with zero within two standard deviations.}
\end{figure}

\begin{figure}[bp]
  \includegraphics[width=\linewidth]{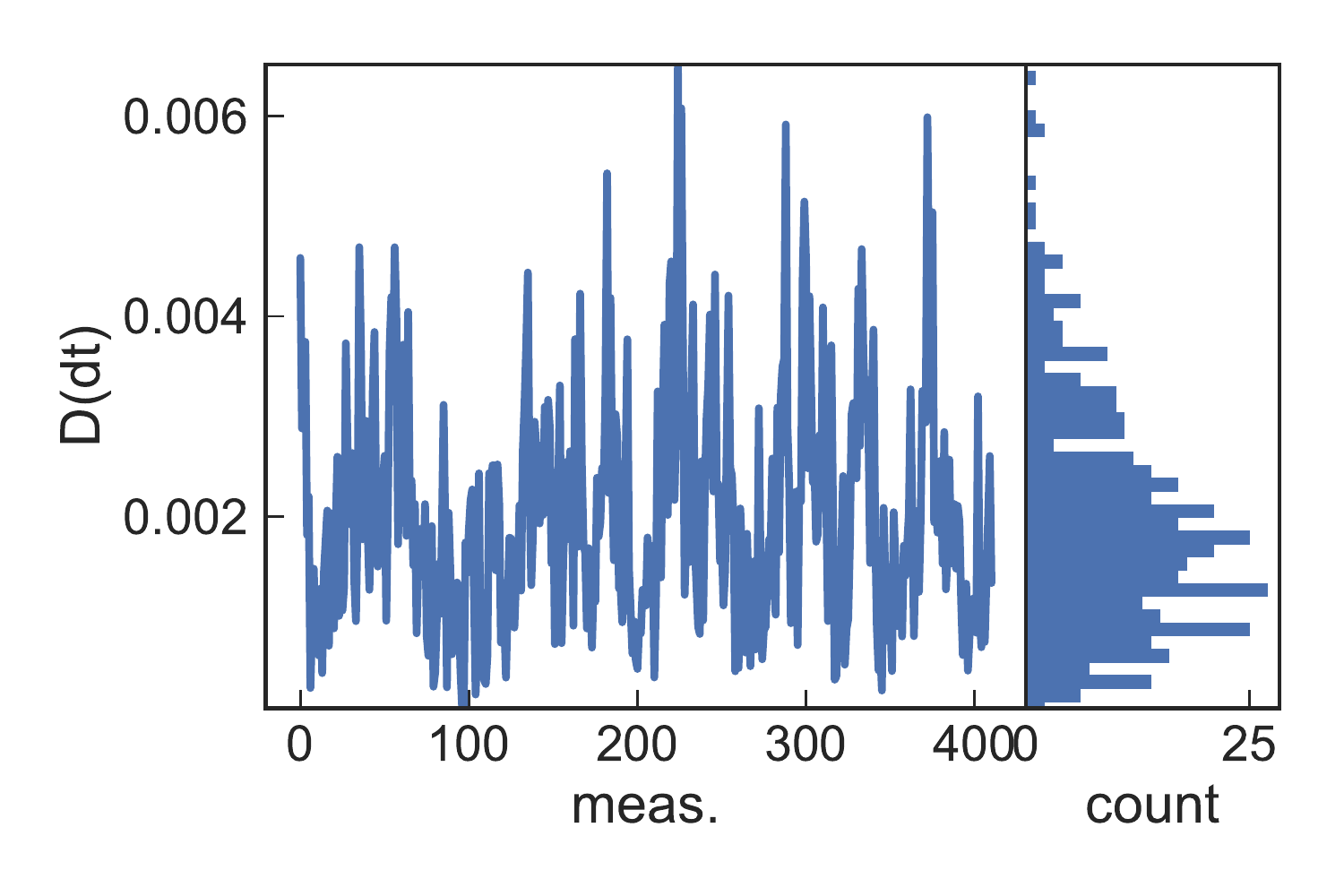} \\
  \includegraphics[width=\linewidth]{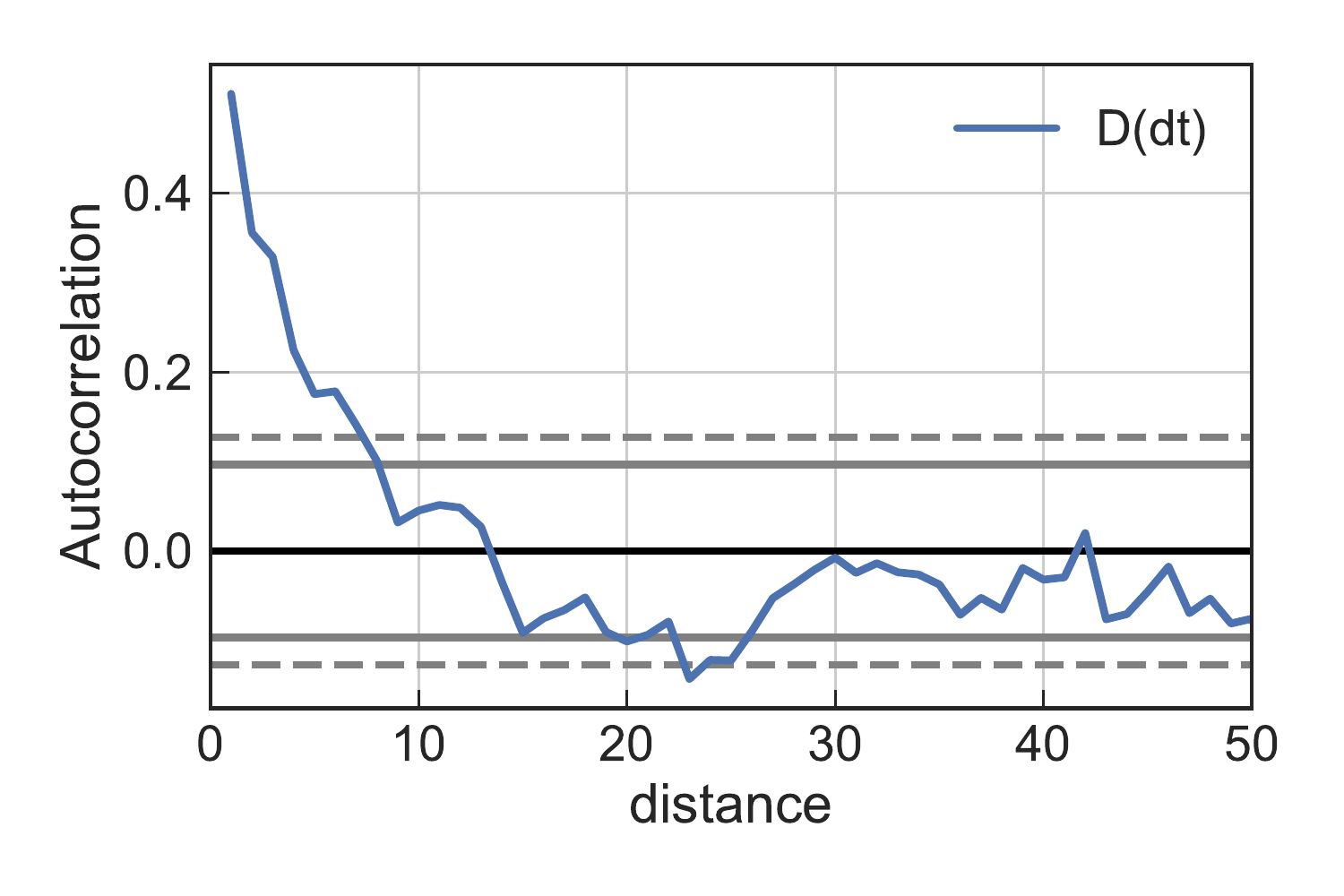}
  \caption{\label{fig:Dt_autocorr_ensA_s0}Autocorrelation function, history and histogram for the finite time difference of the disconnected correlator $D(dt) = D(t + 1) - D(t)$ on the first stream of the $am_f = 0.00125$ $N_f = 8$ ensemble at a fixed representative time separation $t = 9$.}
\end{figure}

The disconnected correlator $D(t)$ in \eq{eq:doubletrace} is effectively the time correlation of \pbp and we study its autocorrelation function along the measurement time series at a fixed time separation.
In \fig{fig:D_autocorr_ensA_s0} we show the $D(t = 9)$ history and histogram for the first stream of the $am_f = 0.00125$ $N_f = 8$ ensemble, together with its autocorrelation function.
The autocorrelation function is compatible with zero within two standard deviations (dashed grey lines) when the distance between measurements is approximately 20, and the integrated autocorrelation time is approximately 10.
We notice that the autocorrelation function is significantly smaller when we consider the finite time difference of correlators $D(dt) = D(t + 1) - D(t)$ that we fit to extract the $0^{++}$ meson mass, as can be seen in \fig{fig:Dt_autocorr_ensA_s0}, with an integrated autocorrelation time of approximately 3.
We find that similar observations hold for all ensembles.

\begin{figure}[tbp]
  \includegraphics[width=\linewidth]{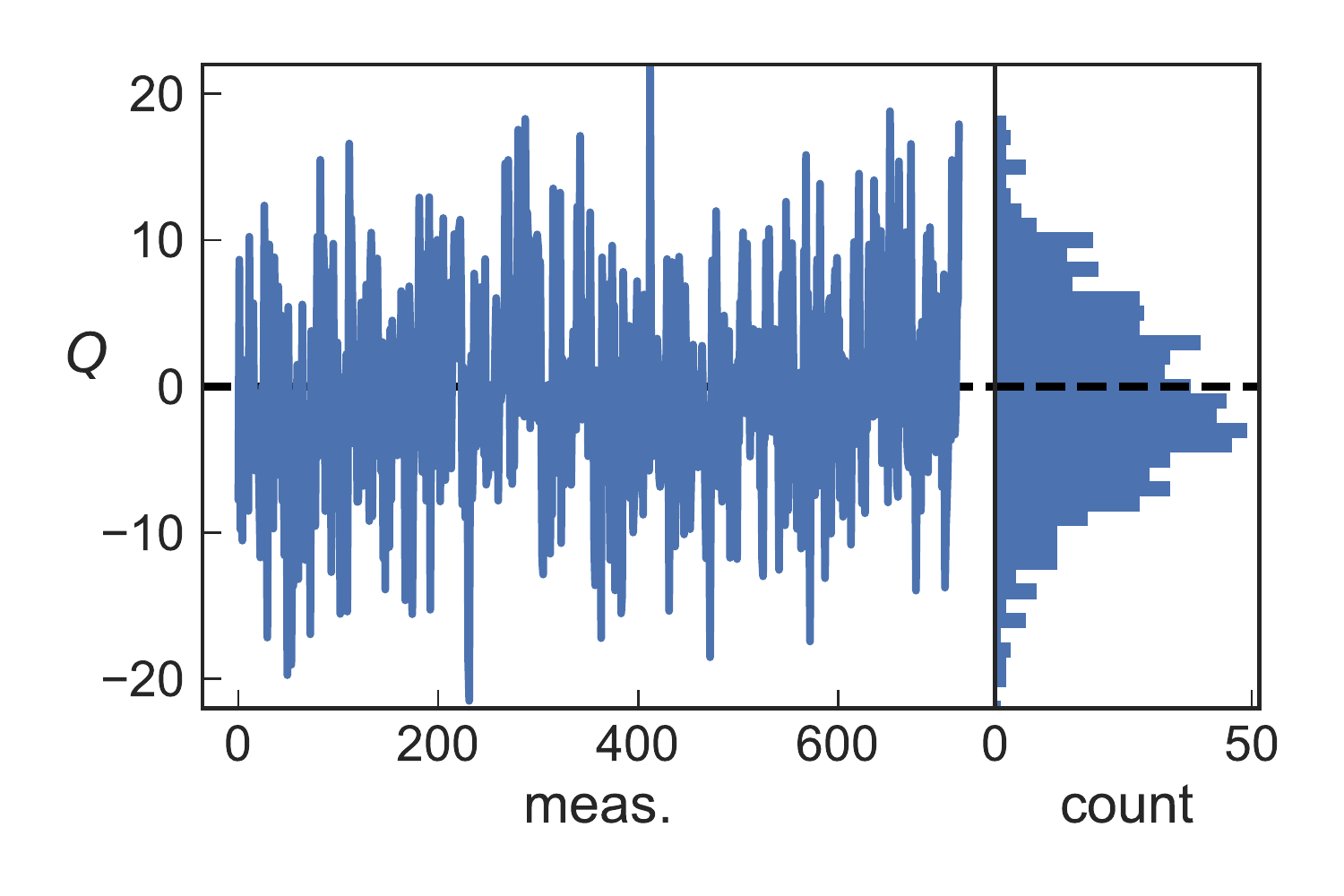} \\
  \includegraphics[width=\linewidth]{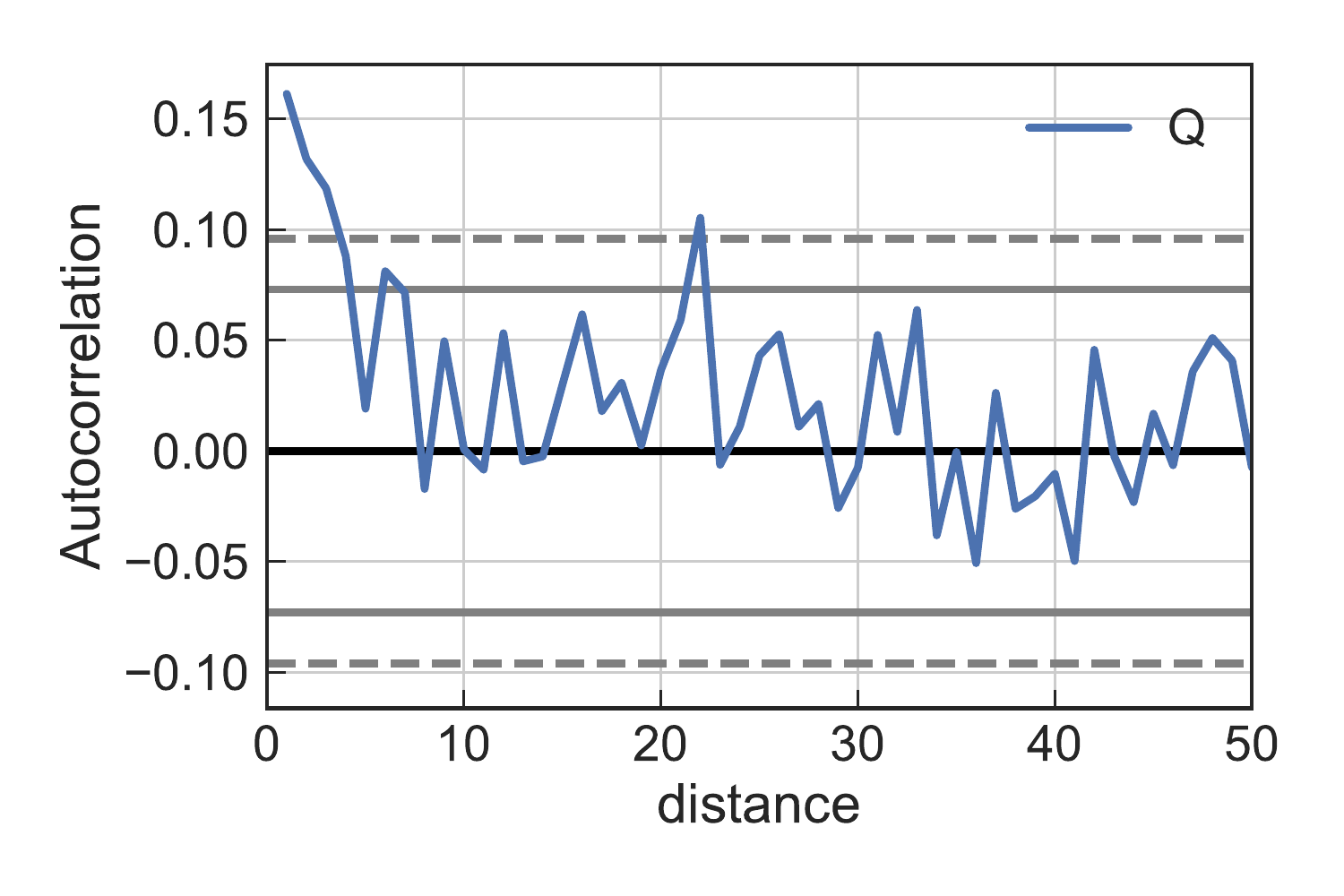}
  \caption{\label{fig:Q_ensA_s0}Topological charge evolution, histogram and autocorrelation function for the first stream of the $am_f = 0.00125$ $N_f = 8$ ensemble.}
\end{figure}

\begin{figure}[tbp]
  \includegraphics[width=\linewidth]{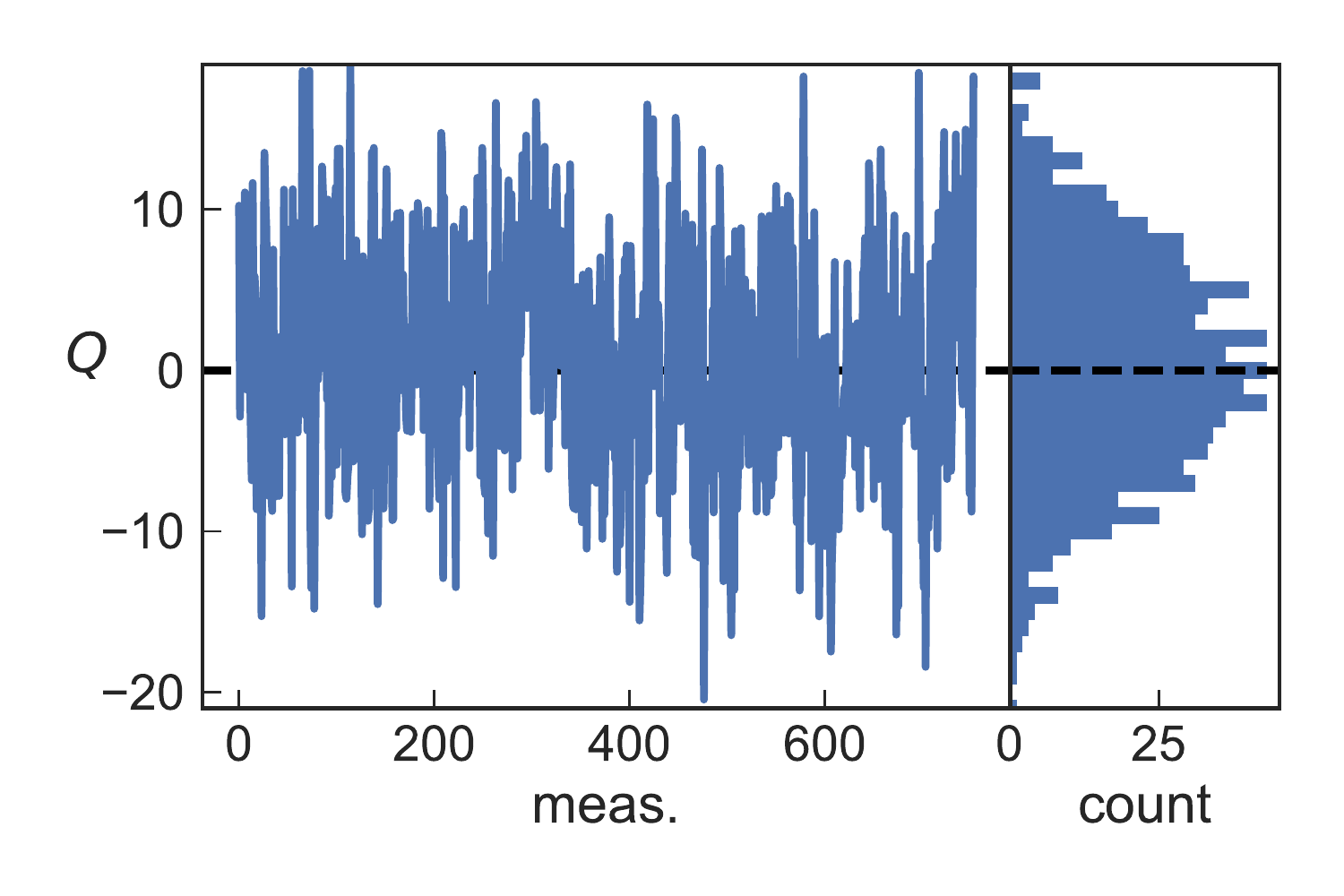} \\
  \includegraphics[width=\linewidth]{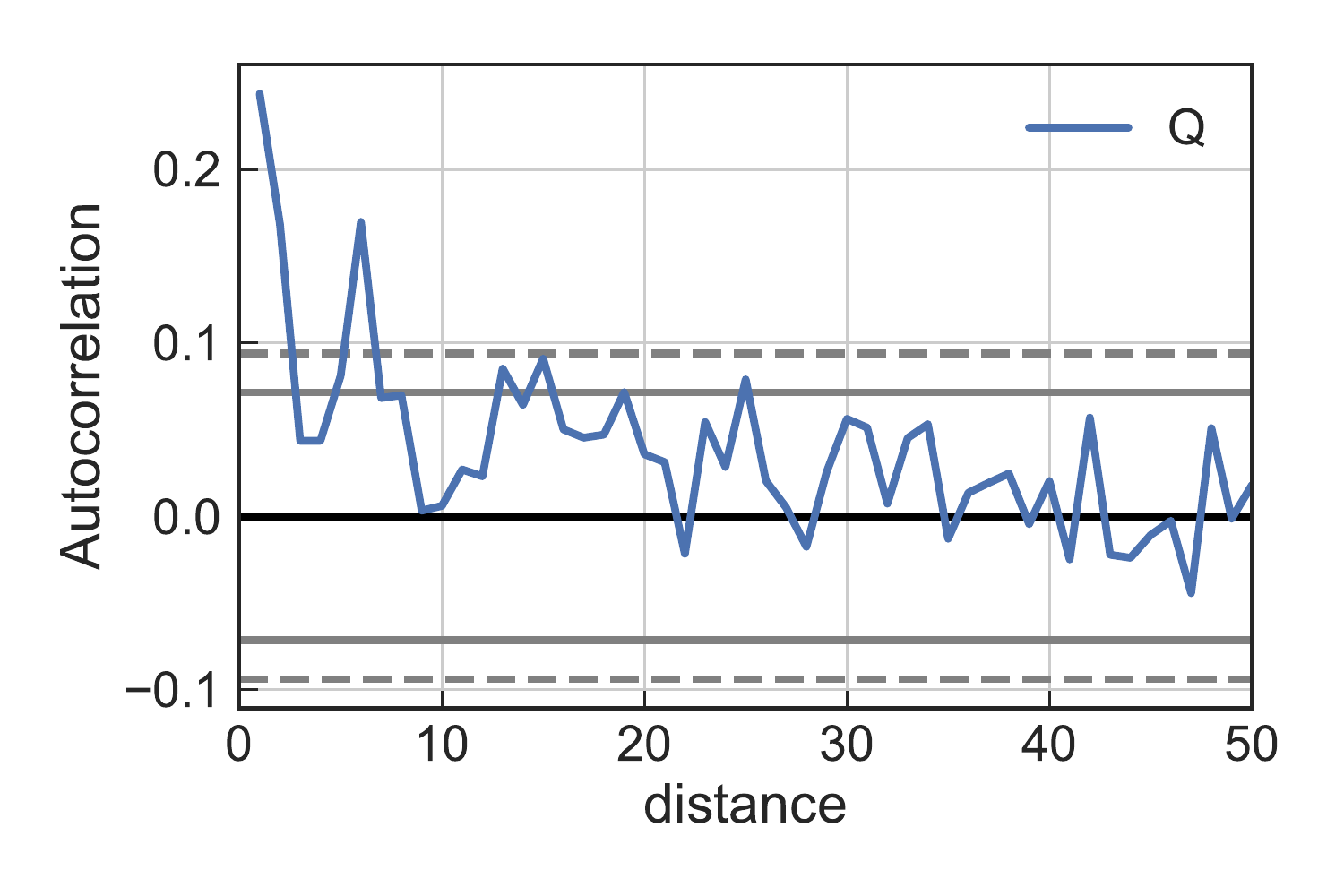}
  \caption{\label{fig:Q_ensA_s1}Topological charge evolution, histogram and autocorrelation function for the second stream of the $am_f = 0.00125$ $N_f = 8$ ensemble.}
\end{figure}

\begin{figure}[tbp]
  \includegraphics[width=\linewidth]{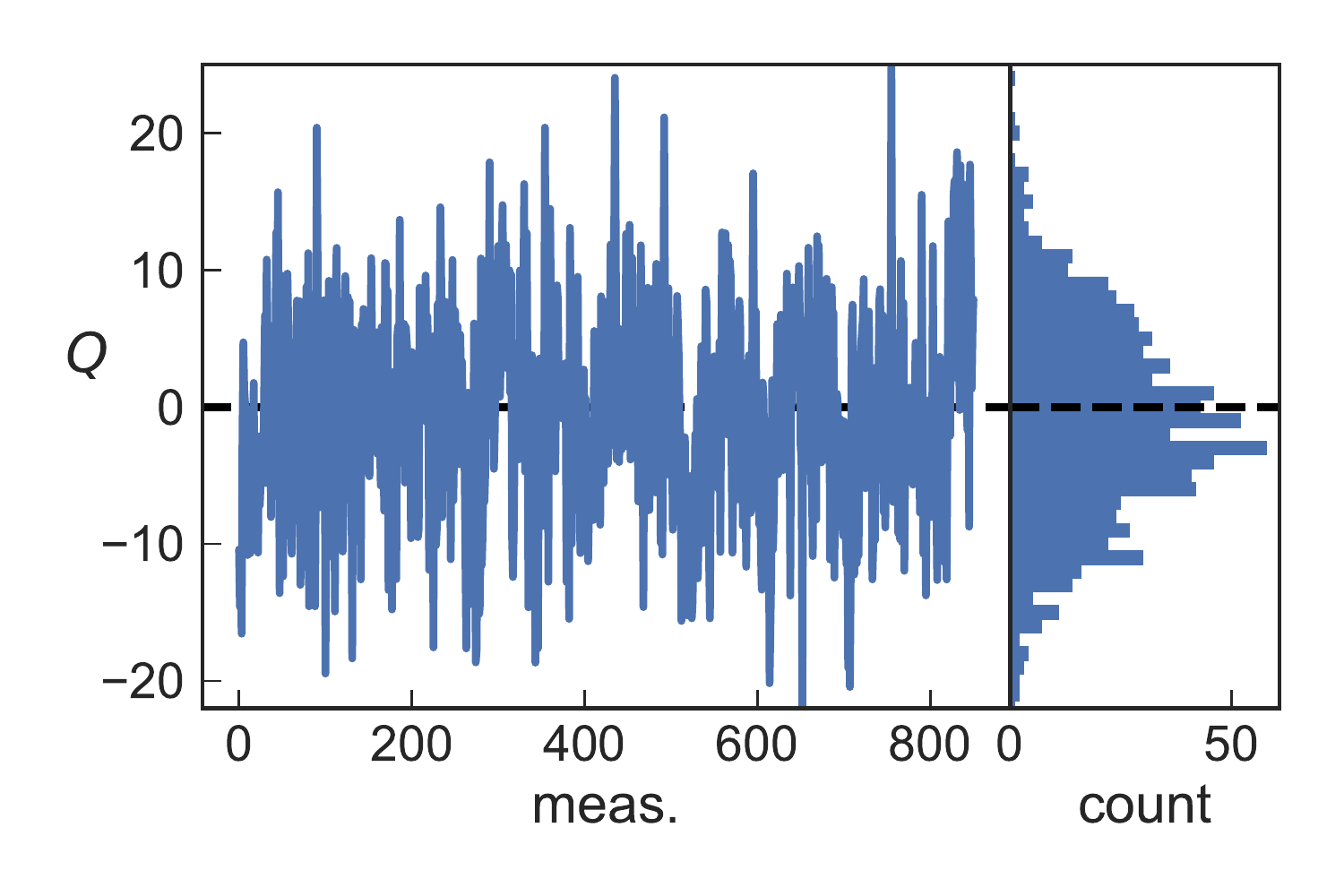} \\
  \includegraphics[width=\linewidth]{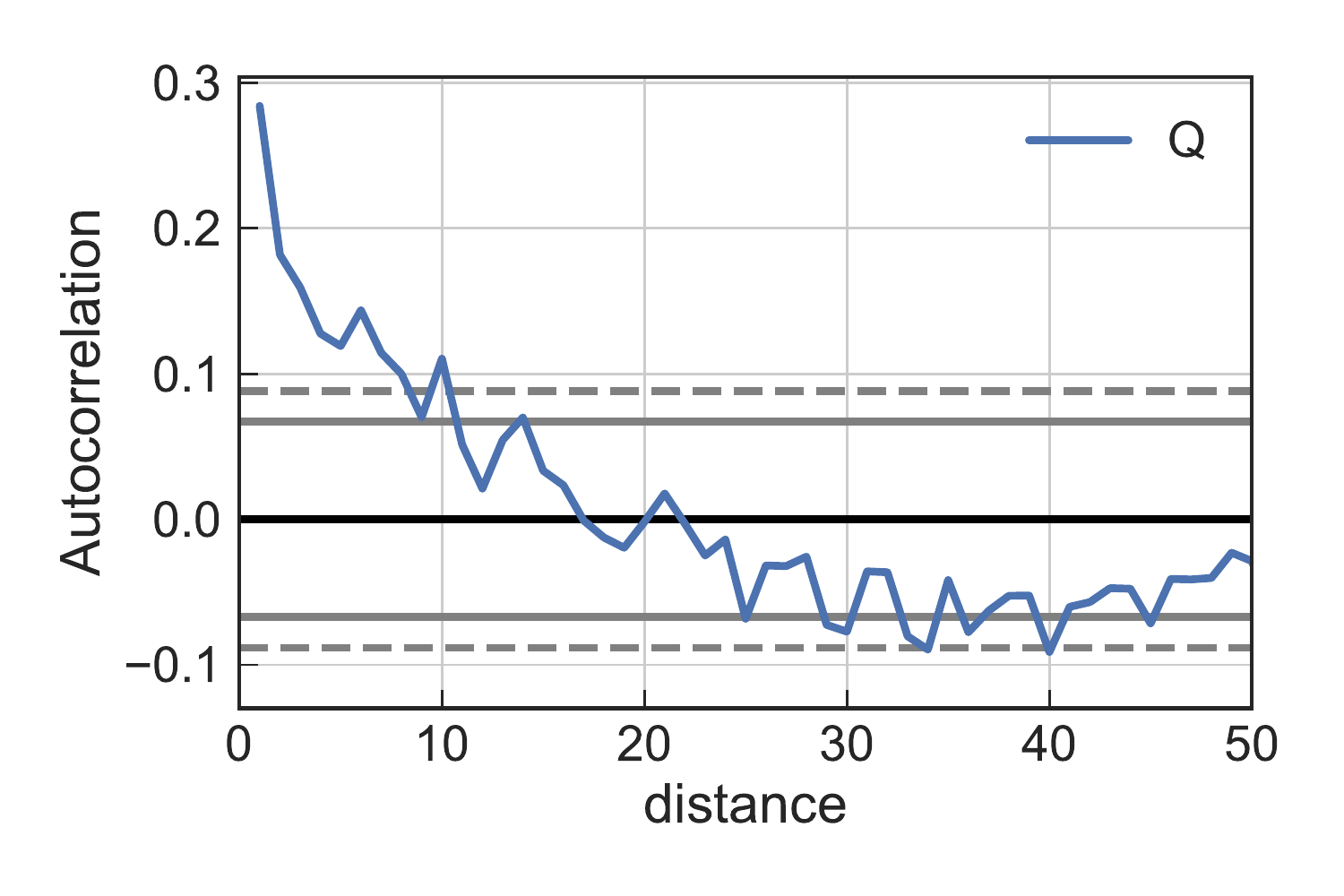}
  \caption{\label{fig:Q_ensA_s2}Topological charge evolution, histogram and autocorrelation function for the third stream of the $am_f = 0.00125$ $N_f = 8$ ensemble.}
\end{figure}

\begin{figure}[tbp]
  \includegraphics[width=\linewidth]{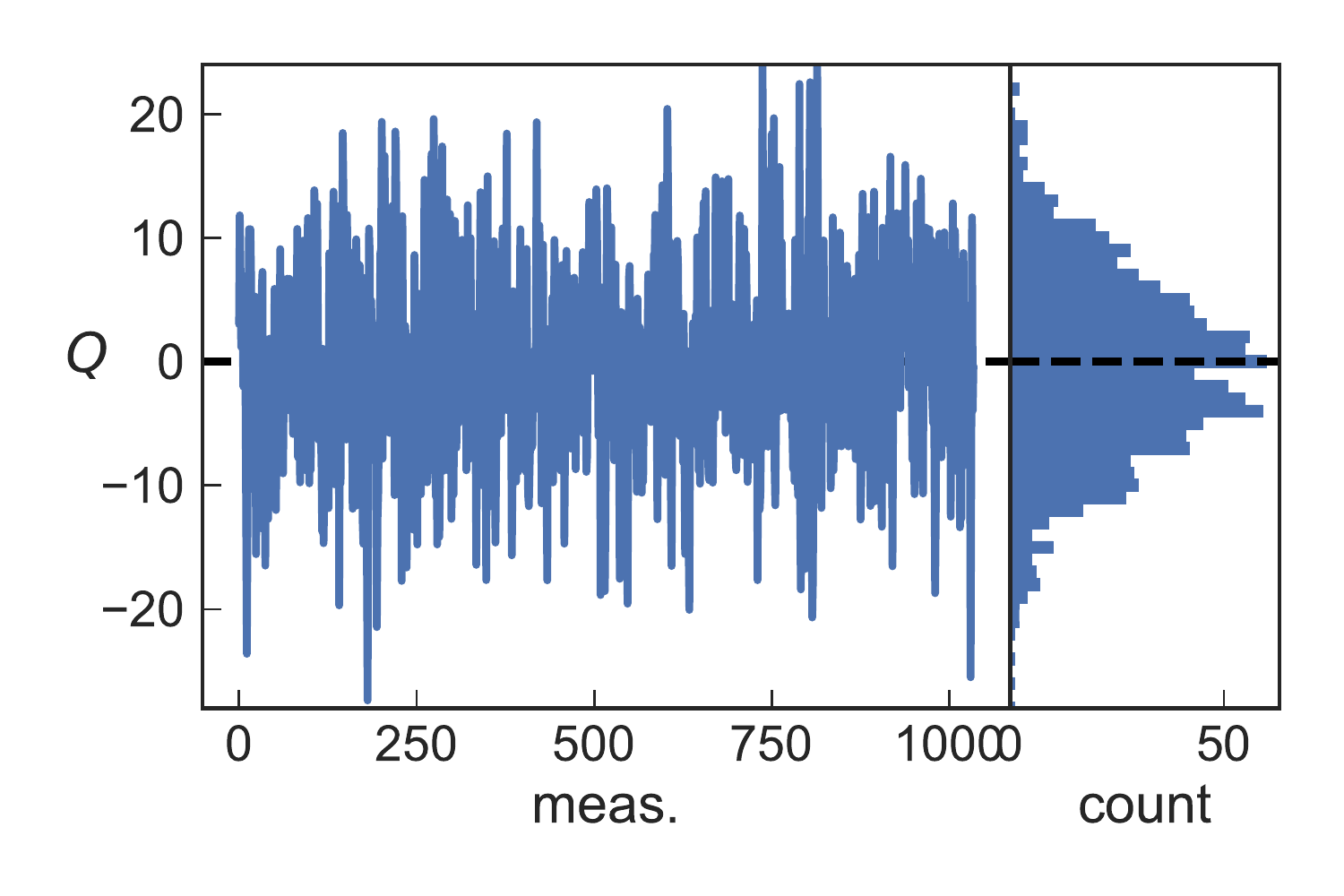} \\
  \includegraphics[width=\linewidth]{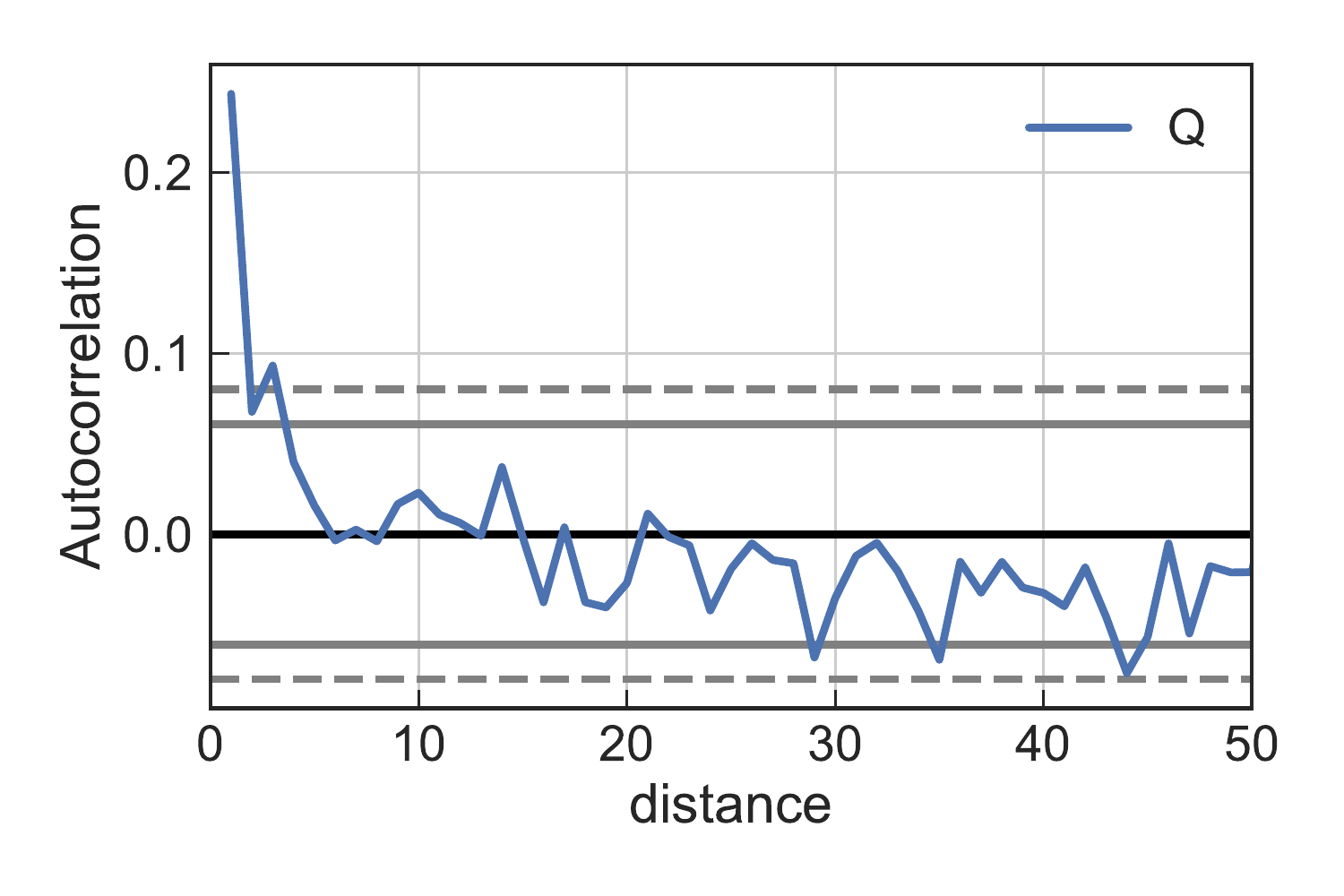}
  \caption{\label{fig:Q_ensW}Topological charge evolution, histogram and autocorrelation function for the $am_f = 0.003$ $N_f = 4$ ensemble.}
\end{figure}

On all the configurations where we measure the spectrum, we also measure the topological charge, using the clover definition of the discretized $F_{\mu\nu}\widetilde F^{\mu\nu}$ gauge tensor.
For each ensemble we use the Wilson flow to smooth the configurations and obtain the topological charge $Q$ at flow time $t_w$ satisfying $\sqrt{8t_w} = L / 2$, where $L$ is the number of points in the spatial directions.
For the lightest-mass ensembles with $N_f = 8$ and $N_f = 4$ (which have the smallest $a / \sqrt{8t_0}$, \cf \fig{fig:Wflow_scale}), Figs.~\ref{fig:Q_ensA_s0} through \ref{fig:Q_ensW} show a well-balanced topological charge, with frequent fluctuations and no sign of topological freezing.
The autocorrelation function for the topological charge is also compatible with zero (within two standard deviations) after a distance of a few measurements.
The topological charge history, histogram and autocorrelation functions for the three Monte Carlo streams at $am_f = 0.00125$ for $N_f = 8$ are shown in Figs.~\ref{fig:Q_ensA_s0} through \ref{fig:Q_ensA_s2}, while the same information for the $am_f = 0.003$ $N_f = 4$ ensemble is shown in \fig{fig:Q_ensW}.

\clearpage

\section{\label{app:sigma}Further details of \si mass determination} 
Here we give additional details on our numerical procedures and results for determination of the $0^{++}$ \si meson mass.
Our procedure for fitting to determine the mass and estimating the fit-range systematic error is described in detail in Secs.~\ref{ssec:spectrum_disc} and \ref{ssec:spectrum_syst} above.
Briefly, our procedure for the \si meson is the same as that for any other state, except for two key differences: first, we fit finite time differences of the correlation functions in order to remove the vacuum contribution; second, we carry out a joint fit of the $D(t)$ and $S(t)$ correlators in order to obtain more stable results for $M_{\si}$.

\begin{figure}[tbp]
  \includegraphics[width=\linewidth]{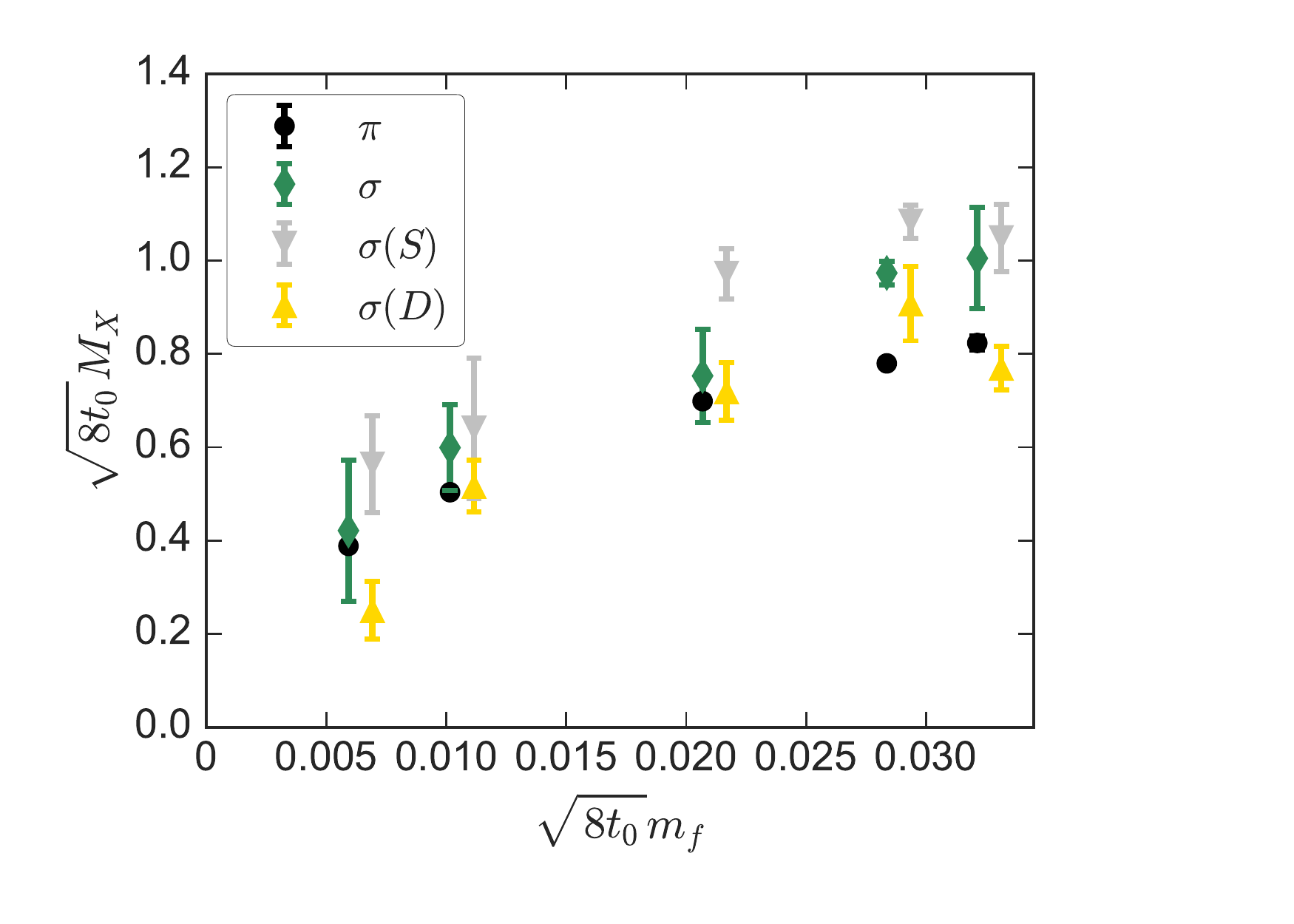}
  \caption{\label{fig:zpp_compare}Comparison in the 8-flavor theory of the $0^{++}$ mass spectrum as determined through joint fits to $S(t)$ and $D(t)$ as in our main analysis, and determined with individual fits to $S(t)$ and $D(t)$.  A horizontal offset is added to the $S(t)$ and $D(t)$ results to make the plot more easily readable.}
\end{figure}

\begin{figure}[tbp]
  \includegraphics[width=\linewidth]{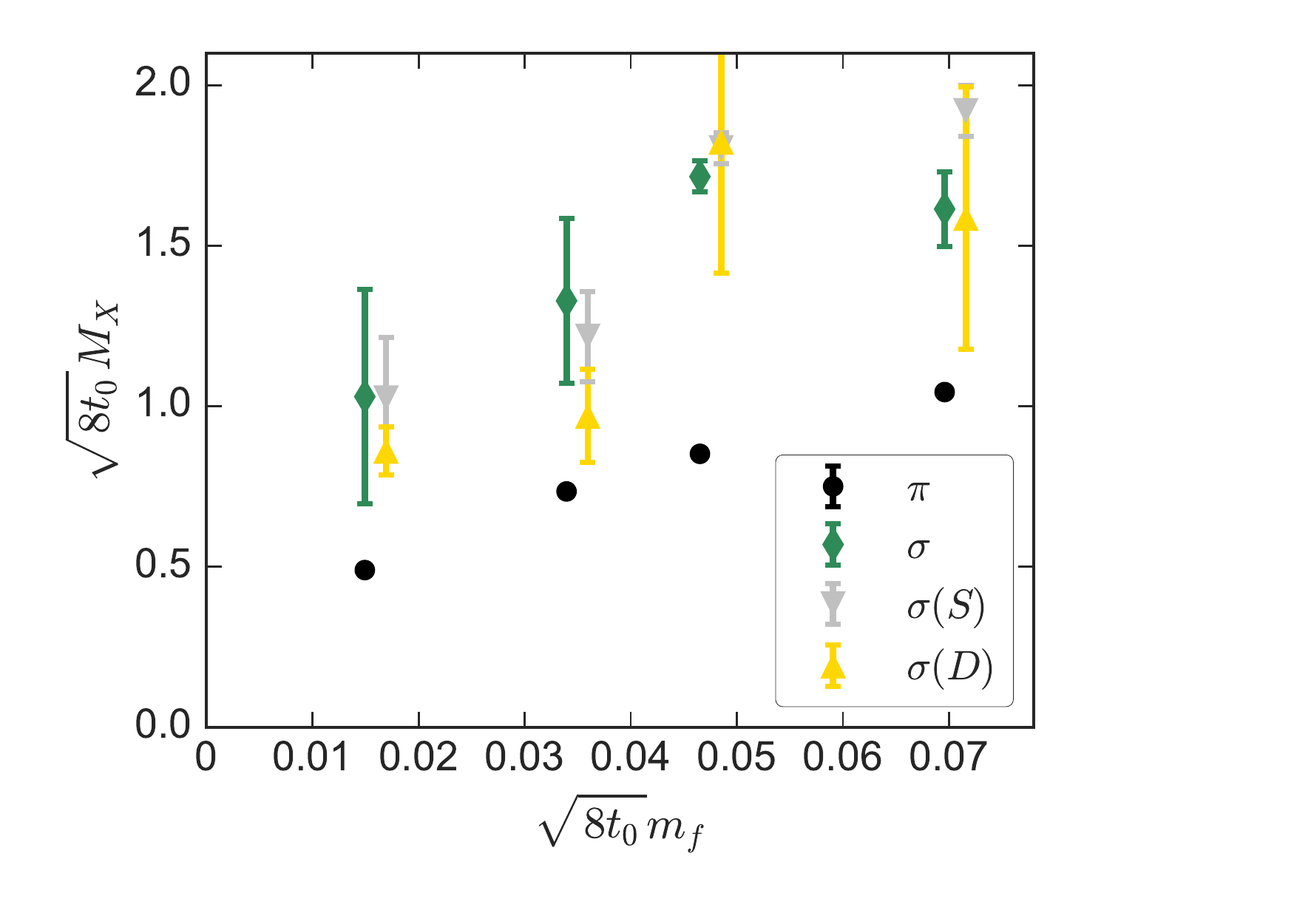}
  \caption{\label{fig:zpp_compare_4f}Comparison in the 4-flavor theory of the $0^{++}$ mass spectrum as determined through joint fits to $S(t)$ and $D(t)$ as in our main analysis, and determined with individual fits to $S(t)$ and $D(t)$.  A horizontal offset is added to the $S(t)$ and $D(t)$ results to make the plot more easily readable.}
\end{figure}

As a test of the joint fit procedure, we can compare our central results for $M_{\si}$ to determinations from the individual correlators $S(t)$ and $D(t)$.
These results are tabulated in \tab{tab:scalar_results} and plotted in Figs.~\ref{fig:zpp_compare} and \ref{fig:zpp_compare_4f}.
There is a clear systematic trend for $M_{\si, S} > M_{\si, D}$, which is expected due to the fact that contamination from states other than the \si enters with positive sign in $S(t)$ but negative sign in $D(t)$; see Eqs.~(\ref{eq:Scorr}) and (\ref{eq:Dcorr}).
The results for $M_{\si}$ from the joint fit are broadly compatible with the individual $S(t)$ and $D(t)$ fits, but offer improved precision compared to the more conservative possibility of taking the difference between $M_{\si, S}$ and $M_{\si, D}$ as a systematic error estimate.

\begin{figure}[tbp]
  \includegraphics[width=\linewidth]{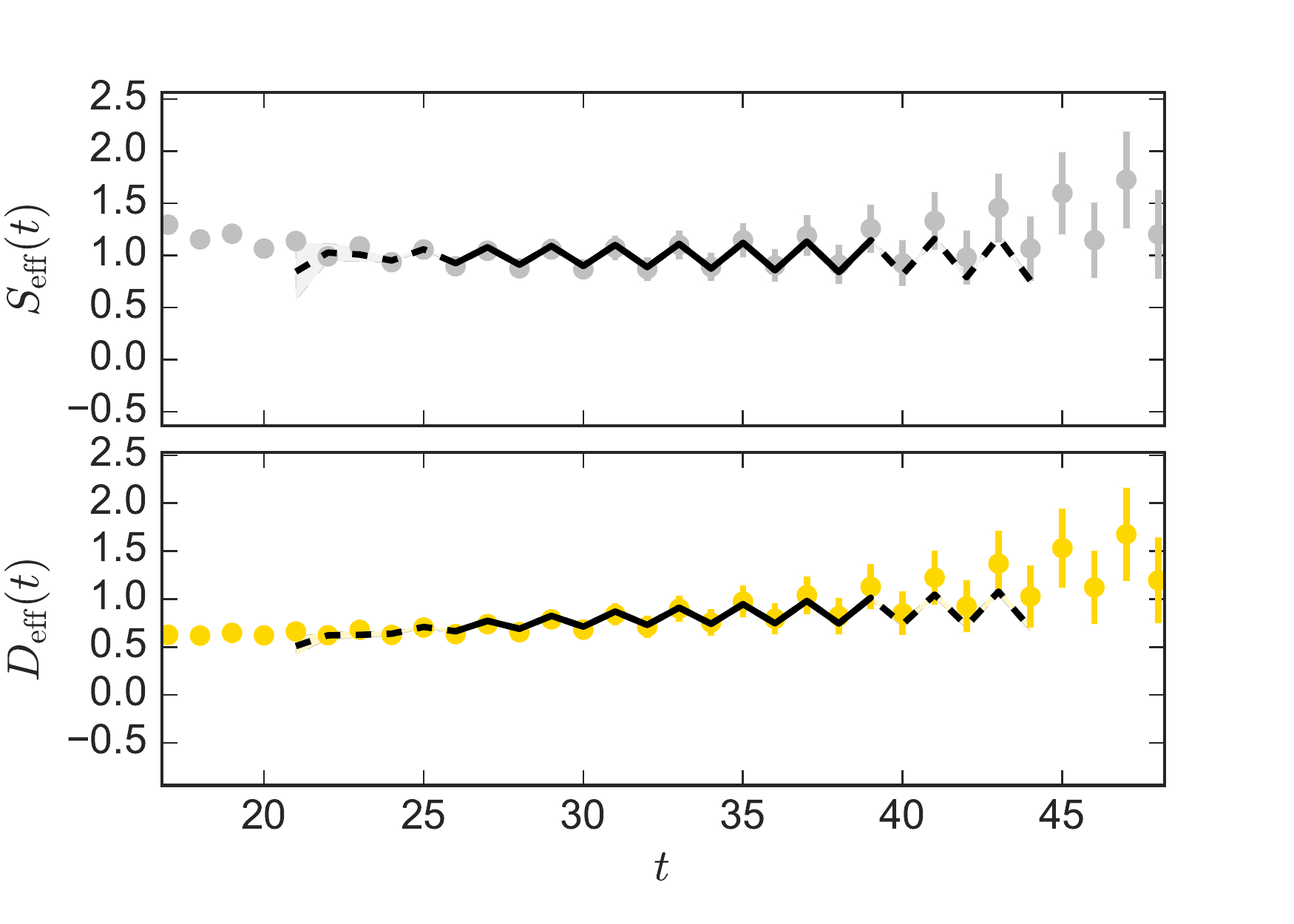} \\
  \includegraphics[width=\linewidth]{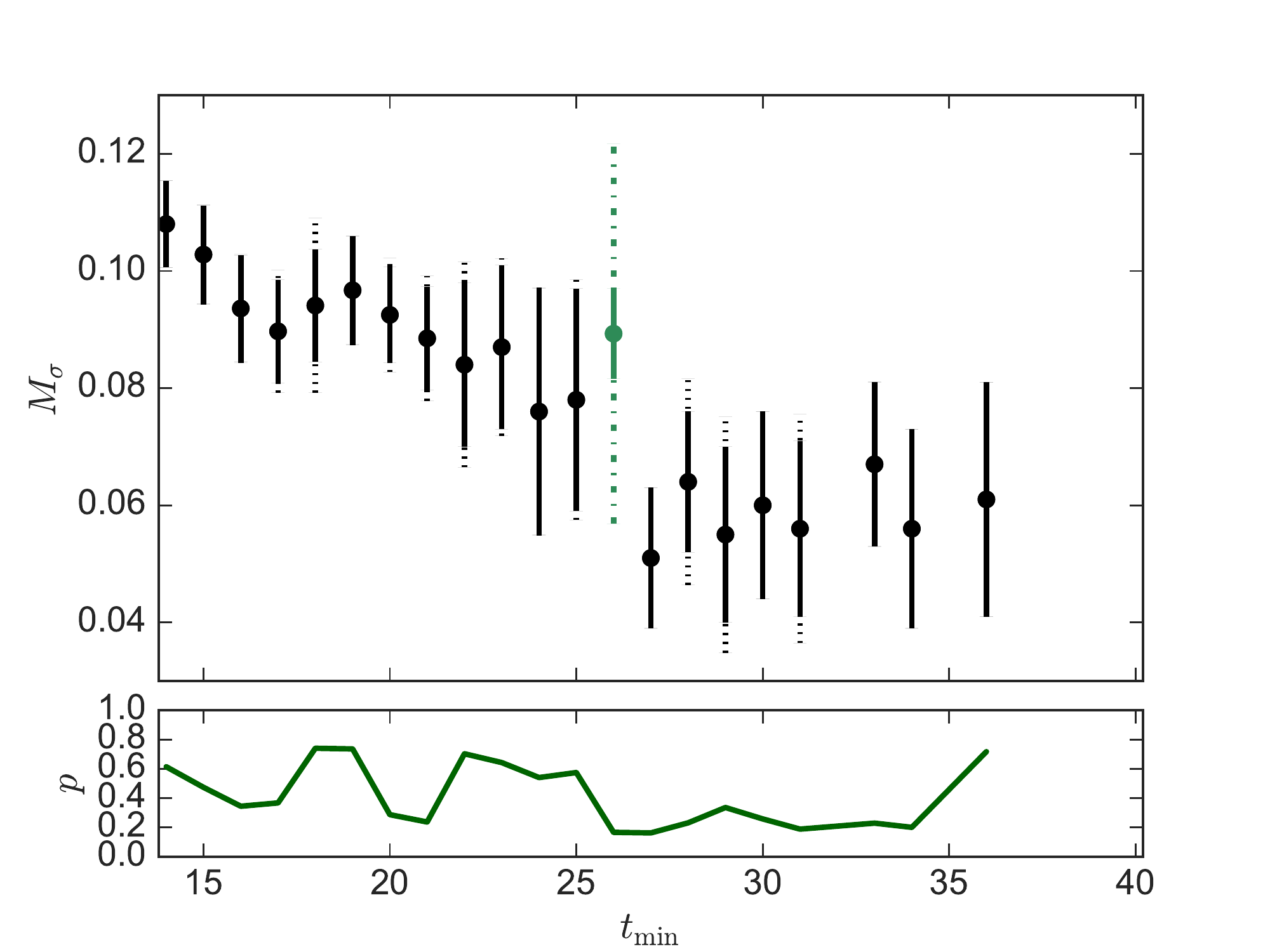}
  \caption{\label{fig:zpp_eff_ensA}Best-fit vs.\ effective correlators (top) and fit-range systematic scan (bottom) for the $0^{++}$ joint fit, $am_f = 0.00125$ $N_f = 8$ ensemble.}
\end{figure}

\begin{figure}[tbp]
  \includegraphics[width=\linewidth]{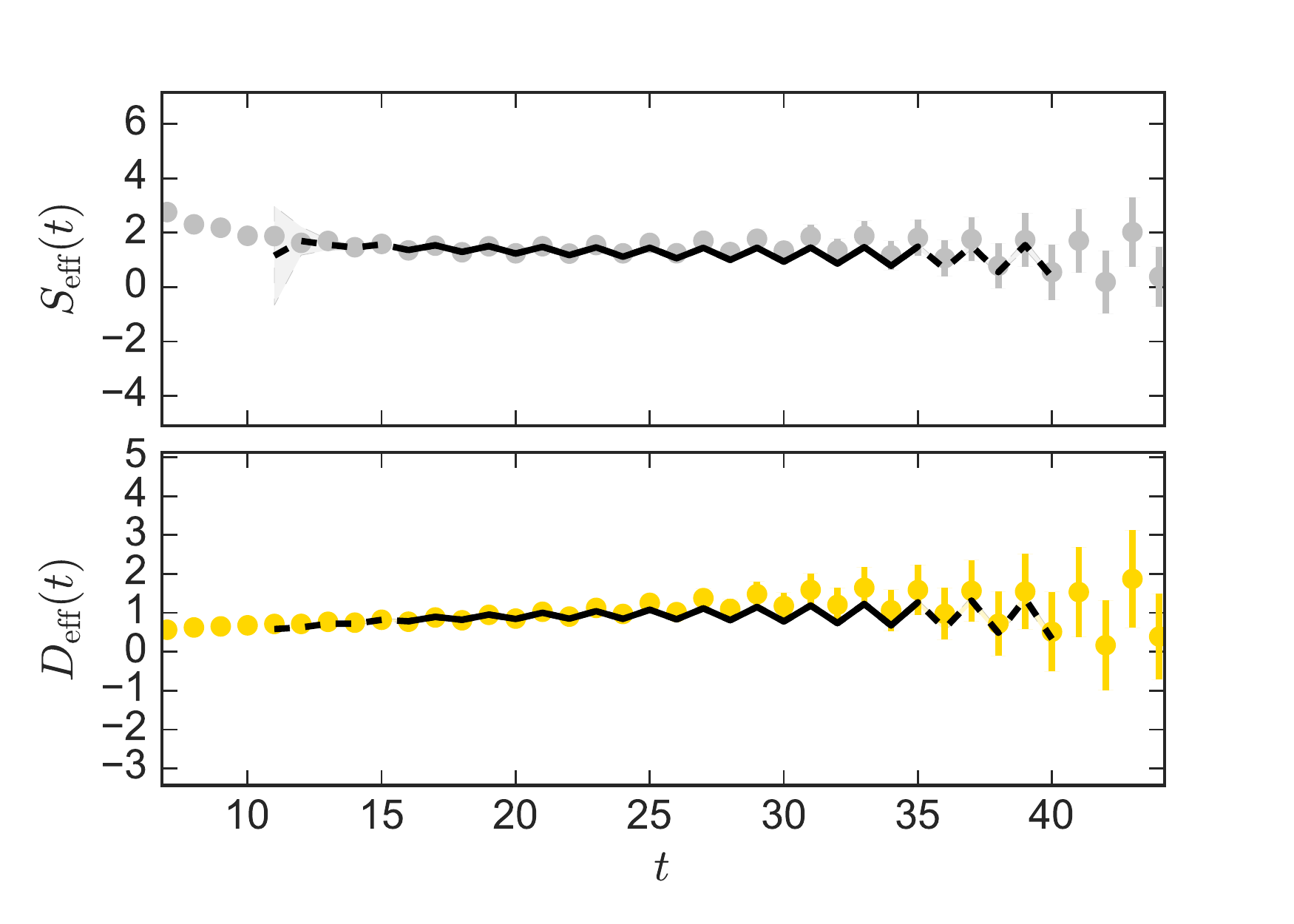} \\
  \includegraphics[width=\linewidth]{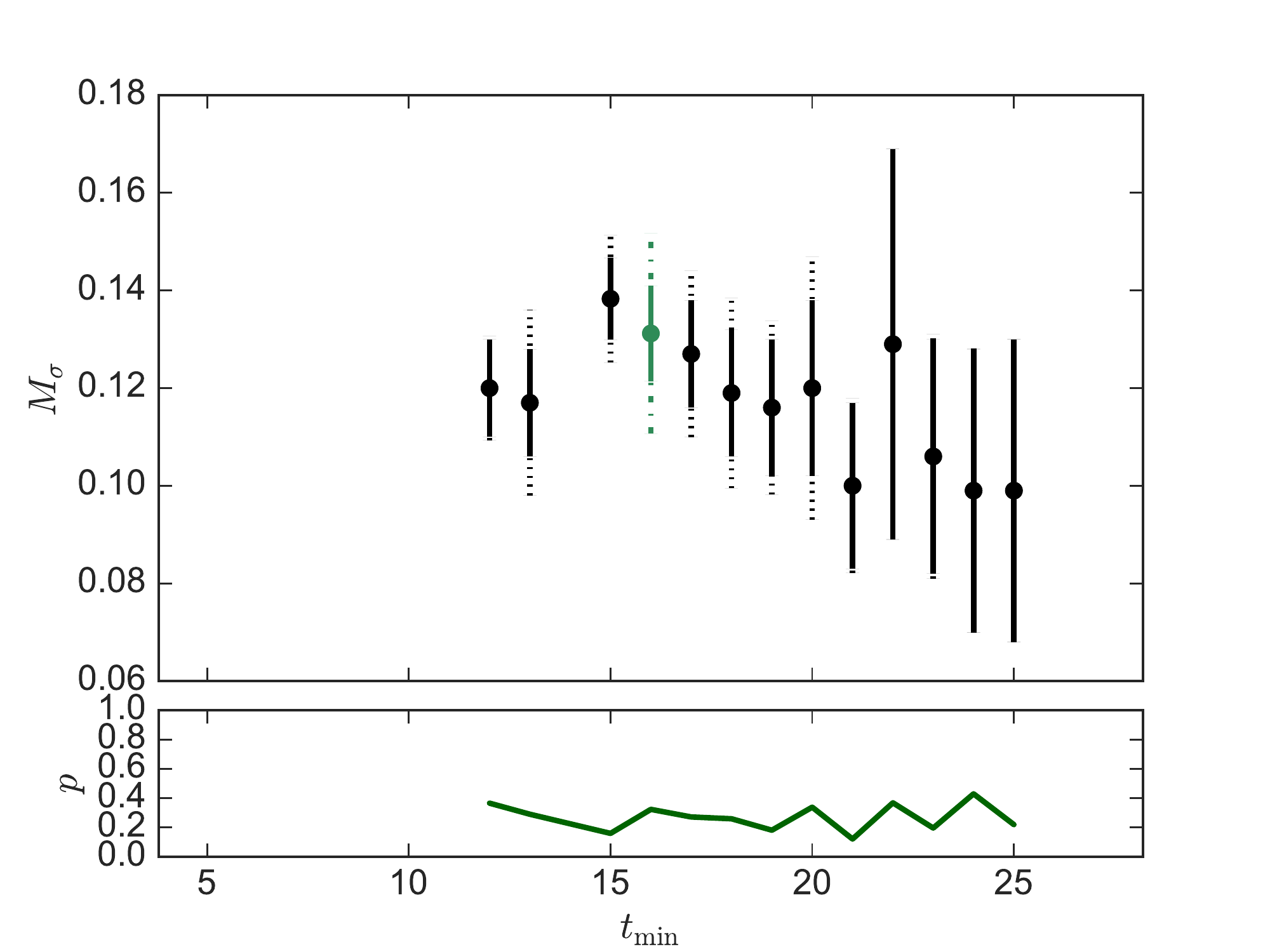}
  \caption{\label{fig:zpp_eff_ensB}Best-fit vs.\ effective correlators (top) and fit-range systematic scan (bottom) for the $0^{++}$ joint fit, $am_f = 0.00222$ $N_f = 8$ ensemble.}
\end{figure}

\begin{figure}[tbp]
  \includegraphics[width=\linewidth]{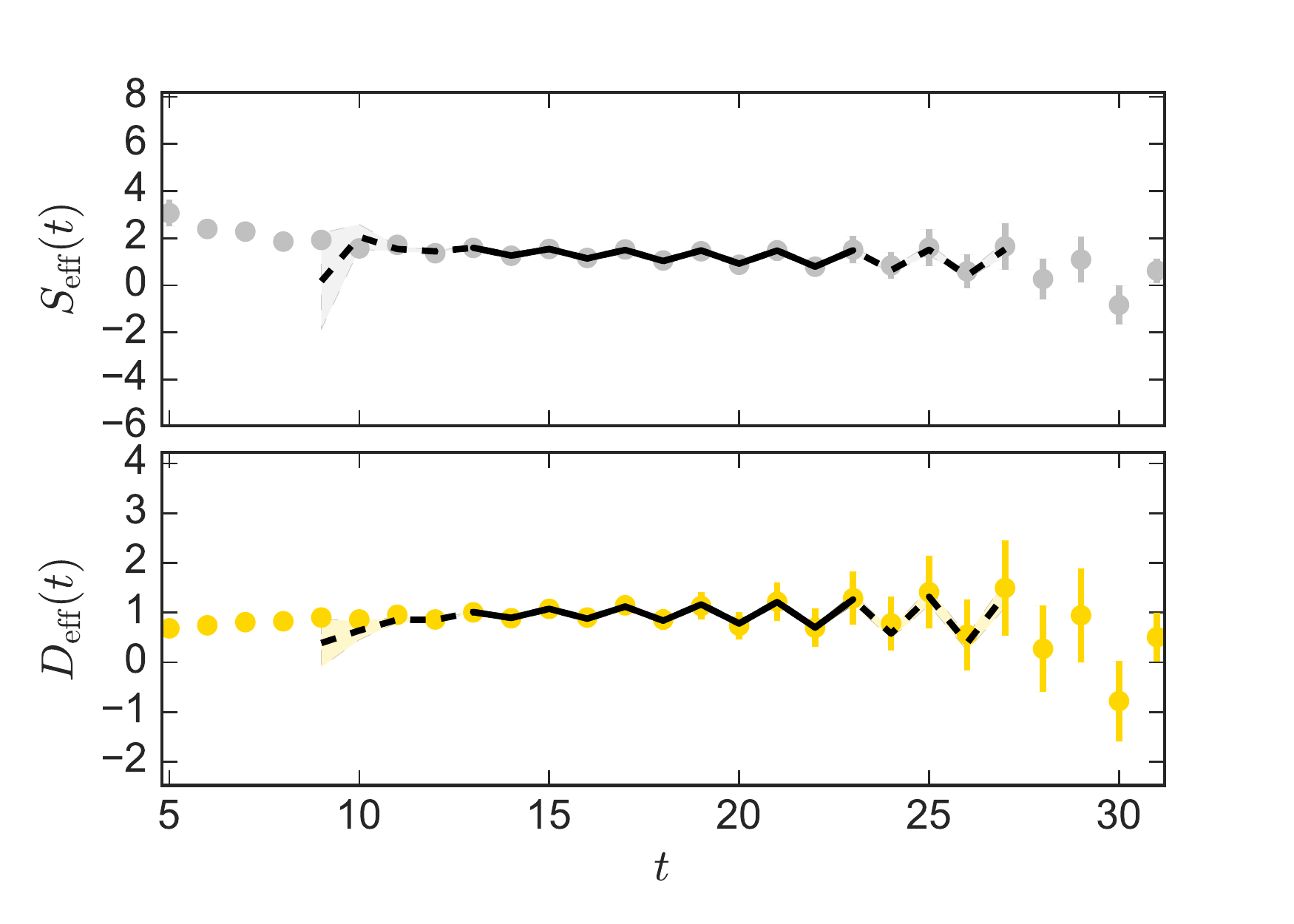} \\
  \includegraphics[width=\linewidth]{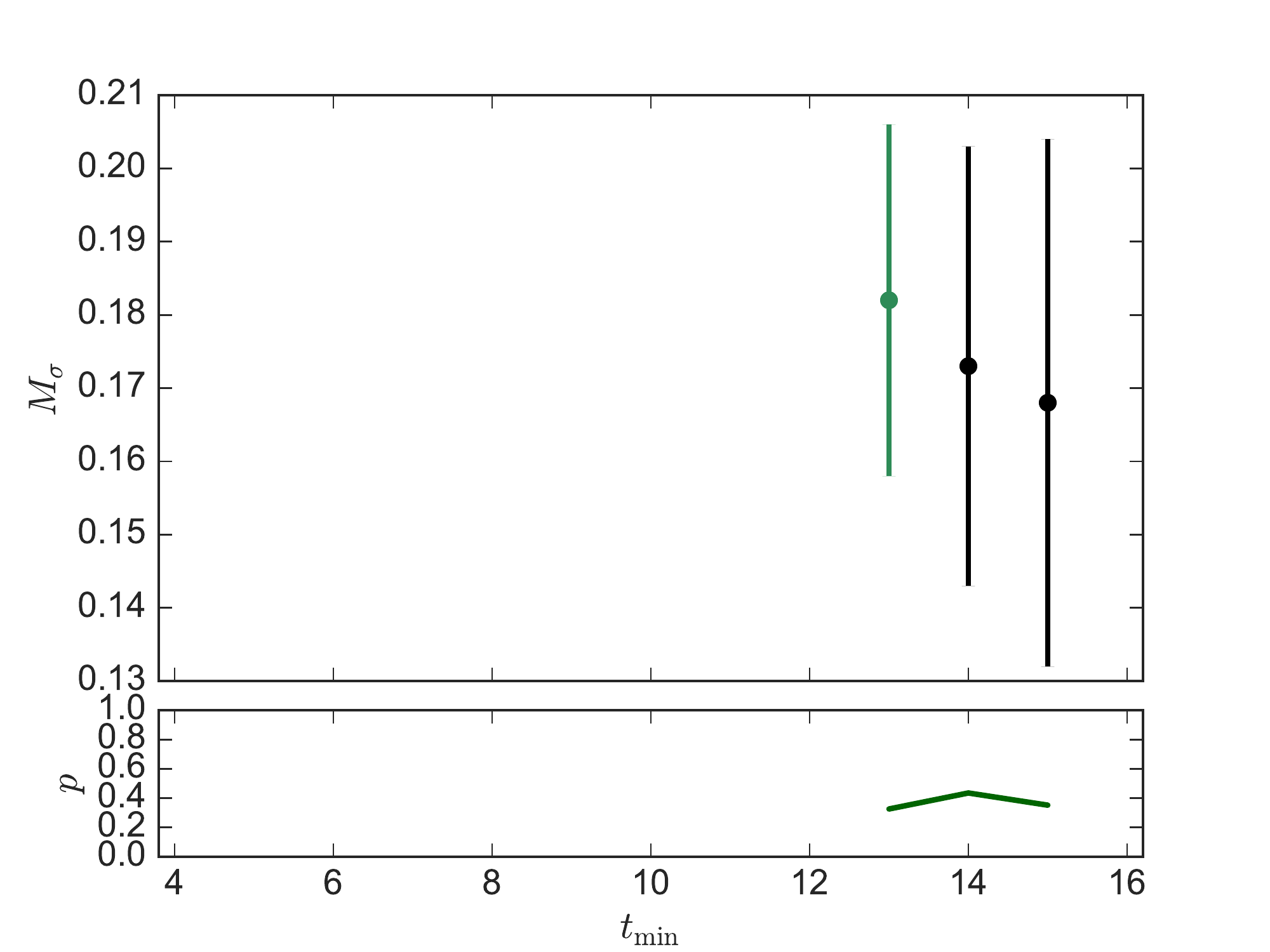}
  \caption{\label{fig:zpp_eff_ensC}Best-fit vs.\ effective correlators (top) and fit-range systematic scan (bottom) for the $0^{++}$ joint fit, $am_f = 0.005$ $N_f = 8$ ensemble.}
\end{figure}

\begin{figure}[tbp]
  \includegraphics[width=\linewidth]{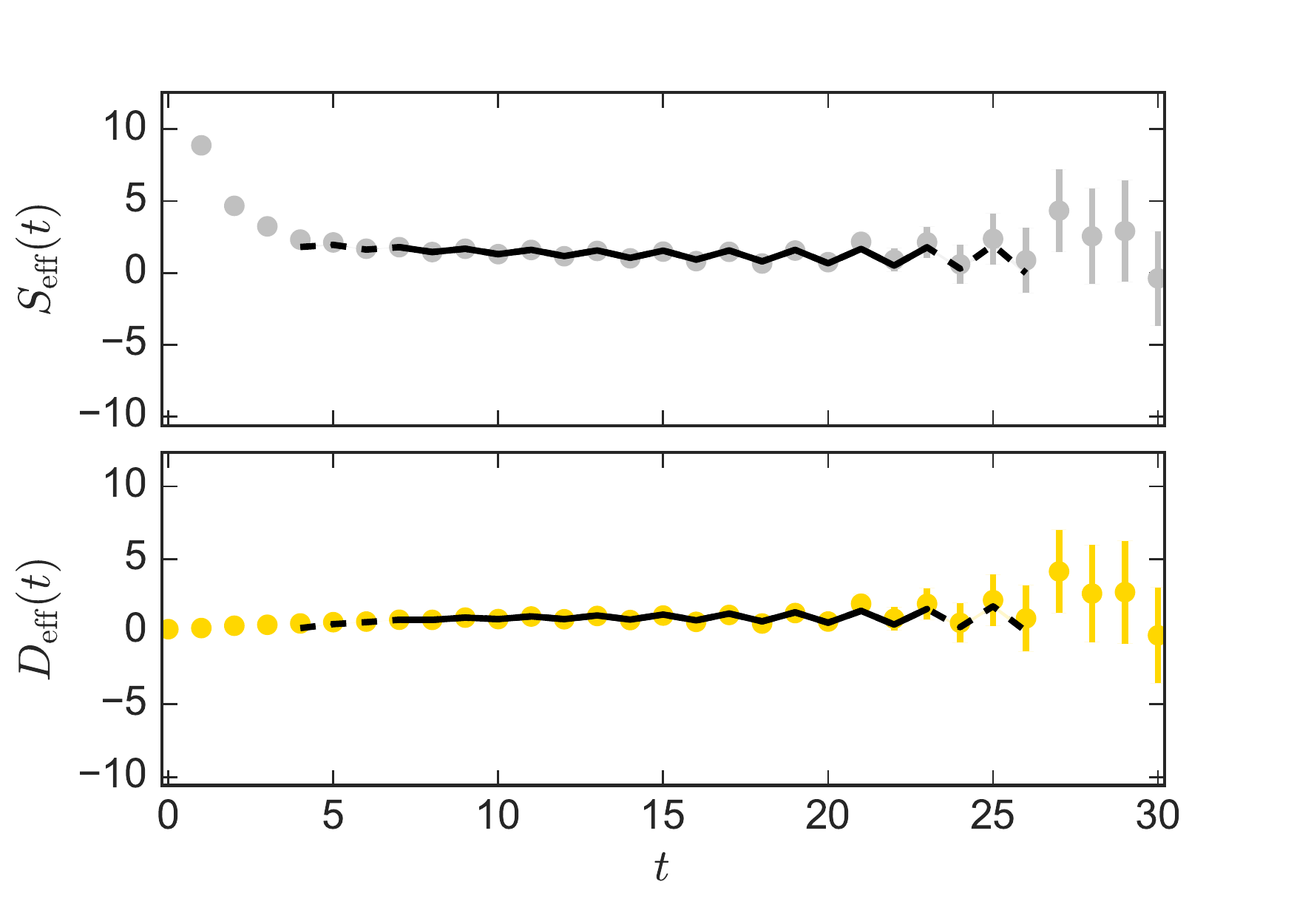} \\
  \includegraphics[width=\linewidth]{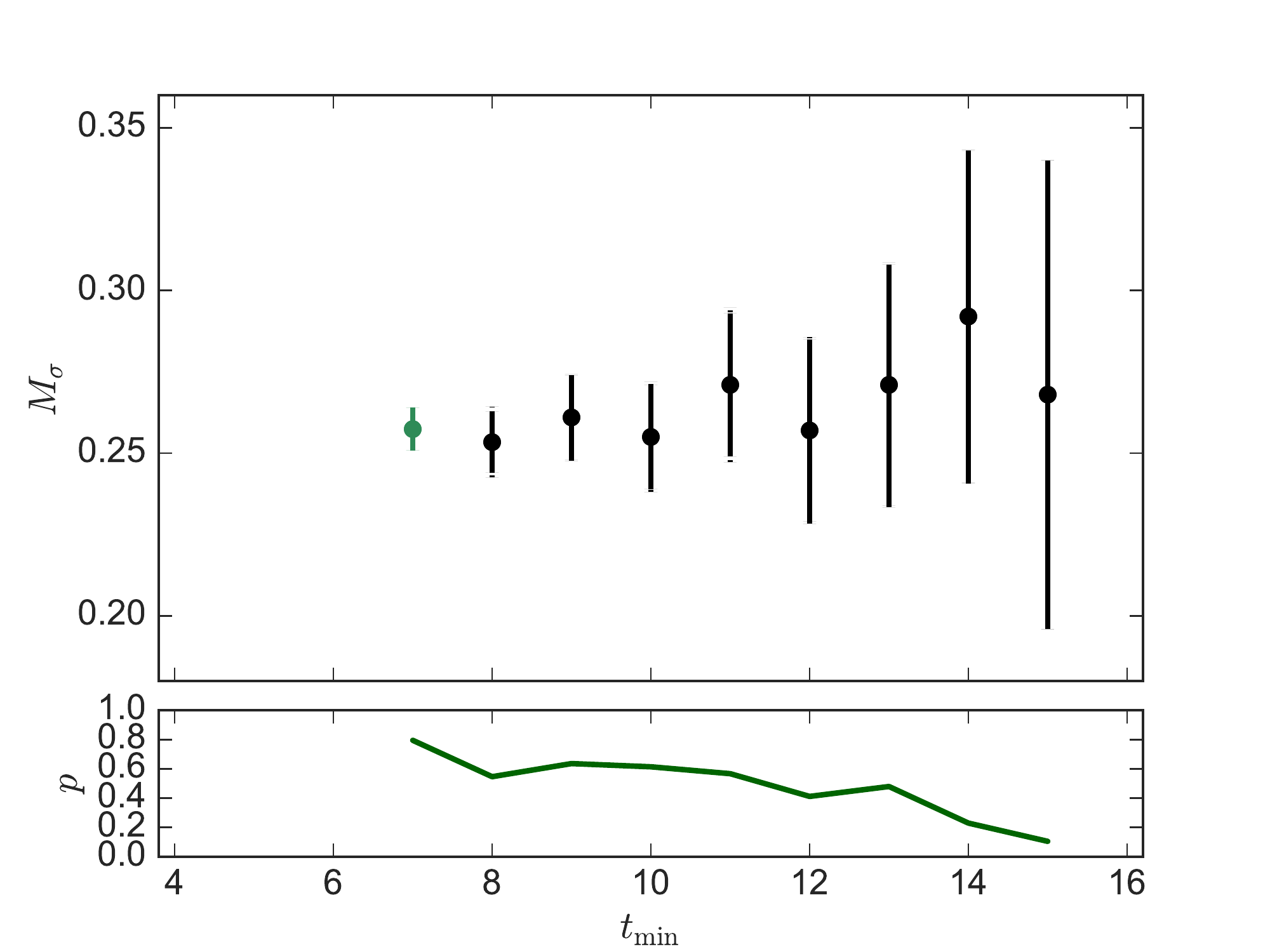}
  \caption{\label{fig:zpp_eff_ensD}Best-fit vs.\ effective correlators (top) and fit-range systematic scan (bottom) for the $0^{++}$ joint fit, $am_f = 0.0075$ $N_f = 8$ ensemble.}
\end{figure}

\begin{figure}[tbp]
  \includegraphics[width=\linewidth]{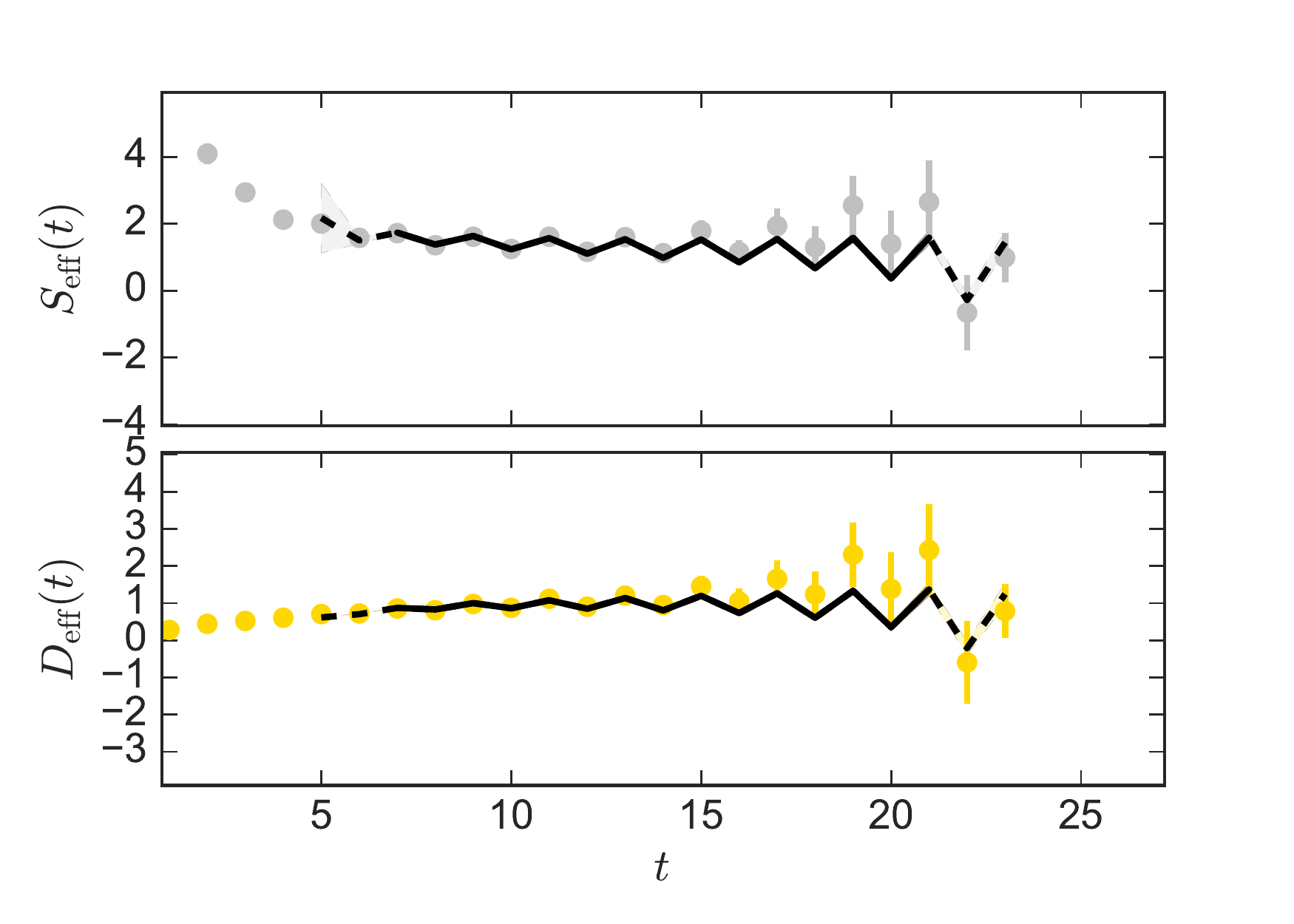} \\
  \includegraphics[width=\linewidth]{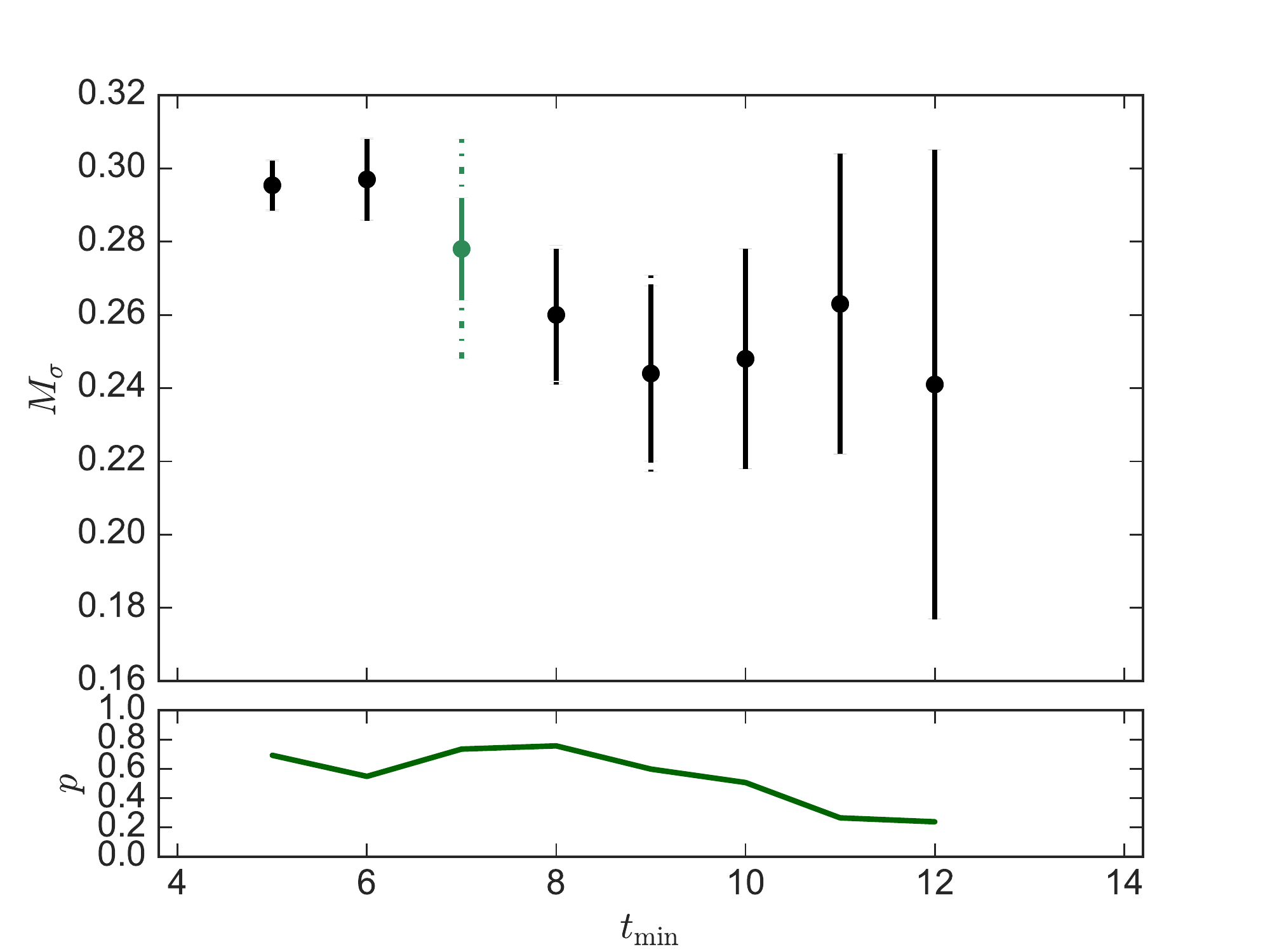}
  \caption{\label{fig:zpp_eff_ensF}Best-fit vs.\ effective correlators (top) and fit-range systematic scan (bottom) for the $0^{++}$ joint fit, $am_f = 0.00889$ $N_f = 8$ ensemble.}
\end{figure}

\begin{figure}[tbp]
  \includegraphics[width=\linewidth]{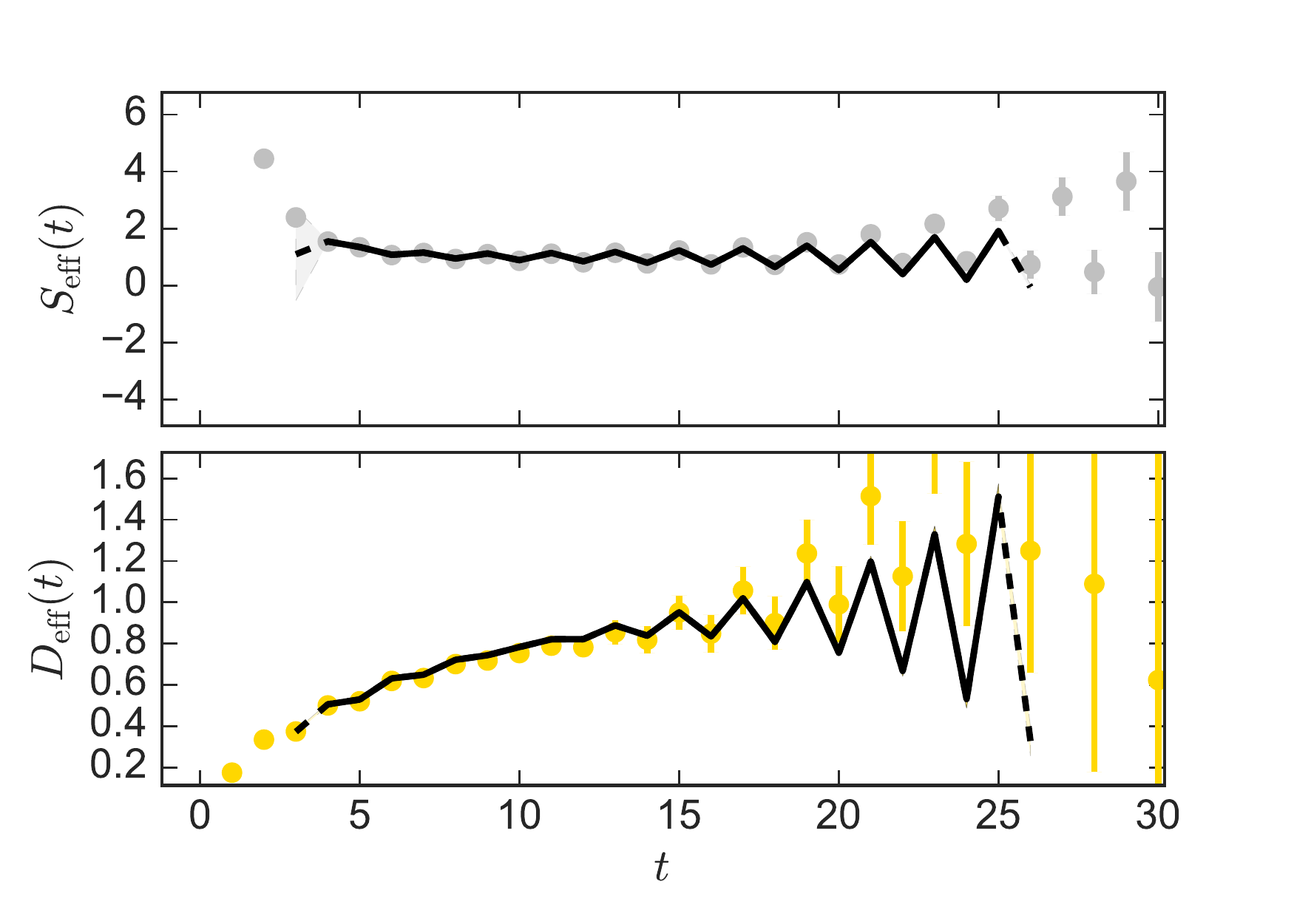} \\
  \includegraphics[width=\linewidth]{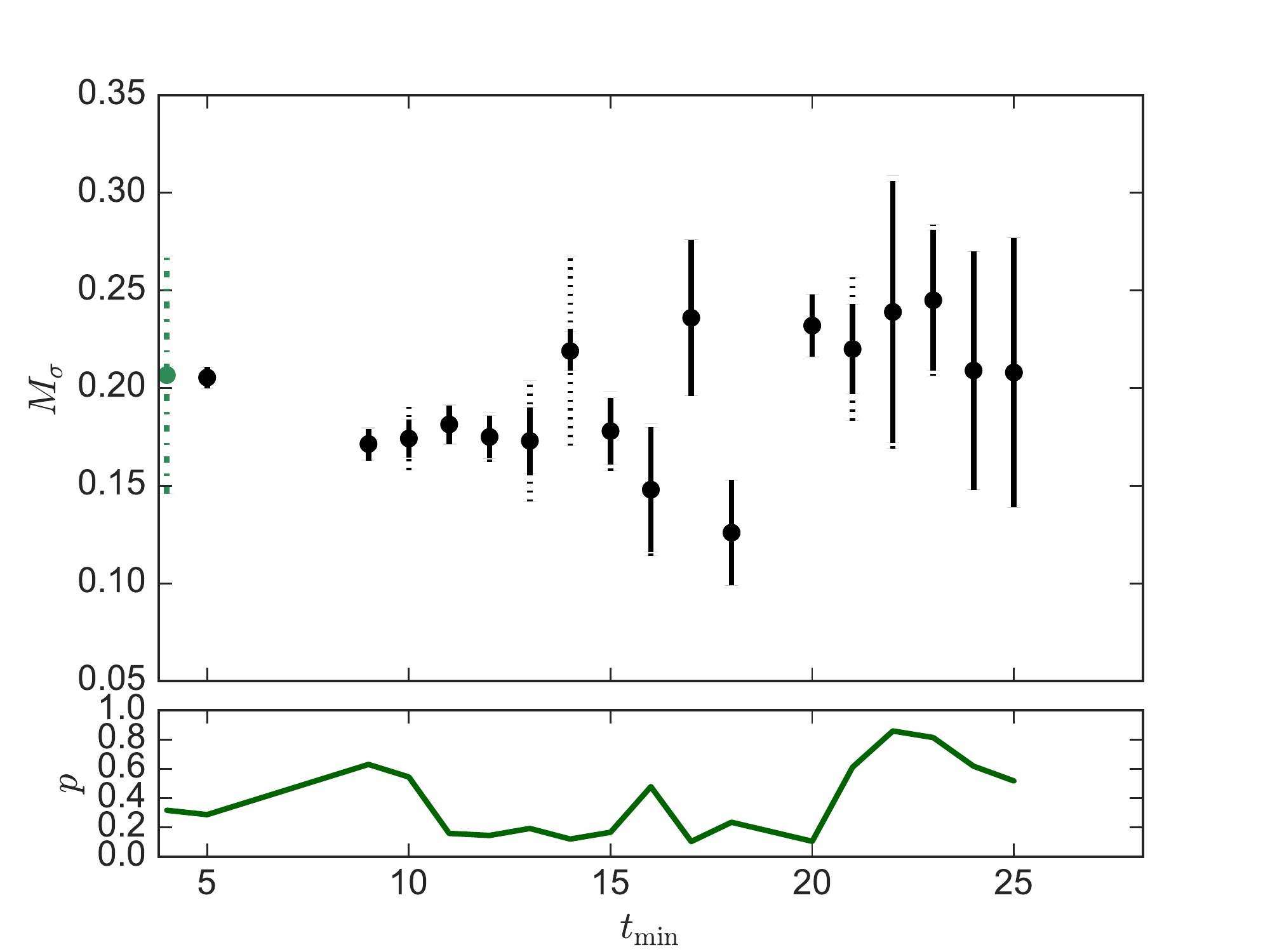}
  \caption{\label{fig:zpp_eff_ensW}Best-fit vs.\ effective correlators (top) and fit-range systematic scan (bottom) for the $0^{++}$ joint fit, $am_f = 0.003$ $N_f = 4$ ensemble.}
\end{figure}

\begin{figure}[tbp]
  \includegraphics[width=\linewidth]{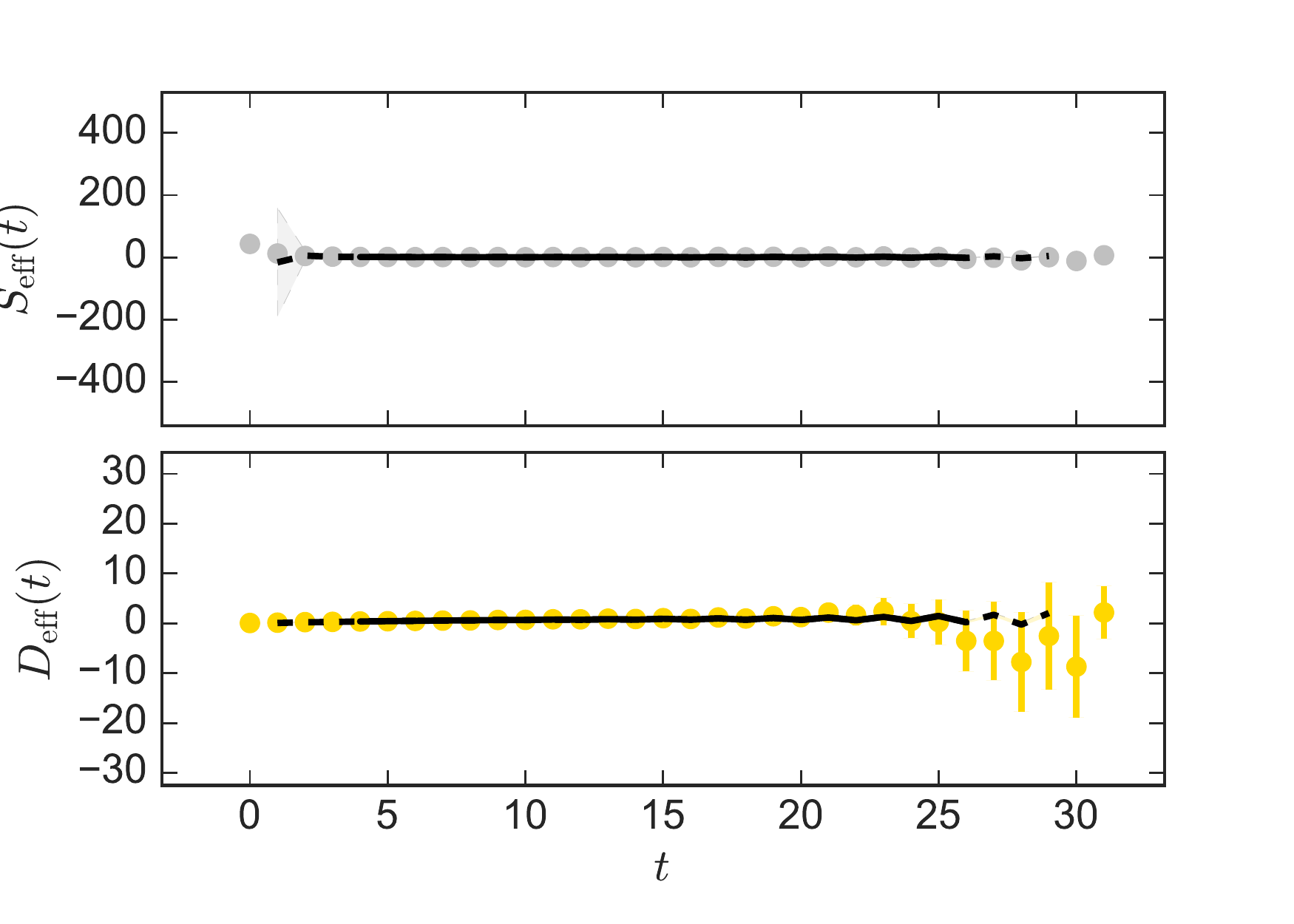} \\
  \includegraphics[width=\linewidth]{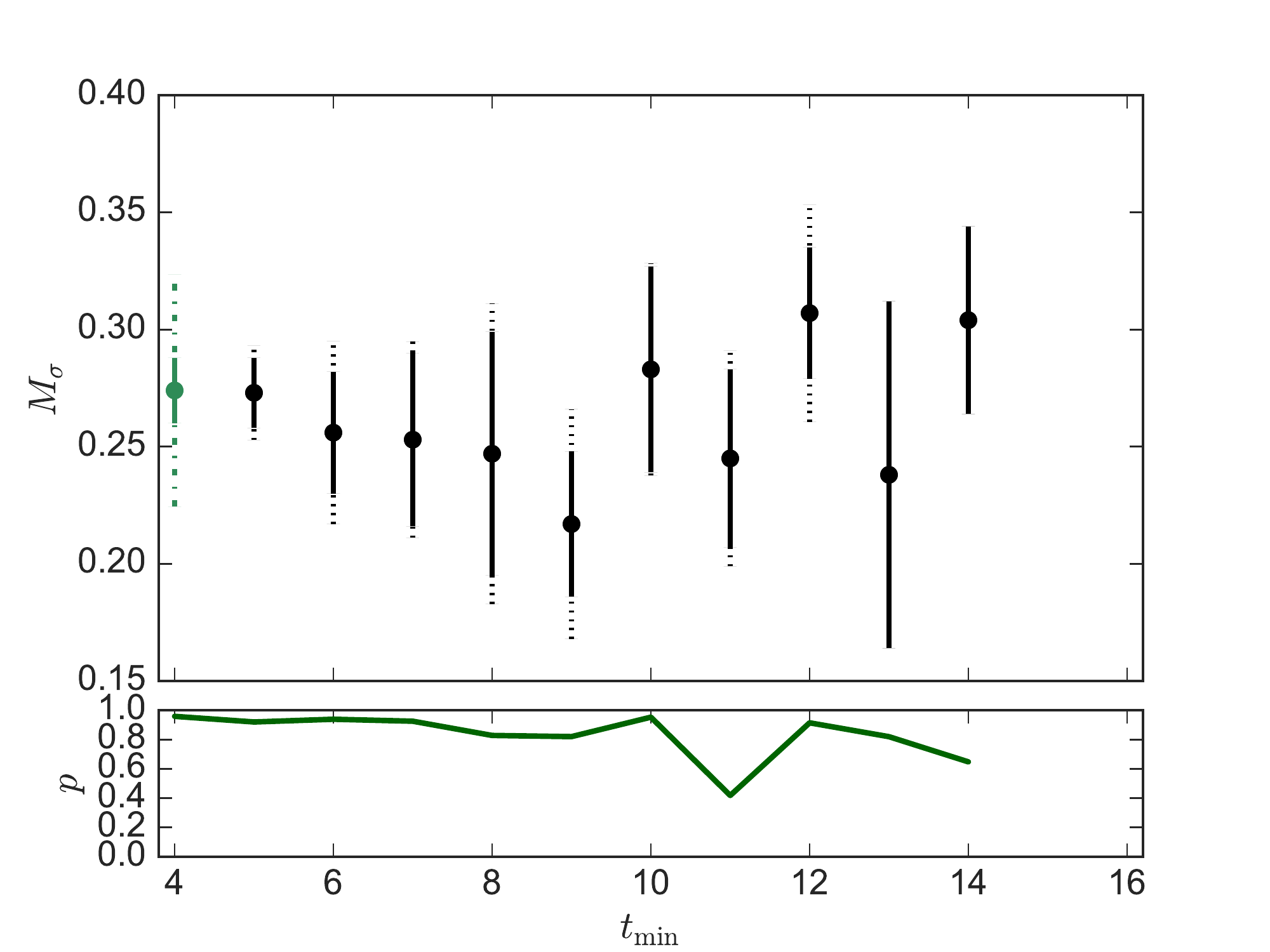}
  \caption{\label{fig:zpp_eff_ensZ}Best-fit vs.\ effective correlators (top) and fit-range systematic scan (bottom) for the $0^{++}$ joint fit, $am_f = 0.007$ $N_f = 4$ ensemble.}
\end{figure}

\begin{figure}[tbp]
  \includegraphics[width=\linewidth]{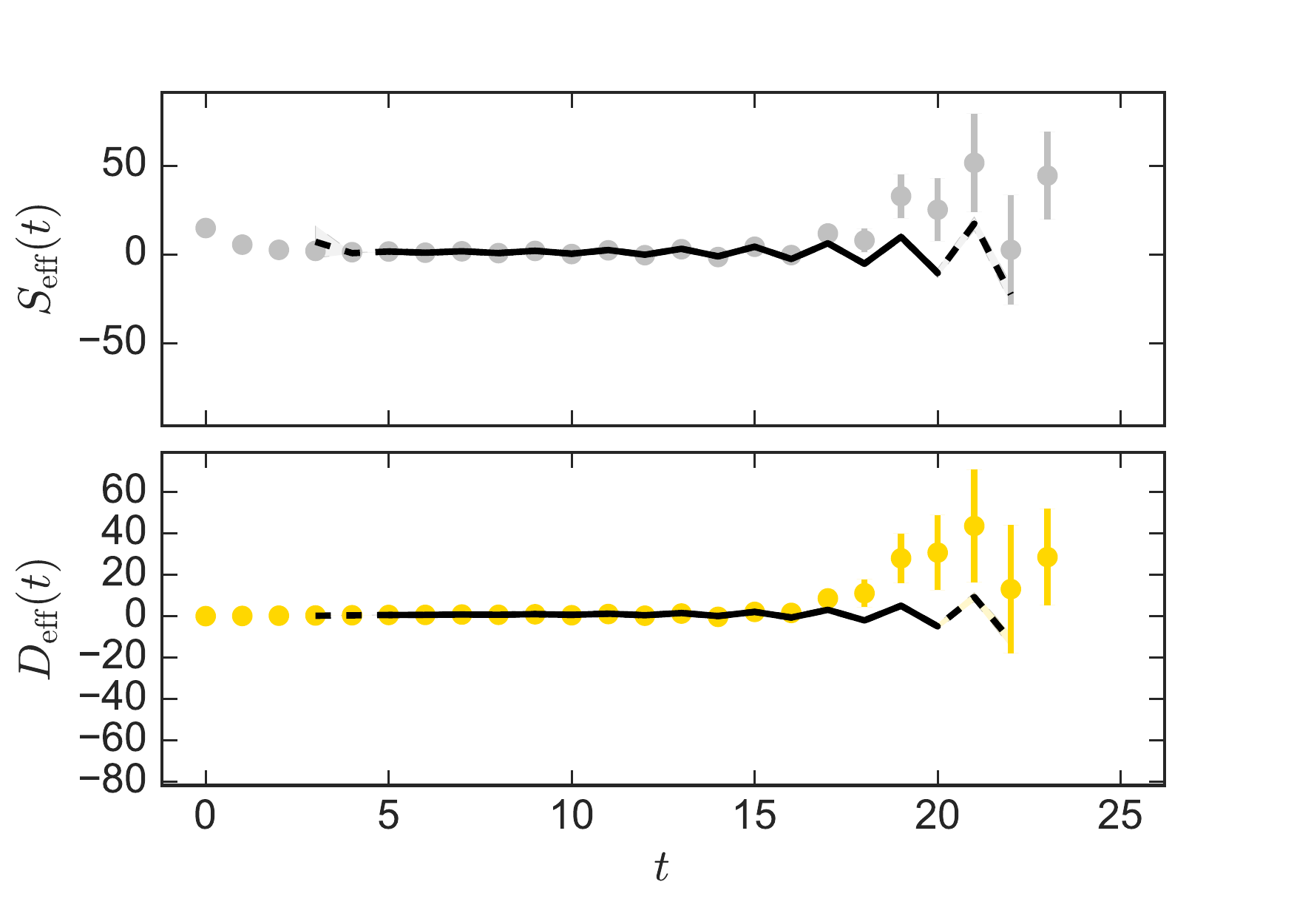} \\
  \includegraphics[width=\linewidth]{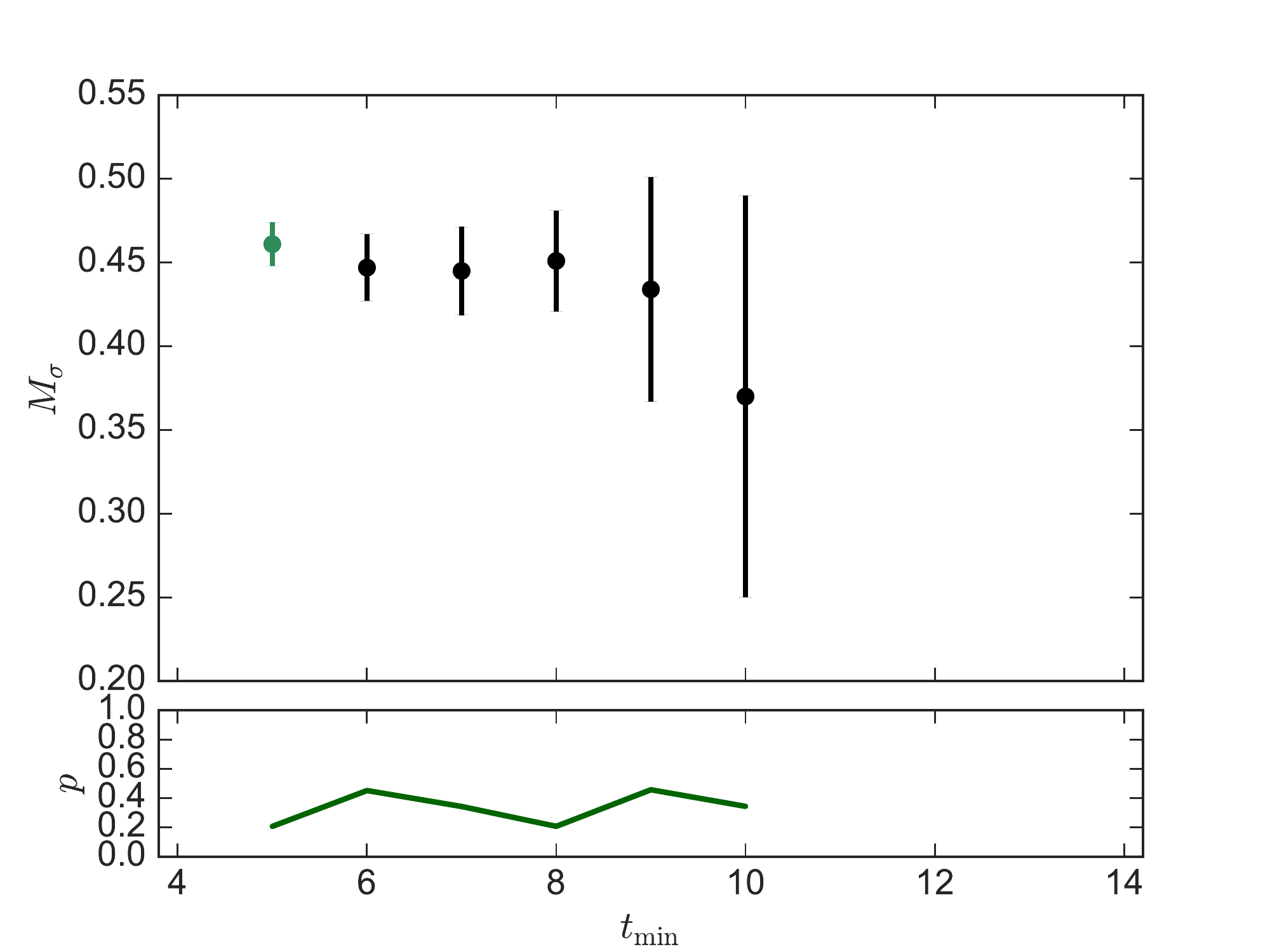}
  \caption{\label{fig:zpp_eff_ensX}Best-fit vs.\ effective correlators (top) and fit-range systematic scan (bottom) for the $0^{++}$ joint fit, $am_f = 0.0125$ $N_f = 4$ ensemble.}
\end{figure}

\begin{figure}[tbp]
  \includegraphics[width=\linewidth]{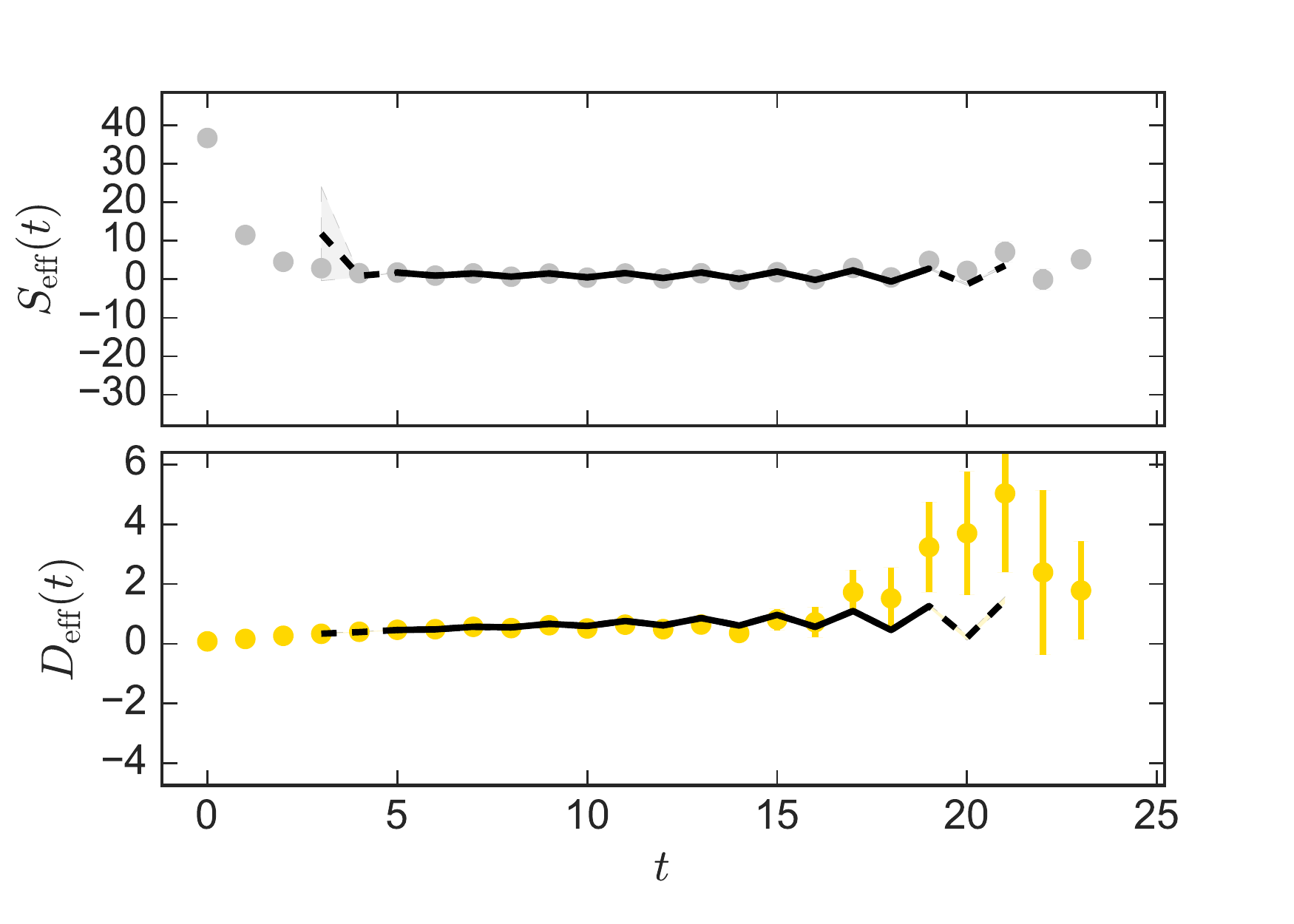} \\
  \includegraphics[width=\linewidth]{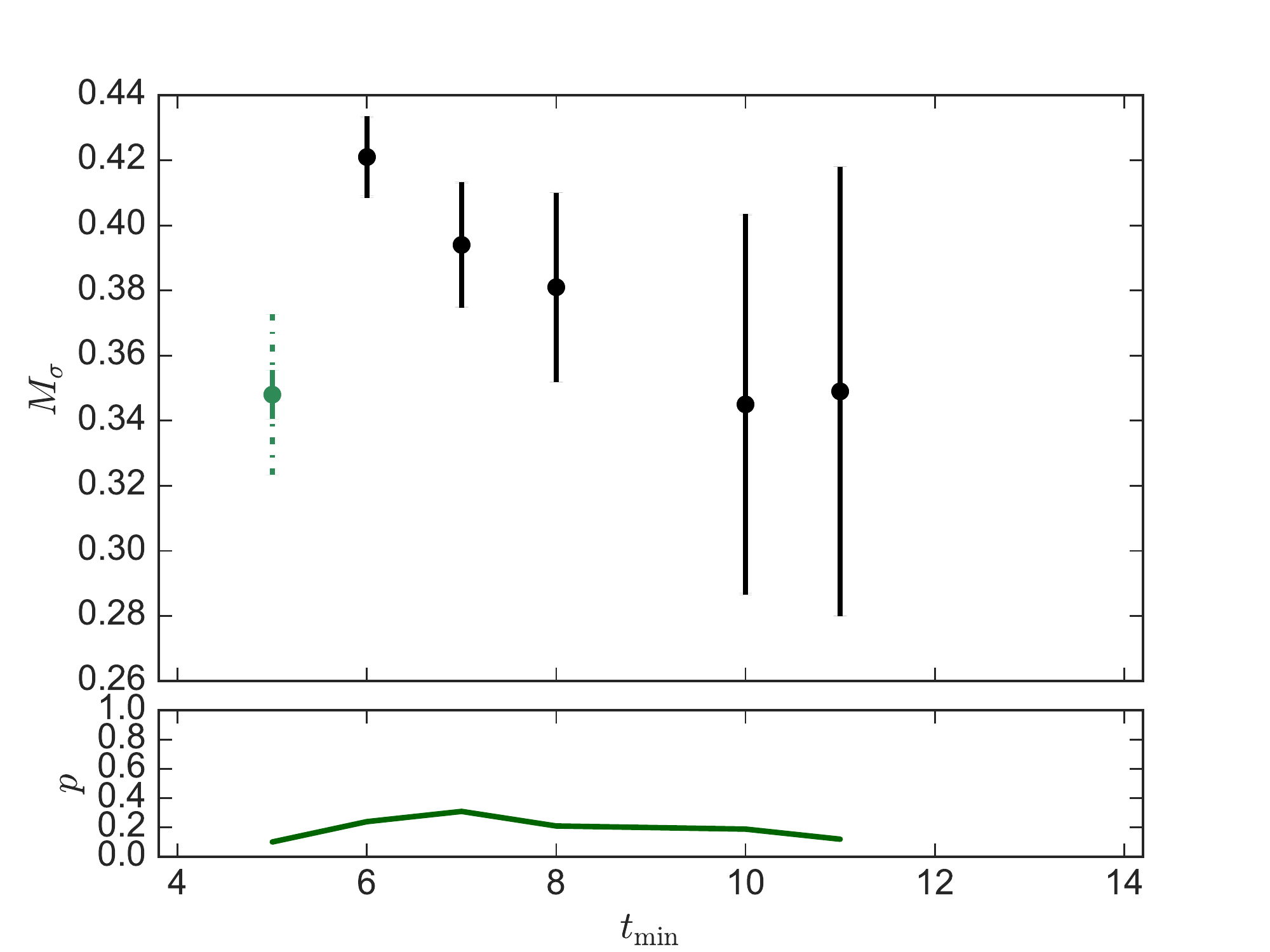}
  \caption{\label{fig:zpp_eff_ensY}Best-fit vs.\ effective correlators (top) and fit-range systematic scan (bottom) for the $0^{++}$ joint fit, $am_f = 0.015$ $N_f = 4$ ensemble.}
\end{figure}

Figures~\ref{fig:zpp_eff_ensA}--\ref{fig:zpp_eff_ensY} show detailed numerical results for the $0^{++}$ joint fit on each 8-flavor and 4-flavor ensemble.
The top panel shows the best fit versus the ``effective correlator'' $C_{\text{eff}}(t) \equiv \frac{C(t)}{A_0} e^{+E_0 t}$, where $A_0$ and $E_0$ are the best-fit ground state amplitude and energy, respectively.
The solid line shows the range of $t$ values used in the best fit, while the dashed lines show the extrapolation of the fit.
These plots demonstrate that our best-fit models describe the correlator data well, even when extrapolated beyond the fit range.
The lower panel in each figure shows the $t_{\text{min}}$ dependence of the joint fit, used to determine the fit-range systematic error.

Finally, we note that the comparison between the 4-flavor and 8-flavor theories provides one of the more important cross-checks on our $0^{++}$ mass extraction.
The qualitatively different results at $N_f = 4$, with the \si found to be much heavier than the $\pi$, are obtained with the same procedure, lattice action, etc.
This rules out the possibility that due to some unaccounted-for systematic effect, we are simply finding $M_{\pi}$ itself from our measurements in the \si channel.

\begin{table*}[tbp]
  \centering
  \renewcommand\arraystretch{1.2}  
  \addtolength{\tabcolsep}{3 pt}   
  \begin{tabular}{cccS[table-format=1.5]|S[table-format=1.7]S[table-format=1.7]S[table-format=1.7]}
    \hline
    $N_f$ & $\be_F$ & $L^3\X N_t$  & $am_f$  & $M_{\si}\sqrt{8t_0}$ & $M_{\si, S}\sqrt{8t_0}$ & $M_{\si, D}\sqrt{8t_0}$ \\
    \hline
    8     & 4.8     & $64^3\X 128$ & 0.00125 & 0.42(15)             & 0.56(10)                & 0.251(62)               \\
    8     & 4.8     & $48^3\X 96$  & 0.00222 & 0.599(91)            & 0.64(15)                & 0.517(55)               \\
    8     & 4.8     & $32^3\X 64$  & 0.005   & 0.753(99)            & 0.972(54)               & 0.720(62)               \\
    8     & 4.8     & $32^3\X 64$  & 0.0075  & 0.973(25)            & 1.084(36)               & 0.908(79)               \\
    8     & 4.8     & $24^3\X 48$  & 0.00889 & 1.01(11)             & 1.048(72)               & 0.770(47)               \\
    \hline
    4     & 6.4     & $24^3\X 48$  & 0.0125  & 1.716(48)            & 1.805(48)               & 1.82(41)                \\
    4     & 6.6     & $48^3\X 96$  & 0.003   & 1.03(33)             & 1.02(19)                & 0.861(75)               \\
    4     & 6.6     & $32^3\X 64$  & 0.007   & 1.33(26)             & 1.22(14)                & 0.97(15)                \\
    4     & 6.6     & $24^3\X 48$  & 0.015   & 1.61(12)             & 1.921(79)               & 1.59(41)                \\
    \hline
  \end{tabular}
  \caption{\label{tab:scalar_results}Results for scalar masses from each of our ensembles.  $M_{\si, S}$ and $M_{\si, D}$ represent determinations from individual fits to the $D(t)$ and $S(t)$ correlators, as described in the text.  All uncertainties include both statistical and fit-range systematic error.}
\end{table*}

\clearpage

\section{\label{app:disc_FV}Discretization and finite-volume effects} 
In this appendix, we give further detail on possible systematic effects due to the finite lattice spacing and lattice volume in our calculations.
We begin with the lattice spacing.
Our $N_f = 8$ lattice calculations are carried out at a single value of the bare gauge coupling.
If we adopt a mass-independent lattice scale setting prescription, i.e., $a = a(\be_F)$, then we are unable to carry out a continuum extrapolation.
Adopting a mass-dependent scheme $a = a(\be_F, m_f)$ is another possibility, for example using the significant $m_f$ dependence observed in $t_0$ (see \fig{fig:Wflow_scale}).
In this case the continuum and chiral ($m_f \to 0$) extrapolations are intertwined.
We defer such an analysis to future work in which the mass dependence is more closely investigated, since this requires the choice of an appropriate chiral effective theory.

\begin{figure}[tbp]
  \includegraphics[width=\linewidth]{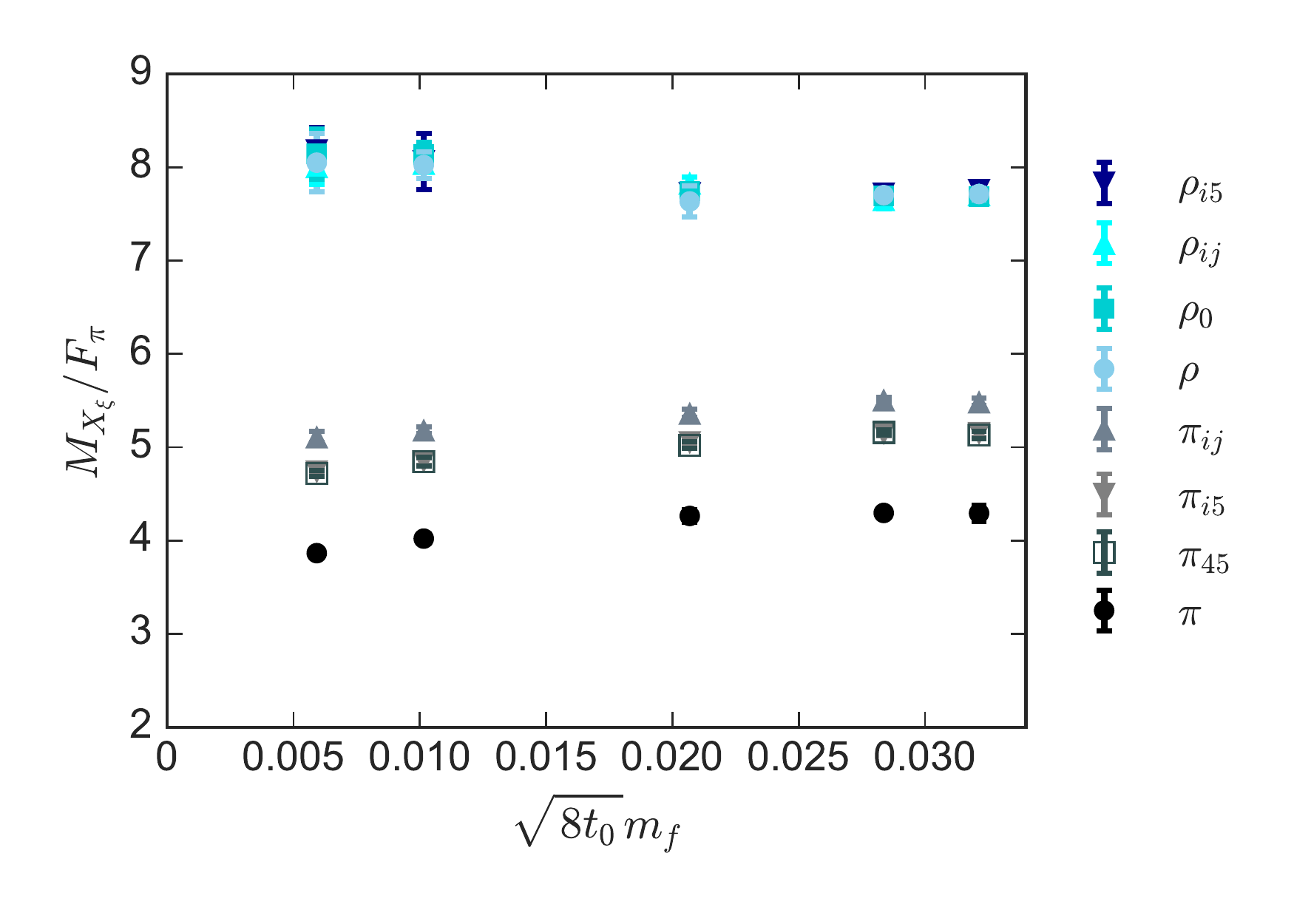}
  \caption{\label{fig:taste_split}Taste-split spectrum for the 8-flavor theory in both the pseudoscalar and vector channels.  The splitting between pion states due to lattice artifacts is significant, on the order of 20\%--30\%.  No significant splitting is visible for the vector mesons.}
\end{figure}

\begin{figure}[tbp]
  \includegraphics[width=\linewidth]{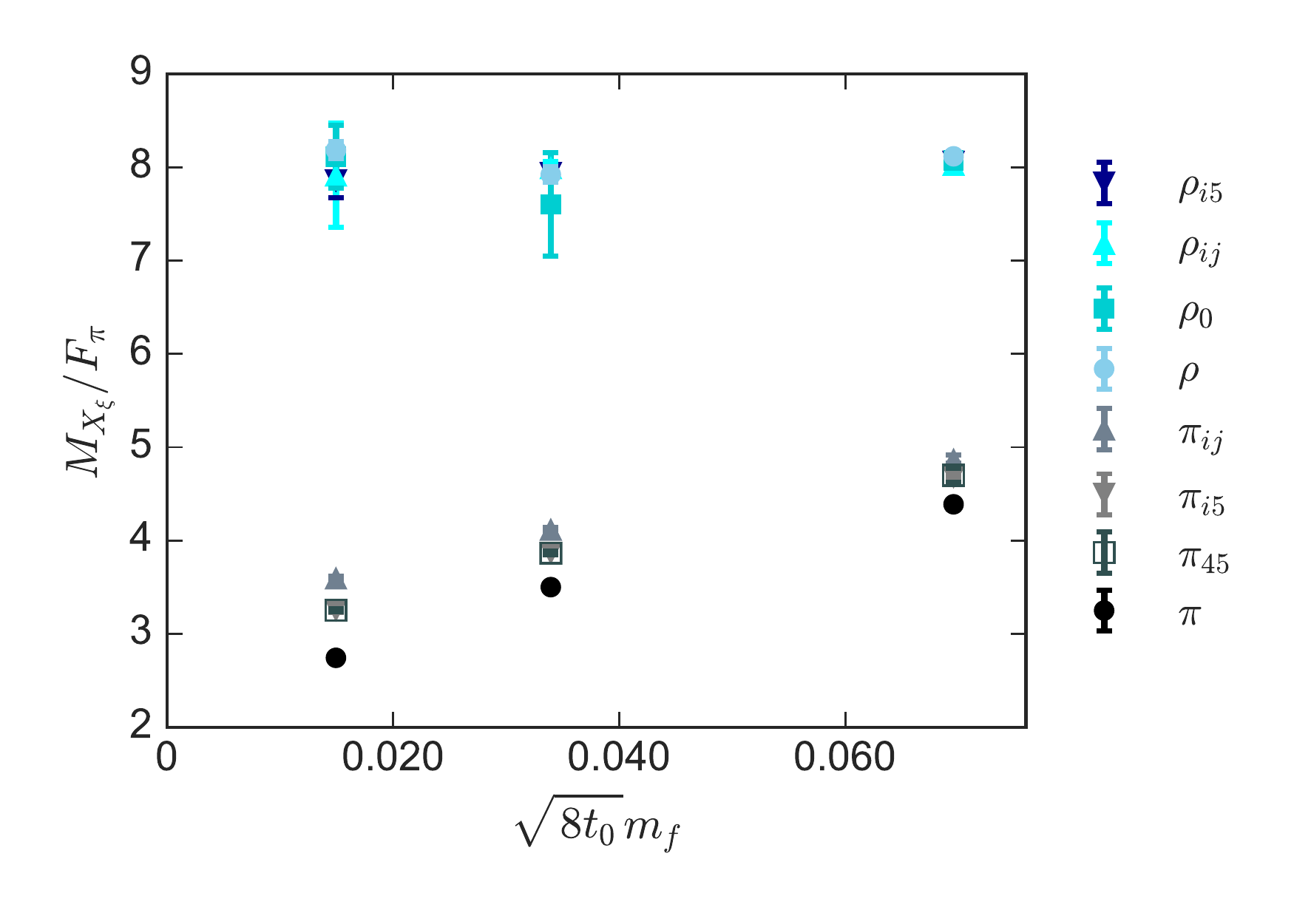}
  \caption{\label{fig:taste_split-4f}Taste-split spectrum for the 4-flavor theory in both the pseudoscalar and vector channels.  The results are qualitatively similar to what is observed for the 8-flavor theory (\protect\fig{fig:taste_split}), although the pion taste splitting is somewhat smaller here.}
\end{figure}

A more direct test for discretization effects involves ``taste breaking'', splitting between the masses of staggered hadrons due to the non-zero lattice spacing.
\tab{tab:taste_results} and Figs.~\ref{fig:taste_split}--\ref{fig:taste_split-4f} show our results for various tastes of the $\pi$ and $\rho$ mesons in both the $N_f = 8$ and $N_f = 4$ theories.
Note that we consider only states which are staggered-flavor singlet, i.e., in the 8-flavor theory we sum over both degenerate staggered species.
No significant taste breaking is seen in the $\rho$ mesons.
The pions, which are particularly sensitive to chiral symmetry breaking, show more significant taste-breaking effects, with masses on the order of 20\%--30\% heavier than the Goldstone pion.
These results are comparable to the 15\%--20\% splittings seen by the MILC Collaboration in QCD for a lattice spacing of $a \approx 0.12$~fm~\cite{Aubin:2004wf} using asqtad-improved staggered fermions.
We note the partial restoration of taste symmetry at finite $a$ predicted by Lee and Sharpe~\cite{Lee:1999zxa} is present in our results, with the $\pi_{i5}$ and $\pi_{45}$ tastes having degenerate masses.

\begin{figure}[tbp]
  \includegraphics[width=\linewidth]{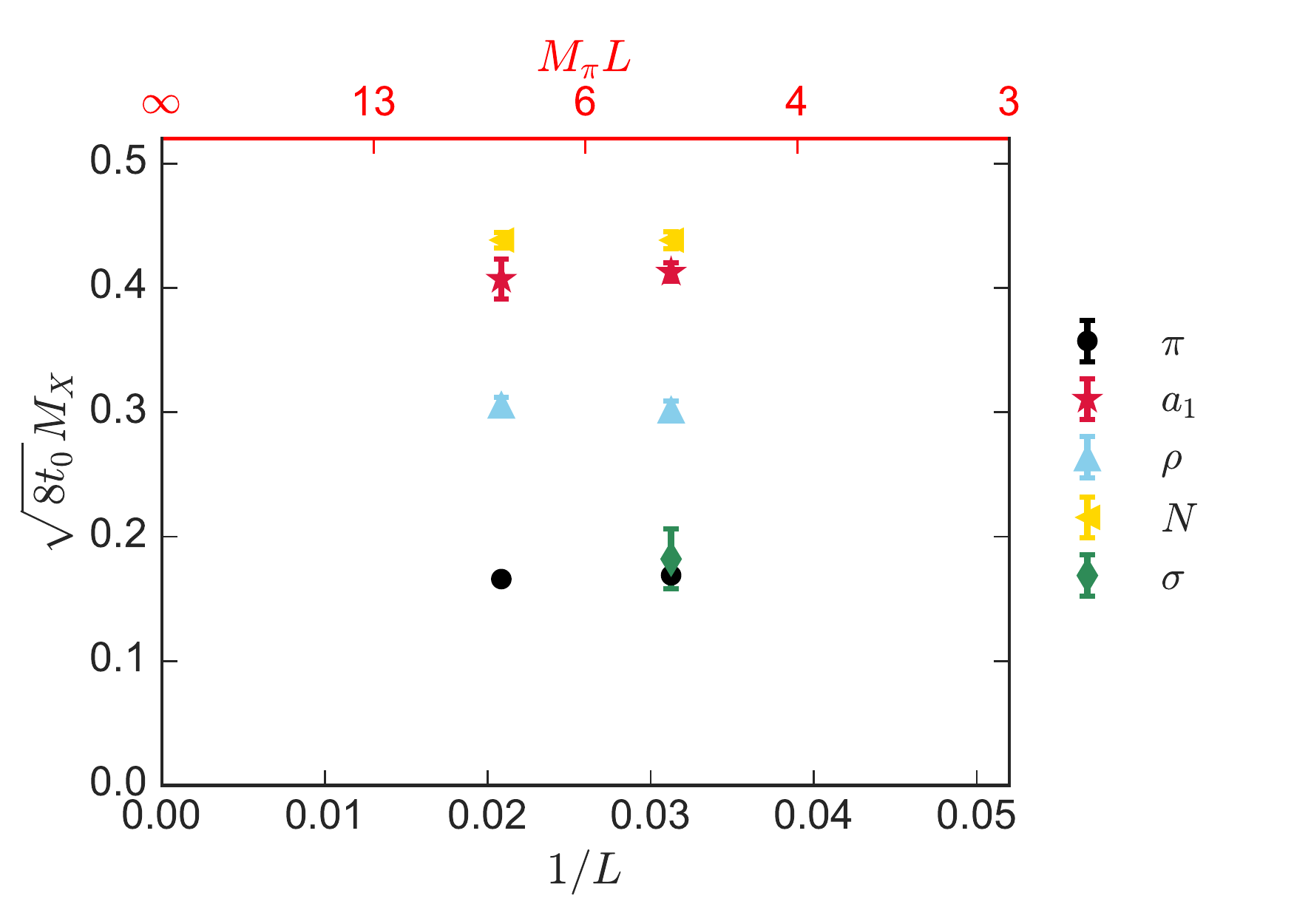}
  \caption{\label{fig:spectrum_finvol_005}The spectrum of the 8-flavor theory vs.\ lattice size $1 / L$ for $am_f = 0.005$.  The lower horizontal axis shows the quantity $1 / L$, while the upper axis shows the finite-volume figure of merit $M_{\pi} L$, with $M_{\pi}$ taken from the largest volume available.  No significant volume dependence is visible.}
\end{figure}

\begin{figure}[tbp]
  \includegraphics[width=\linewidth]{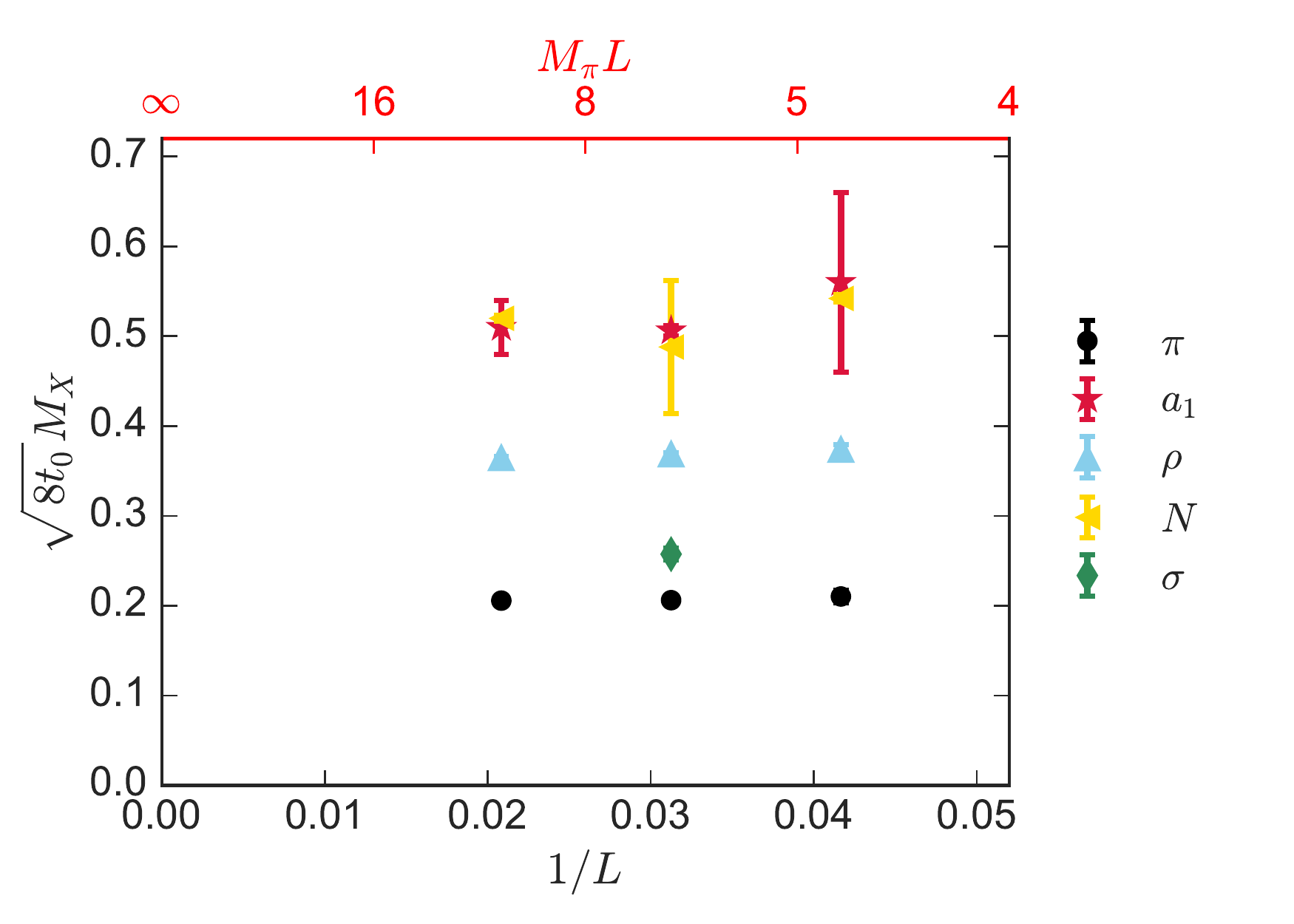}
  \caption{\label{fig:spectrum_finvol_0075}The spectrum of the 8-flavor theory vs.\ lattice size $1 / L$ for $am_f = 0.0075$, plotted as in \protect\fig{fig:spectrum_finvol_005}.  No significant volume dependence is visible.}
\end{figure}

\begin{figure}[tbp]
  \includegraphics[width=\linewidth]{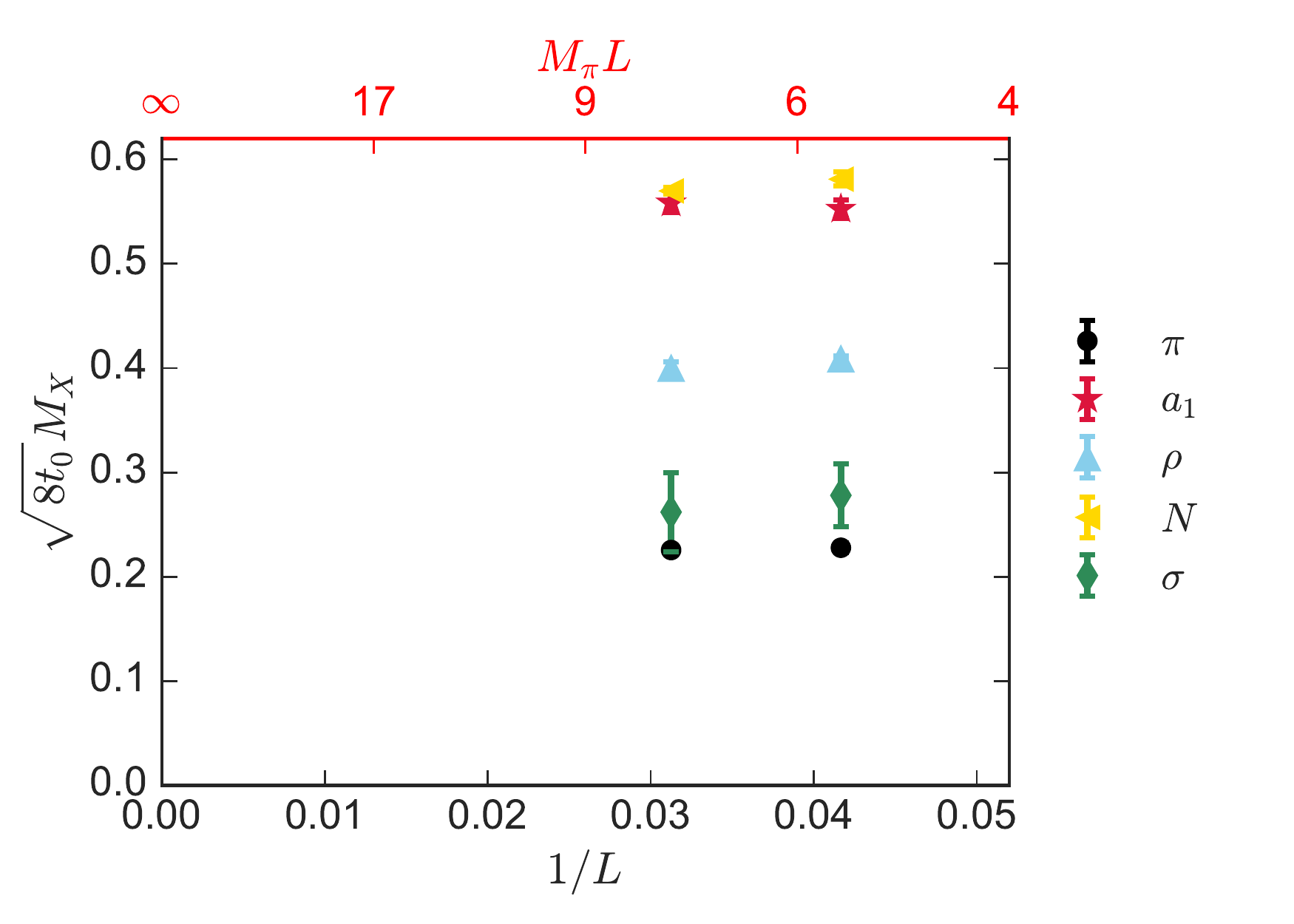}
  \caption{\label{fig:spectrum_finvol_00889}The spectrum of the 8-flavor theory vs.\ lattice size $1 / L$ for $am_f = 0.00889$, plotted as in \protect\fig{fig:spectrum_finvol_005}.  No significant volume dependence is visible.}
\end{figure}

We now turn to finite-volume corrections.
In order to explicitly test for such effects, we analyze a number of additional $N_f = 8$ ensembles that are matched to specific ensembles in our main analysis but have smaller or larger lattice volumes.
These ensembles are specified in \tab{tab:sim_pars_FV}.
Results of our spectrum analysis on these ensembles are given in \tab{tab:spec_results_FV} and shown in Figs.~\ref{fig:spectrum_finvol_005}, \ref{fig:spectrum_finvol_0075} and \ref{fig:spectrum_finvol_00889}.
The disconnected diagrams needed to reconstruct the \si correlator were measured only on a subset of these ensembles, so in some cases we do not compute its mass.

No significant finite-volume dependence is seen for any of the states in our spectrum, including the light \si scalar, although we only have an explicit test of the latter at the heavy fermion mass $am_f = 0.00889$.
In general, direct finite-volume tests are only available on the ensembles with heavier fermion mass, but we note that the volume of the lighter-mass ensembles scales up such that the figure of merit $M_{\pi} L \geq 5.3$ on all ensembles used in the main analysis.

\begin{table*}[tbp]
  \centering
  \renewcommand\arraystretch{1.2}  
  \addtolength{\tabcolsep}{3 pt}   
  \begin{tabular}{ccS[table-format=1.5]|S[table-format=1.9]S[table-format=1.9]S[table-format=1.8]S[table-format=1.8]S[table-format=1.8]S[table-format=1.8]}
    \hline
    $N_f$ & $L^3\X N_t$  & $am_f$  & $M_{\pi}\sqrt{8t_0}$ & $M_{\pi, i5}\sqrt{8t_0}$ & $M_{\pi, ij}\sqrt{8t_0}$ & $M_{\rho, i5}\sqrt{8t_0}$ & $M_{\rho, ij}\sqrt{8t_0}$ & $M_{\rho, 4}\sqrt{8t_0}$ \\
    \hline
    8     & $64^3\X 128$ & 0.00125 & 0.3885(30)           & 0.4759(13)               & 0.5137(52)               & 0.822(25)                 & 0.805(18)                 & 0.819(27)                \\
    8     & $48^3\X 96$  & 0.00222 & 0.5036(55)           & 0.6070(30)               & 0.6493(14)               & 1.010(37)                 & 1.0072(69)                & 1.019(16)                \\
    8     & $32^3\X 64$  & 0.005   & 0.6988(99)           & 0.8273(13)               & 0.8790(33)               & 1.265(15)                 & 1.2833(58)                & 1.2693(87)               \\
    8     & $32^3\X 64$  & 0.0075  & 0.7798(25)           & 0.93408(57)              & 0.9996(22)               & 1.400(17)                 & 1.389(18)                 & 1.3959(38)               \\
    8     & $24^3\X 48$  & 0.00889 & 0.824(16)            & 0.9873(13)               & 1.0525(23)               & 1.4869(94)                & 1.4786(69)                & 1.4743(87)               \\
    \hline
    4     & $24^3\X 48$  & 0.0125  & 0.8518(31)           & 0.98025(56)              & 1.0663(17)               & 1.784(13)                 & 1.7729(97)                & 1.786(16)                \\
    4     & $48^3\X 96$  & 0.003   & 0.48965(55)          & 0.58011(70)              & 0.6425(18)               & 1.402(31)                 & 1.41(10)                  & 1.448(60)                \\
    4     & $32^3\X 64$  & 0.007   & 0.7346(33)           & 0.81086(92)              & 0.8644(23)               & 1.665(16)                 & 1.677(13)                 & 1.624(73)                \\
    4     & $24^3\X 48$  & 0.015   & 1.044(11)            & 1.11317(93)              & 1.1593(93)               & 1.916(26)                 & 1.911(22)                 & 1.9210(88)               \\
    \hline
  \end{tabular}
  \caption{\label{tab:taste_results}Results for taste-split masses from each of our ensembles.  All uncertainties include both statistical and fit-range systematic errors.  The Goldstone pion mass is included as the first column for easier comparison with the taste-split masses.}
\end{table*}

\begin{table*}[tbp]
  \centering
  \renewcommand\arraystretch{1.2}  
  \addtolength{\tabcolsep}{3 pt}   
  \begin{tabular}{S[table-format=1.5]cc|S[table-format=5]cS[table-format=4]S[table-format=3]}
    \hline
    $am_f$  & $L^3\X N_t$          & $\tau$ & {MDTU} & Sep. & {\# Est.} & {Bins} \\ 
    \hline
    0.005   & $\mathbf{48^3\X 96}$ & 1.0    &  3520  & 40   &  89       &  89    \\ 
            &                      &        &  3120  & 40   &  79       &  79    \\ 
    \hline
    0.0075  & $\mathit{24^3\X 48}$ & 1.0    &  9670  & 10   & 968       & 242    \\ 
    \hline
    0.0075  & $\mathbf{48^3\X 96}$ & 1.0    & 19400  & 40   & 486       & 243    \\ 
            &                      &        & 19320  & 40   & 484       & 242    \\ 
    \hline
    0.00889 & $\mathbf{32^3\X 64}$ & 1.0    &  5960  & 40   & 150       &  75    \\ 
            &                      &        &  5960  & 40   & 150       &  75    \\ 
            &                      &        &  5960  & 40   & 150       &  75    \\ 
            &                      &        &  5960  & 40   & 150       &  75    \\ 
    \hline
  \end{tabular}
  \caption{\label{tab:sim_pars_FV}Additional $N_f = 8$ ensembles used to explore finite-volume effects, all with $\be_F = 4.8$.  A detailed description of the columns is given in \protect\tab{tab:sim_pars}.  Volumes indicated in bold (italics) are larger (smaller) than the ensembles included in the main analysis.}
\end{table*}

\begin{table*}[tbp]
  \centering
  \renewcommand\arraystretch{1.2}  
  \addtolength{\tabcolsep}{3 pt}   
  \begin{tabular}{S[table-format=1.5]c|S[table-format=1.9]S[table-format=1.6]S[table-format=1.8]S[table-format=1.7]S[table-format=1.7]}
    \hline
    $am_f$  & $L^3\X N_t$          & $M_{\pi}\sqrt{8t_0}$ & $M_{\si}\sqrt{8t_0}$ & $M_{\rho}\sqrt{8t_0}$ & $M_{a_1}\sqrt{8t_0}$ & $M_N \sqrt{8t_0}$ \\
    \hline
    0.005   & $\mathbf{48^3\X 96}$ & 0.68622(46)          & \textemdash          & 1.267(23)             & 1.684(66)            & 1.814(25)         \\
    0.0075  & $\mathit{24^3\X 48}$ & 0.795(26)            & \textemdash          & 1.417(17)             & 2.12(38)             & 2.050(15)         \\
    0.0075  & $\mathbf{48^3\X 96}$ & 0.77796(79)          & \textemdash          & 1.3822(28)            & 1.93(11)             & 1.966(12)         \\
    0.00889 & $\mathbf{32^3\X 64}$ & 0.8156(87)           & 0.95(14)             & 1.448(20)             & 2.019(34)            & 2.060(12)         \\
    \hline
  \end{tabular}
  \caption{\label{tab:spec_results_FV}Results for masses from each of our $N_f = 8$ $\be_F = 4.8$ finite-volume test ensembles.  The uncertainties include both statistical and fit-range systematic errors.}
\end{table*}

\raggedright
\bibliographystyle{apsrev}
\bibliography{KS_nHYP_8f}

\begin{thebibliography}{117}
\expandafter\ifx\csname natexlab\endcsname\relax\def\natexlab#1{#1}\fi
\expandafter\ifx\csname bibnamefont\endcsname\relax
  \def\bibnamefont#1{#1}\fi
\expandafter\ifx\csname bibfnamefont\endcsname\relax
  \def\bibfnamefont#1{#1}\fi
\expandafter\ifx\csname citenamefont\endcsname\relax
  \def\citenamefont#1{#1}\fi
\expandafter\ifx\csname url\endcsname\relax
  \def\url#1{\texttt{#1}}\fi
\expandafter\ifx\csname urlprefix\endcsname\relax\def\urlprefix{URL }\fi
\providecommand{\bibinfo}[2]{#2}
\providecommand{\eprint}[2][]{\url{#2}}

\bibitem[{\citenamefont{Aad et~al.}(2012)}]{Aad:2012tfa}
\bibinfo{collaboration}{ATLAS Collaboration}, \bibinfo{journal}{Phys.
  Lett.} \textbf{\bibinfo{volume}{B716}}, \bibinfo{pages}{1}
  (\bibinfo{year}{2012}), \eprint{1207.7214}.

\bibitem[{\citenamefont{Chatrchyan et~al.}(2012)}]{Chatrchyan:2012ufa}
\bibinfo{collaboration}{CMS Collaboration},
  \bibinfo{journal}{Phys. Lett.} \textbf{\bibinfo{volume}{B716}},
  \bibinfo{pages}{30} (\bibinfo{year}{2012}), \eprint{1207.7235}.

\bibitem[{\citenamefont{Aad et~al.}(2016)}]{Khachatryan:2016vau}
\bibinfo{collaboration}{ATLAS and CMS Collaborations},
  \bibinfo{journal}{JHEP} \textbf{\bibinfo{volume}{1608}}, \bibinfo{pages}{045}
  (\bibinfo{year}{2016}), \eprint{1606.02266}.

\bibitem[{\citenamefont{DeGrand}(2016)}]{DeGrand:2015zxa}
\bibinfo{author}{\bibfnamefont{T.}~\bibnamefont{DeGrand}},
  \bibinfo{journal}{Rev. Mod. Phys.} \textbf{\bibinfo{volume}{88}},
  \bibinfo{pages}{015001} (\bibinfo{year}{2016}), \eprint{1510.05018}.

\bibitem[{\citenamefont{Nogradi and Patella}(2016)}]{Nogradi:2016qek}
\bibinfo{author}{\bibfnamefont{D.}~\bibnamefont{Nogradi}} \bibnamefont{and}
  \bibinfo{author}{\bibfnamefont{A.}~\bibnamefont{Patella}},
  \bibinfo{journal}{Int. J. Mod. Phys.} \textbf{\bibinfo{volume}{A31}},
  \bibinfo{pages}{1643003} (\bibinfo{year}{2016}), \eprint{1607.07638}.

\bibitem[{\citenamefont{Svetitsky}(2018)}]{Svetitsky:2017xqk}
\bibinfo{author}{\bibfnamefont{B.}~\bibnamefont{Svetitsky}},
  \bibinfo{journal}{EPJ Web Conf.} \textbf{\bibinfo{volume}{175}},
  \bibinfo{pages}{01017} (\bibinfo{year}{2018}), \eprint{1708.04840}.

\bibitem[{\citenamefont{Aoki et~al.}(2014{\natexlab{a}})\citenamefont{Aoki,
  Aoyama, Kurachi, Maskawa, Nagai, Ohki, Rinaldi, Shibata, Yamawaki, and
  Yamazaki}}]{Aoki:2013pca}
\bibinfo{author}{\bibfnamefont{Y.}~\bibnamefont{Aoki}},
  \bibinfo{author}{\bibfnamefont{T.}~\bibnamefont{Aoyama}},
  \bibinfo{author}{\bibfnamefont{M.}~\bibnamefont{Kurachi}},
  \bibinfo{author}{\bibfnamefont{T.}~\bibnamefont{Maskawa}},
  \bibinfo{author}{\bibfnamefont{K.-i.} \bibnamefont{Nagai}},
  \bibinfo{author}{\bibfnamefont{H.}~\bibnamefont{Ohki}},
  \bibinfo{author}{\bibfnamefont{E.}~\bibnamefont{Rinaldi}},
  \bibinfo{author}{\bibfnamefont{A.}~\bibnamefont{Shibata}},
  \bibinfo{author}{\bibfnamefont{K.}~\bibnamefont{Yamawaki}} \bibnamefont{and}
  \bibinfo{author}{\bibfnamefont{T.}~\bibnamefont{Yamazaki}}
  (\bibinfo{collaboration}{LatKMI Collaboration}), in
  \emph{\bibinfo{booktitle}{{Strong Coupling Gauge Theories in the LHC
  Perspective (SCGT12)}}} (\bibinfo{year}{2014}{\natexlab{a}}),
  \bibinfo{pages}{80--88}, \eprint{1302.4577}.

\bibitem[{\citenamefont{Appelquist et~al.}(2013)\citenamefont{Appelquist,
  Brower, Catterall, Fleming, Giedt, Hasenfratz, Kuti, Neil, and
  Schaich}}]{Appelquist:2013sia}
\bibinfo{author}{\bibfnamefont{T.}~\bibnamefont{Appelquist}},
  \bibinfo{author}{\bibfnamefont{R.}~\bibnamefont{Brower}},
  \bibinfo{author}{\bibfnamefont{S.}~\bibnamefont{Catterall}},
  \bibinfo{author}{\bibfnamefont{G.}~\bibnamefont{Fleming}},
  \bibinfo{author}{\bibfnamefont{J.}~\bibnamefont{Giedt}},
  \bibinfo{author}{\bibfnamefont{A.}~\bibnamefont{Hasenfratz}},
  \bibinfo{author}{\bibfnamefont{J.}~\bibnamefont{Kuti}},
  \bibinfo{author}{\bibfnamefont{E.}~\bibnamefont{Neil}} \bibnamefont{and}
  \bibinfo{author}{\bibfnamefont{D.}~\bibnamefont{Schaich}}, in
  \emph{\bibinfo{booktitle}{{Community Summer Study on the Future of U.S.\
  Particle Physics}}} (\bibinfo{year}{2013}), \eprint{1309.1206}.

\bibitem[{\citenamefont{Aoki et~al.}(2013{\natexlab{a}})\citenamefont{Aoki,
  Aoyama, Kurachi, Maskawa, Nagai, Ohki, Rinaldi, Shibata, Yamawaki, and
  Yamazaki}}]{Aoki:2013zsa}
\bibinfo{author}{\bibfnamefont{Y.}~\bibnamefont{Aoki}},
  \bibinfo{author}{\bibfnamefont{T.}~\bibnamefont{Aoyama}},
  \bibinfo{author}{\bibfnamefont{M.}~\bibnamefont{Kurachi}},
  \bibinfo{author}{\bibfnamefont{T.}~\bibnamefont{Maskawa}},
  \bibinfo{author}{\bibfnamefont{K.-i.} \bibnamefont{Nagai}},
  \bibinfo{author}{\bibfnamefont{H.}~\bibnamefont{Ohki}},
  \bibinfo{author}{\bibfnamefont{E.}~\bibnamefont{Rinaldi}},
  \bibinfo{author}{\bibfnamefont{A.}~\bibnamefont{Shibata}},
  \bibinfo{author}{\bibfnamefont{K.}~\bibnamefont{Yamawaki}} \bibnamefont{and}
  \bibinfo{author}{\bibfnamefont{T.}~\bibnamefont{Yamazaki}}
  (\bibinfo{collaboration}{LatKMI Collaboration}), \bibinfo{journal}{Phys. Rev.
  Lett.} \textbf{\bibinfo{volume}{111}}, \bibinfo{pages}{162001}
  (\bibinfo{year}{2013}{\natexlab{a}}), \eprint{1305.6006}.

\bibitem[{\citenamefont{Aoki et~al.}(2014{\natexlab{b}})\citenamefont{Aoki,
  Aoyama, Kurachi, Maskawa, Miura, Nagai, Ohki, Rinaldi, Shibata, Yamawaki
  et~al.}}]{Aoki:2013hla}
\bibinfo{author}{\bibfnamefont{Y.}~\bibnamefont{Aoki}},
  \bibinfo{author}{\bibfnamefont{T.}~\bibnamefont{Aoyama}},
  \bibinfo{author}{\bibfnamefont{M.}~\bibnamefont{Kurachi}},
  \bibinfo{author}{\bibfnamefont{T.}~\bibnamefont{Maskawa}},
  \bibinfo{author}{\bibfnamefont{K.}~\bibnamefont{Miura}},
  \bibinfo{author}{\bibfnamefont{K.-i.} \bibnamefont{Nagai}},
  \bibinfo{author}{\bibfnamefont{H.}~\bibnamefont{Ohki}},
  \bibinfo{author}{\bibfnamefont{E.}~\bibnamefont{Rinaldi}},
  \bibinfo{author}{\bibfnamefont{A.}~\bibnamefont{Shibata}},
  \bibinfo{author}{\bibfnamefont{K.}~\bibnamefont{Yamawaki}}
  \bibnamefont{and} \bibinfo{author}{\bibfnamefont{T.}~\bibnamefont{Yamazaki}}
  (\bibinfo{collaboration}{LatKMI Collaboration}),
  \bibinfo{journal}{PoS} \textbf{\bibinfo{volume}{LATTICE2013}},
  \bibinfo{pages}{077} (\bibinfo{year}{2014}{\natexlab{b}}),
  \eprint{1311.6885}.

\bibitem[{\citenamefont{Aoki et~al.}(2014{\natexlab{c}})\citenamefont{Aoki,
  Aoyama, Kurachi, Maskawa, Miura, Nagai, Ohki, Rinaldi, Shibata, Yamawaki
  et~al.}}]{Aoki:2014oha}
\bibinfo{author}{\bibfnamefont{Y.}~\bibnamefont{Aoki}},
  \bibinfo{author}{\bibfnamefont{T.}~\bibnamefont{Aoyama}},
  \bibinfo{author}{\bibfnamefont{M.}~\bibnamefont{Kurachi}},
  \bibinfo{author}{\bibfnamefont{T.}~\bibnamefont{Maskawa}},
  \bibinfo{author}{\bibfnamefont{K.}~\bibnamefont{Miura}},
  \bibinfo{author}{\bibfnamefont{K.-i.} \bibnamefont{Nagai}},
  \bibinfo{author}{\bibfnamefont{H.}~\bibnamefont{Ohki}},
  \bibinfo{author}{\bibfnamefont{E.}~\bibnamefont{Rinaldi}},
  \bibinfo{author}{\bibfnamefont{A.}~\bibnamefont{Shibata}},
  \bibinfo{author}{\bibfnamefont{K.}~\bibnamefont{Yamawaki}}
  \bibnamefont{and} \bibinfo{author}{\bibfnamefont{T.}~\bibnamefont{Yamazaki}}
  (\bibinfo{collaboration}{LatKMI Collaboration}),
  \bibinfo{journal}{Phys. Rev.} \textbf{\bibinfo{volume}{D89}},
  \bibinfo{pages}{111502} (\bibinfo{year}{2014}{\natexlab{c}}),
  \eprint{1403.5000}.

\bibitem[{\citenamefont{Athenodorou et~al.}(2015)\citenamefont{Athenodorou,
  Bennett, Bergner, and Lucini}}]{Athenodorou:2014eua}
\bibinfo{author}{\bibfnamefont{A.}~\bibnamefont{Athenodorou}},
  \bibinfo{author}{\bibfnamefont{E.}~\bibnamefont{Bennett}},
  \bibinfo{author}{\bibfnamefont{G.}~\bibnamefont{Bergner}} \bibnamefont{and}
  \bibinfo{author}{\bibfnamefont{B.}~\bibnamefont{Lucini}},
  \bibinfo{journal}{Phys. Rev.} \textbf{\bibinfo{volume}{D91}},
  \bibinfo{pages}{114508} (\bibinfo{year}{2015}), \eprint{1412.5994}.

\bibitem[{\citenamefont{Fodor et~al.}(2015{\natexlab{a}})\citenamefont{Fodor,
  Holland, Kuti, Mondal, Nogradi, and Wong}}]{Fodor:2015vwa}
\bibinfo{author}{\bibfnamefont{Z.}~\bibnamefont{Fodor}},
  \bibinfo{author}{\bibfnamefont{K.}~\bibnamefont{Holland}},
  \bibinfo{author}{\bibfnamefont{J.}~\bibnamefont{Kuti}},
  \bibinfo{author}{\bibfnamefont{S.}~\bibnamefont{Mondal}},
  \bibinfo{author}{\bibfnamefont{D.}~\bibnamefont{Nogradi}} \bibnamefont{and}
  \bibinfo{author}{\bibfnamefont{C.~H.} \bibnamefont{Wong}},
  \bibinfo{journal}{PoS} \textbf{\bibinfo{volume}{LATTICE2014}},
  \bibinfo{pages}{244} (\bibinfo{year}{2015}{\natexlab{a}}),
  \eprint{1502.00028}.

\bibitem[{\citenamefont{Rinaldi}(2017)}]{Rinaldi:2015axa}
\bibinfo{author}{\bibfnamefont{E.}~\bibnamefont{Rinaldi}}
  (\bibinfo{collaboration}{LSD Collaboration}), \bibinfo{journal}{Int. J. Mod.
  Phys.} \textbf{\bibinfo{volume}{A32}}, \bibinfo{pages}{1747002}
  (\bibinfo{year}{2017}), \eprint{1510.06771}.

\bibitem[{\citenamefont{Brower et~al.}(2016)\citenamefont{Brower, Hasenfratz,
  Rebbi, Weinberg, and Witzel}}]{Brower:2015owo}
\bibinfo{author}{\bibfnamefont{R.~C.} \bibnamefont{Brower}},
  \bibinfo{author}{\bibfnamefont{A.}~\bibnamefont{Hasenfratz}},
  \bibinfo{author}{\bibfnamefont{C.}~\bibnamefont{Rebbi}},
  \bibinfo{author}{\bibfnamefont{E.}~\bibnamefont{Weinberg}} \bibnamefont{and}
  \bibinfo{author}{\bibfnamefont{O.}~\bibnamefont{Witzel}},
  \bibinfo{journal}{Phys. Rev.} \textbf{\bibinfo{volume}{D93}},
  \bibinfo{pages}{075028} (\bibinfo{year}{2016}), \eprint{1512.02576}.

\bibitem[{\citenamefont{Fodor et~al.}(2016)\citenamefont{Fodor, Holland, Kuti,
  Mondal, Nogradi, and Wong}}]{Fodor:2016pls}
\bibinfo{author}{\bibfnamefont{Z.}~\bibnamefont{Fodor}},
  \bibinfo{author}{\bibfnamefont{K.}~\bibnamefont{Holland}},
  \bibinfo{author}{\bibfnamefont{J.}~\bibnamefont{Kuti}},
  \bibinfo{author}{\bibfnamefont{S.}~\bibnamefont{Mondal}},
  \bibinfo{author}{\bibfnamefont{D.}~\bibnamefont{Nogradi}} \bibnamefont{and}
  \bibinfo{author}{\bibfnamefont{C.~H.} \bibnamefont{Wong}},
  \bibinfo{journal}{PoS} \textbf{\bibinfo{volume}{LATTICE2015}},
  \bibinfo{pages}{219} (\bibinfo{year}{2016}), \eprint{1605.08750}.

\bibitem[{\citenamefont{Hasenfratz et~al.}(2017)\citenamefont{Hasenfratz,
  Rebbi, and Witzel}}]{Hasenfratz:2016gut}
\bibinfo{author}{\bibfnamefont{A.}~\bibnamefont{Hasenfratz}},
  \bibinfo{author}{\bibfnamefont{C.}~\bibnamefont{Rebbi}} \bibnamefont{and}
  \bibinfo{author}{\bibfnamefont{O.}~\bibnamefont{Witzel}},
  \bibinfo{journal}{Phys. Lett.} \textbf{\bibinfo{volume}{B773}},
  \bibinfo{pages}{86} (\bibinfo{year}{2017}), \eprint{1609.01401}.

\bibitem[{\citenamefont{Del~Debbio et~al.}(2016)\citenamefont{Del~Debbio,
  Lucini, Patella, Pica, and Rago}}]{DelDebbio:2015byq}
\bibinfo{author}{\bibfnamefont{L.}~\bibnamefont{Del~Debbio}},
  \bibinfo{author}{\bibfnamefont{B.}~\bibnamefont{Lucini}},
  \bibinfo{author}{\bibfnamefont{A.}~\bibnamefont{Patella}},
  \bibinfo{author}{\bibfnamefont{C.}~\bibnamefont{Pica}} \bibnamefont{and}
  \bibinfo{author}{\bibfnamefont{A.}~\bibnamefont{Rago}},
  \bibinfo{journal}{Phys. Rev.} \textbf{\bibinfo{volume}{D93}},
  \bibinfo{pages}{054505} (\bibinfo{year}{2016}), \eprint{1512.08242}.

\bibitem[{\citenamefont{Appelquist et~al.}(2016)\citenamefont{Appelquist,
  Brower, Fleming, Hasenfratz, Jin, Kiskis, Neil, Osborn, Rebbi, Rinaldi
  et~al.}}]{Appelquist:2016viq}
\bibinfo{author}{\bibfnamefont{T.}~\bibnamefont{Appelquist}},
  \bibinfo{author}{\bibfnamefont{R.~C.} \bibnamefont{Brower}},
  \bibinfo{author}{\bibfnamefont{G.~T.} \bibnamefont{Fleming}},
  \bibinfo{author}{\bibfnamefont{A.}~\bibnamefont{Hasenfratz}},
  \bibinfo{author}{\bibfnamefont{X.-Y.} \bibnamefont{Jin}},
  \bibinfo{author}{\bibfnamefont{J.}~\bibnamefont{Kiskis}},
  \bibinfo{author}{\bibfnamefont{E.~T.} \bibnamefont{Neil}},
  \bibinfo{author}{\bibfnamefont{J.~C.} \bibnamefont{Osborn}},
  \bibinfo{author}{\bibfnamefont{C.}~\bibnamefont{Rebbi}},
  \bibinfo{author}{\bibfnamefont{E.}~\bibnamefont{Rinaldi}},
  \bibinfo{author}{\bibfnamefont{D.}~\bibnamefont{Schaich}},
  \bibinfo{author}{\bibfnamefont{P.}~\bibnamefont{Vranas}},
  \bibinfo{author}{\bibfnamefont{E.}~\bibnamefont{Weinberg}}
  \bibnamefont{and} \bibinfo{author}{\bibfnamefont{O.}~\bibnamefont{Witzel}}
  (\bibinfo{collaboration}{LSD Collaboration}),
  \bibinfo{journal}{Phys. Rev.} \textbf{\bibinfo{volume}{D93}},
  \bibinfo{pages}{114514} (\bibinfo{year}{2016}), \eprint{1601.04027}.

\bibitem[{\citenamefont{Aoki et~al.}(2017{\natexlab{a}})\citenamefont{Aoki,
  Aoyama, Bennett, Kurachi, Maskawa, Miura, Nagai, Ohki, Rinaldi, Shibata
  et~al.}}]{Aoki:2016wnc}
\bibinfo{author}{\bibfnamefont{Y.}~\bibnamefont{Aoki}},
  \bibinfo{author}{\bibfnamefont{T.}~\bibnamefont{Aoyama}},
  \bibinfo{author}{\bibfnamefont{E.}~\bibnamefont{Bennett}},
  \bibinfo{author}{\bibfnamefont{M.}~\bibnamefont{Kurachi}},
  \bibinfo{author}{\bibfnamefont{T.}~\bibnamefont{Maskawa}},
  \bibinfo{author}{\bibfnamefont{K.}~\bibnamefont{Miura}},
  \bibinfo{author}{\bibfnamefont{K.-i.} \bibnamefont{Nagai}},
  \bibinfo{author}{\bibfnamefont{H.}~\bibnamefont{Ohki}},
  \bibinfo{author}{\bibfnamefont{E.}~\bibnamefont{Rinaldi}},
  \bibinfo{author}{\bibfnamefont{A.}~\bibnamefont{Shibata}},
  \bibinfo{author}{\bibfnamefont{K.}~\bibnamefont{Yamawaki}} \bibnamefont{and}
  \bibinfo{author}{\bibfnamefont{T.}~\bibnamefont{Yamazaki}}
  (\bibinfo{collaboration}{LatKMI Collaboration}),
  \bibinfo{journal}{Phys. Rev.} \textbf{\bibinfo{volume}{D96}},
  \bibinfo{pages}{014508} (\bibinfo{year}{2017}{\natexlab{a}}),
  \eprint{1610.07011}.

\bibitem[{\citenamefont{Gasbarro and Fleming}(2017)}]{Gasbarro:2017fmi}
\bibinfo{author}{\bibfnamefont{A.~D.} \bibnamefont{Gasbarro}} \bibnamefont{and}
  \bibinfo{author}{\bibfnamefont{G.~T.} \bibnamefont{Fleming}},
  \bibinfo{journal}{PoS} \textbf{\bibinfo{volume}{LATTICE2016}},
  \bibinfo{pages}{242} (\bibinfo{year}{2017}), \eprint{1702.00480}.

\bibitem[{\citenamefont{Athenodorou et~al.}(2016)\citenamefont{Athenodorou,
  Bennett, Bergner, Elander, Lin, Lucini, and Piai}}]{Athenodorou:2017dbf}
\bibinfo{author}{\bibfnamefont{A.}~\bibnamefont{Athenodorou}},
  \bibinfo{author}{\bibfnamefont{E.}~\bibnamefont{Bennett}},
  \bibinfo{author}{\bibfnamefont{G.}~\bibnamefont{Bergner}},
  \bibinfo{author}{\bibfnamefont{D.}~\bibnamefont{Elander}},
  \bibinfo{author}{\bibfnamefont{C.~J.~D.} \bibnamefont{Lin}},
  \bibinfo{author}{\bibfnamefont{B.}~\bibnamefont{Lucini}} \bibnamefont{and}
  \bibinfo{author}{\bibfnamefont{M.}~\bibnamefont{Piai}},
  \bibinfo{journal}{PoS} \textbf{\bibinfo{volume}{LATTICE2016}},
  \bibinfo{pages}{232} (\bibinfo{year}{2016}), \eprint{1702.06452}.

\bibitem[{\citenamefont{Fodor et~al.}(2018)\citenamefont{Fodor, Holland, Kuti,
  Nogradi, and Wong}}]{Fodor:2017nlp}
\bibinfo{author}{\bibfnamefont{Z.}~\bibnamefont{Fodor}},
  \bibinfo{author}{\bibfnamefont{K.}~\bibnamefont{Holland}},
  \bibinfo{author}{\bibfnamefont{J.}~\bibnamefont{Kuti}},
  \bibinfo{author}{\bibfnamefont{D.}~\bibnamefont{Nogradi}} \bibnamefont{and}
  \bibinfo{author}{\bibfnamefont{C.~H.} \bibnamefont{Wong}},
  \bibinfo{journal}{EPJ Web Conf.} \textbf{\bibinfo{volume}{175}},
  \bibinfo{pages}{08015} (\bibinfo{year}{2018}), \eprint{1712.08594}.

\bibitem[{\citenamefont{Patrignani et~al.}(2016)}]{Patrignani:2016xqp}
\bibinfo{collaboration}{Particle Data Group},
  \bibinfo{journal}{Chin. Phys.} \textbf{\bibinfo{volume}{C40}},
  \bibinfo{pages}{100001} (\bibinfo{year}{2016}).

\bibitem[{\citenamefont{Azatov}(2017)}]{Azatov:2017gas}
\bibinfo{author}{\bibfnamefont{A.}~\bibnamefont{Azatov}},
  \bibinfo{journal}{PoS} \textbf{\bibinfo{volume}{EPS-HEP2017}},
  \bibinfo{pages}{255} (\bibinfo{year}{2017}).

\bibitem[{\citenamefont{Hasenfratz et~al.}(2015)\citenamefont{Hasenfratz,
  Schaich, and Veernala}}]{Hasenfratz:2014rna}
\bibinfo{author}{\bibfnamefont{A.}~\bibnamefont{Hasenfratz}},
  \bibinfo{author}{\bibfnamefont{D.}~\bibnamefont{Schaich}} \bibnamefont{and}
  \bibinfo{author}{\bibfnamefont{A.}~\bibnamefont{Veernala}},
  \bibinfo{journal}{JHEP} \textbf{\bibinfo{volume}{1506}}, \bibinfo{pages}{143}
  (\bibinfo{year}{2015}), \eprint{1410.5886}.

\bibitem[{\citenamefont{Fodor et~al.}(2015{\natexlab{b}})\citenamefont{Fodor,
  Holland, Kuti, Mondal, Nogradi, and Wong}}]{Fodor:2015baa}
\bibinfo{author}{\bibfnamefont{Z.}~\bibnamefont{Fodor}},
  \bibinfo{author}{\bibfnamefont{K.}~\bibnamefont{Holland}},
  \bibinfo{author}{\bibfnamefont{J.}~\bibnamefont{Kuti}},
  \bibinfo{author}{\bibfnamefont{S.}~\bibnamefont{Mondal}},
  \bibinfo{author}{\bibfnamefont{D.}~\bibnamefont{Nogradi}} \bibnamefont{and}
  \bibinfo{author}{\bibfnamefont{C.~H.} \bibnamefont{Wong}},
  \bibinfo{journal}{JHEP} \textbf{\bibinfo{volume}{1506}}, \bibinfo{pages}{019}
  (\bibinfo{year}{2015}{\natexlab{b}}), \eprint{1503.01132}.

\bibitem[{\citenamefont{Appelquist et~al.}(2014)\citenamefont{Appelquist,
  Brower, Fleming, Kiskis, Lin, Neil, Osborn, Rebbi, Rinaldi, Schaich
  et~al.}}]{Appelquist:2014zsa}
\bibinfo{author}{\bibfnamefont{T.}~\bibnamefont{Appelquist}},
  \bibinfo{author}{\bibfnamefont{R.~C.} \bibnamefont{Brower}},
  \bibinfo{author}{\bibfnamefont{G.~T.} \bibnamefont{Fleming}},
  \bibinfo{author}{\bibfnamefont{J.}~\bibnamefont{Kiskis}},
  \bibinfo{author}{\bibfnamefont{M.~F.} \bibnamefont{Lin}},
  \bibinfo{author}{\bibfnamefont{E.~T.} \bibnamefont{Neil}},
  \bibinfo{author}{\bibfnamefont{J.~C.} \bibnamefont{Osborn}},
  \bibinfo{author}{\bibfnamefont{C.}~\bibnamefont{Rebbi}},
  \bibinfo{author}{\bibfnamefont{E.}~\bibnamefont{Rinaldi}},
  \bibinfo{author}{\bibfnamefont{D.}~\bibnamefont{Schaich}},
  \bibinfo{author}{\bibfnamefont{C.}~\bibnamefont{Schroeder}},
  \bibinfo{author}{\bibfnamefont{S.}~\bibnamefont{Syritsyn}},
  \bibinfo{author}{\bibfnamefont{G.}~\bibnamefont{Voronov}},
  \bibinfo{author}{\bibfnamefont{P.}~\bibnamefont{Vranas}},
  \bibinfo{author}{\bibfnamefont{E.}~\bibnamefont{Weinberg}}
  \bibnamefont{and} \bibinfo{author}{\bibfnamefont{O.}~\bibnamefont{Witzel}}
  (\bibinfo{collaboration}{LSD Collaboration}),
  \bibinfo{journal}{Phys. Rev.} \textbf{\bibinfo{volume}{D90}},
  \bibinfo{pages}{114502} (\bibinfo{year}{2014}), \eprint{1405.4752}.

\bibitem[{\citenamefont{Cheng et~al.}(2013)\citenamefont{Cheng, Hasenfratz,
  Petropoulos, and Schaich}}]{Cheng:2013eu}
\bibinfo{author}{\bibfnamefont{A.}~\bibnamefont{Cheng}},
  \bibinfo{author}{\bibfnamefont{A.}~\bibnamefont{Hasenfratz}},
  \bibinfo{author}{\bibfnamefont{G.}~\bibnamefont{Petropoulos}}
  \bibnamefont{and} \bibinfo{author}{\bibfnamefont{D.}~\bibnamefont{Schaich}},
  \bibinfo{journal}{JHEP} \textbf{\bibinfo{volume}{1307}}, \bibinfo{pages}{061}
  (\bibinfo{year}{2013}), \eprint{1301.1355}.

\bibitem[{\citenamefont{Aoki et~al.}(2013{\natexlab{b}})\citenamefont{Aoki,
  Aoyama, Kurachi, Maskawa, Nagai, Ohki, Shibata, Yamawaki, and
  Yamazaki}}]{Aoki:2013xza}
\bibinfo{author}{\bibfnamefont{Y.}~\bibnamefont{Aoki}},
  \bibinfo{author}{\bibfnamefont{T.}~\bibnamefont{Aoyama}},
  \bibinfo{author}{\bibfnamefont{M.}~\bibnamefont{Kurachi}},
  \bibinfo{author}{\bibfnamefont{T.}~\bibnamefont{Maskawa}},
  \bibinfo{author}{\bibfnamefont{K.-i.} \bibnamefont{Nagai}},
  \bibinfo{author}{\bibfnamefont{H.}~\bibnamefont{Ohki}},
  \bibinfo{author}{\bibfnamefont{A.}~\bibnamefont{Shibata}},
  \bibinfo{author}{\bibfnamefont{K.}~\bibnamefont{Yamawaki}} \bibnamefont{and}
  \bibinfo{author}{\bibfnamefont{T.}~\bibnamefont{Yamazaki}}
  (\bibinfo{collaboration}{LatKMI Collaboration}), \bibinfo{journal}{Phys.
  Rev.} \textbf{\bibinfo{volume}{D87}}, \bibinfo{pages}{094511}
  (\bibinfo{year}{2013}{\natexlab{b}}), \eprint{1302.6859}.

\bibitem[{\citenamefont{Schaich}(2013)}]{Schaich:2013eba}
\bibinfo{author}{\bibfnamefont{D.}~\bibnamefont{Schaich}},
  \bibinfo{journal}{PoS} \textbf{\bibinfo{volume}{LATTICE 2013}},
  \bibinfo{pages}{072} (\bibinfo{year}{2013}), \eprint{1310.7006}.

\bibitem[{\citenamefont{Aoki et~al.}(2018)\citenamefont{Aoki, Aoyama, Bennett,
  Kurachi, Maskawa, Miura, Nagai, Ohki, Rinaldi, Shibata
  et~al.}}]{Aoki:2017fnr}
\bibinfo{author}{\bibfnamefont{Y.}~\bibnamefont{Aoki}},
  \bibinfo{author}{\bibfnamefont{T.}~\bibnamefont{Aoyama}},
  \bibinfo{author}{\bibfnamefont{E.}~\bibnamefont{Bennett}},
  \bibinfo{author}{\bibfnamefont{M.}~\bibnamefont{Kurachi}},
  \bibinfo{author}{\bibfnamefont{T.}~\bibnamefont{Maskawa}},
  \bibinfo{author}{\bibfnamefont{K.}~\bibnamefont{Miura}},
  \bibinfo{author}{\bibfnamefont{K.-i.} \bibnamefont{Nagai}},
  \bibinfo{author}{\bibfnamefont{H.}~\bibnamefont{Ohki}},
  \bibinfo{author}{\bibfnamefont{E.}~\bibnamefont{Rinaldi}},
  \bibinfo{author}{\bibfnamefont{A.}~\bibnamefont{Shibata}},
  \bibinfo{author}{\bibfnamefont{K.}~\bibnamefont{Yamawaki}} \bibnamefont{and}
  \bibinfo{author}{\bibfnamefont{T.}~\bibnamefont{Yamazaki}}
  (\bibinfo{collaboration}{LatKMI Collaboration}),
  \bibinfo{journal}{EPJ Web Conf.}
  \textbf{\bibinfo{volume}{175}}, \bibinfo{pages}{08023}
  (\bibinfo{year}{2018}), \eprint{1710.06549}.

\bibitem[{\citenamefont{Appelquist et~al.}(2008)\citenamefont{Appelquist,
  Fleming, and Neil}}]{Appelquist:2007hu}
\bibinfo{author}{\bibfnamefont{T.}~\bibnamefont{Appelquist}},
  \bibinfo{author}{\bibfnamefont{G.~T.} \bibnamefont{Fleming}}
  \bibnamefont{and} \bibinfo{author}{\bibfnamefont{E.~T.} \bibnamefont{Neil}},
  \bibinfo{journal}{Phys. Rev. Lett.} \textbf{\bibinfo{volume}{100}},
  \bibinfo{pages}{171607} (\bibinfo{year}{2008}), \eprint{0712.0609}.

\bibitem[{\citenamefont{Appelquist et~al.}(2009)\citenamefont{Appelquist,
  Fleming, and Neil}}]{Appelquist:2009ty}
\bibinfo{author}{\bibfnamefont{T.}~\bibnamefont{Appelquist}},
  \bibinfo{author}{\bibfnamefont{G.~T.} \bibnamefont{Fleming}}
  \bibnamefont{and} \bibinfo{author}{\bibfnamefont{E.~T.} \bibnamefont{Neil}},
  \bibinfo{journal}{Phys. Rev.} \textbf{\bibinfo{volume}{D79}},
  \bibinfo{pages}{076010} (\bibinfo{year}{2009}), \eprint{0901.3766}.

\bibitem[{\citenamefont{Deuzeman et~al.}(2008)\citenamefont{Deuzeman, Lombardo,
  and Pallante}}]{Deuzeman:2008sc}
\bibinfo{author}{\bibfnamefont{A.}~\bibnamefont{Deuzeman}},
  \bibinfo{author}{\bibfnamefont{M.~P.} \bibnamefont{Lombardo}}
  \bibnamefont{and} \bibinfo{author}{\bibfnamefont{E.}~\bibnamefont{Pallante}},
  \bibinfo{journal}{Phys. Lett.} \textbf{\bibinfo{volume}{B670}},
  \bibinfo{pages}{41} (\bibinfo{year}{2008}), \eprint{0804.2905}.

\bibitem[{\citenamefont{Miura and Lombardo}(2013)}]{Miura:2012zqa}
\bibinfo{author}{\bibfnamefont{K.}~\bibnamefont{Miura}} \bibnamefont{and}
  \bibinfo{author}{\bibfnamefont{M.~P.} \bibnamefont{Lombardo}},
  \bibinfo{journal}{Nucl. Phys.} \textbf{\bibinfo{volume}{B871}},
  \bibinfo{pages}{52} (\bibinfo{year}{2013}), \eprint{1212.0955}.

\bibitem[{\citenamefont{Jin and Mawhinney}(2010)}]{Jin:2010vm}
\bibinfo{author}{\bibfnamefont{X.-Y.} \bibnamefont{Jin}} \bibnamefont{and}
  \bibinfo{author}{\bibfnamefont{R.~D.} \bibnamefont{Mawhinney}},
  \bibinfo{journal}{PoS} \textbf{\bibinfo{volume}{Lattice 2010}},
  \bibinfo{pages}{055} (\bibinfo{year}{2010}), \eprint{1011.1511}.

\bibitem[{\citenamefont{Schaich et~al.}(2012)\citenamefont{Schaich, Cheng,
  Hasenfratz, and Petropoulos}}]{Schaich:2012fr}
\bibinfo{author}{\bibfnamefont{D.}~\bibnamefont{Schaich}},
  \bibinfo{author}{\bibfnamefont{A.}~\bibnamefont{Cheng}},
  \bibinfo{author}{\bibfnamefont{A.}~\bibnamefont{Hasenfratz}}
  \bibnamefont{and}
  \bibinfo{author}{\bibfnamefont{G.}~\bibnamefont{Petropoulos}},
  \bibinfo{journal}{Proc. Sci.} \textbf{\bibinfo{volume}{Lattice 2012}},
  \bibinfo{pages}{028} (\bibinfo{year}{2012}), \eprint{1207.7164}.

\bibitem[{\citenamefont{Hasenfratz et~al.}(2014)\citenamefont{Hasenfratz,
  Cheng, Petropoulos, and Schaich}}]{Hasenfratz:2013uha}
\bibinfo{author}{\bibfnamefont{A.}~\bibnamefont{Hasenfratz}},
  \bibinfo{author}{\bibfnamefont{A.}~\bibnamefont{Cheng}},
  \bibinfo{author}{\bibfnamefont{G.}~\bibnamefont{Petropoulos}}
  \bibnamefont{and} \bibinfo{author}{\bibfnamefont{D.}~\bibnamefont{Schaich}},
  in \emph{\bibinfo{booktitle}{{Strong Coupling Gauge Theories in the LHC
  Perspective (SCGT12)}}} (\bibinfo{year}{2014}), \bibinfo{pages}{44--50},
  \eprint{1303.7129}.

\bibitem[{\citenamefont{Lombardo et~al.}(2014)\citenamefont{Lombardo, Miura,
  Nunes~da Silva, and Pallante}}]{Lombardo:2014mda}
\bibinfo{author}{\bibfnamefont{M.~P.} \bibnamefont{Lombardo}},
  \bibinfo{author}{\bibfnamefont{K.}~\bibnamefont{Miura}},
  \bibinfo{author}{\bibfnamefont{T.~J.} \bibnamefont{Nunes~da Silva}}
  \bibnamefont{and} \bibinfo{author}{\bibfnamefont{E.}~\bibnamefont{Pallante}},
  \bibinfo{journal}{Int. J. Mod. Phys.} \textbf{\bibinfo{volume}{A29}},
  \bibinfo{pages}{1445007} (\bibinfo{year}{2014}), \eprint{1410.2036}.

\bibitem[{\citenamefont{Schaich et~al.}(2018)\citenamefont{Schaich, Hasenfratz,
  and Rinaldi}}]{Schaich:2015psa}
\bibinfo{author}{\bibfnamefont{D.}~\bibnamefont{Schaich}},
  \bibinfo{author}{\bibfnamefont{A.}~\bibnamefont{Hasenfratz}}
  \bibnamefont{and} \bibinfo{author}{\bibfnamefont{E.}~\bibnamefont{Rinaldi}}
  (\bibinfo{collaboration}{LSD Collaboration}), in
  \emph{\bibinfo{booktitle}{{Sakata Memorial KMI Workshop on Origin of Mass and
  Strong Coupling Gauge Theories (SCGT15), Nagoya, 3--6 March 2015}}}
  (\bibinfo{year}{2018}), \bibinfo{pages}{351--354}, \eprint{1506.08791}.

\bibitem[{\citenamefont{Fodor et~al.}(2009)\citenamefont{Fodor, Holland, Kuti,
  Nogradi, and Schroeder}}]{Fodor:2009wk}
\bibinfo{author}{\bibfnamefont{Z.}~\bibnamefont{Fodor}},
  \bibinfo{author}{\bibfnamefont{K.}~\bibnamefont{Holland}},
  \bibinfo{author}{\bibfnamefont{J.}~\bibnamefont{Kuti}},
  \bibinfo{author}{\bibfnamefont{D.}~\bibnamefont{Nogradi}} \bibnamefont{and}
  \bibinfo{author}{\bibfnamefont{C.}~\bibnamefont{Schroeder}},
  \bibinfo{journal}{Phys. Lett.} \textbf{\bibinfo{volume}{B681}},
  \bibinfo{pages}{353} (\bibinfo{year}{2009}), \eprint{0907.4562}.

\bibitem[{\citenamefont{Cheng et~al.}(2012)\citenamefont{Cheng, Hasenfratz, and
  Schaich}}]{Cheng:2011ic}
\bibinfo{author}{\bibfnamefont{A.}~\bibnamefont{Cheng}},
  \bibinfo{author}{\bibfnamefont{A.}~\bibnamefont{Hasenfratz}}
  \bibnamefont{and} \bibinfo{author}{\bibfnamefont{D.}~\bibnamefont{Schaich}},
  \bibinfo{journal}{Phys. Rev.} \textbf{\bibinfo{volume}{D85}},
  \bibinfo{pages}{094509} (\bibinfo{year}{2012}), \eprint{1111.2317}.

\bibitem[{\citenamefont{Aoki et~al.}(2016{\natexlab{a}})\citenamefont{Aoki,
  Aoyama, Bennett, Kurachi, Maskawa, Miura, Nagai, Ohki, Rinaldi, Shibata
  et~al.}}]{Aoki:2016fxd}
\bibinfo{author}{\bibfnamefont{Y.}~\bibnamefont{Aoki}},
  \bibinfo{author}{\bibfnamefont{T.}~\bibnamefont{Aoyama}},
  \bibinfo{author}{\bibfnamefont{E.}~\bibnamefont{Bennett}},
  \bibinfo{author}{\bibfnamefont{M.}~\bibnamefont{Kurachi}},
  \bibinfo{author}{\bibfnamefont{T.}~\bibnamefont{Maskawa}},
  \bibinfo{author}{\bibfnamefont{K.}~\bibnamefont{Miura}},
  \bibinfo{author}{\bibfnamefont{K.-i.} \bibnamefont{Nagai}},
  \bibinfo{author}{\bibfnamefont{H.}~\bibnamefont{Ohki}},
  \bibinfo{author}{\bibfnamefont{E.}~\bibnamefont{Rinaldi}},
  \bibinfo{author}{\bibfnamefont{A.}~\bibnamefont{Shibata}},
  \bibinfo{author}{\bibfnamefont{K.}~\bibnamefont{Yamawaki}} \bibnamefont{and}
  \bibinfo{author}{\bibfnamefont{T.}~\bibnamefont{Yamazaki}}
  (\bibinfo{collaboration}{LatKMI Collaboration}),
  \bibinfo{journal}{PoS} \textbf{\bibinfo{volume}{LATTICE 2015}},
  \bibinfo{pages}{213} (\bibinfo{year}{2016}{\natexlab{a}}),
  \eprint{1601.02287}.

\bibitem[{\citenamefont{Aoki et~al.}(2016{\natexlab{b}})\citenamefont{Aoki,
  Aoyama, Bennett, Kurachi, Maskawa, Miura, Nagai, Ohki, Rinaldi, Shibata
  et~al.}}]{Aoki:2016bfp}
\bibinfo{author}{\bibfnamefont{Y.}~\bibnamefont{Aoki}},
  \bibinfo{author}{\bibfnamefont{T.}~\bibnamefont{Aoyama}},
  \bibinfo{author}{\bibfnamefont{E.}~\bibnamefont{Bennett}},
  \bibinfo{author}{\bibfnamefont{M.}~\bibnamefont{Kurachi}},
  \bibinfo{author}{\bibfnamefont{T.}~\bibnamefont{Maskawa}},
  \bibinfo{author}{\bibfnamefont{K.}~\bibnamefont{Miura}},
  \bibinfo{author}{\bibfnamefont{K.-i.} \bibnamefont{Nagai}},
  \bibinfo{author}{\bibfnamefont{H.}~\bibnamefont{Ohki}},
  \bibinfo{author}{\bibfnamefont{E.}~\bibnamefont{Rinaldi}},
  \bibinfo{author}{\bibfnamefont{A.}~\bibnamefont{Shibata}},
  \bibinfo{author}{\bibfnamefont{K.}~\bibnamefont{Yamawaki}} \bibnamefont{and}
  \bibinfo{author}{\bibfnamefont{T.}~\bibnamefont{Yamazaki}}
  (\bibinfo{collaboration}{LatKMI Collaboration}),
  \bibinfo{journal}{PoS} \textbf{\bibinfo{volume}{LATTICE 2015}},
  \bibinfo{pages}{245} (\bibinfo{year}{2016}{\natexlab{b}}),
  \eprint{1602.00796}.

\bibitem[{\citenamefont{Hasenfratz et~al.}(2012)\citenamefont{Hasenfratz,
  Cheng, Petropoulos, and Schaich}}]{Hasenfratz:2012fp}
\bibinfo{author}{\bibfnamefont{A.}~\bibnamefont{Hasenfratz}},
  \bibinfo{author}{\bibfnamefont{A.}~\bibnamefont{Cheng}},
  \bibinfo{author}{\bibfnamefont{G.}~\bibnamefont{Petropoulos}}
  \bibnamefont{and} \bibinfo{author}{\bibfnamefont{D.}~\bibnamefont{Schaich}},
  \bibinfo{journal}{PoS} \textbf{\bibinfo{volume}{Lattice 2012}},
  \bibinfo{pages}{034} (\bibinfo{year}{2012}), \eprint{1207.7162}.

\bibitem[{\citenamefont{Hasenfratz}(2010)}]{Hasenfratz:2010fi}
\bibinfo{author}{\bibfnamefont{A.}~\bibnamefont{Hasenfratz}},
  \bibinfo{journal}{Phys. Rev.} \textbf{\bibinfo{volume}{D82}},
  \bibinfo{pages}{014506} (\bibinfo{year}{2010}), \eprint{1004.1004}.

\bibitem[{\citenamefont{Hasenfratz}(2012)}]{Hasenfratz:2011xn}
\bibinfo{author}{\bibfnamefont{A.}~\bibnamefont{Hasenfratz}},
  \bibinfo{journal}{Phys. Rev. Lett.} \textbf{\bibinfo{volume}{108}},
  \bibinfo{pages}{061601} (\bibinfo{year}{2012}), \eprint{1106.5293}.

\bibitem[{\citenamefont{Petropoulos et~al.}(2012)\citenamefont{Petropoulos,
  Cheng, Hasenfratz, and Schaich}}]{Petropoulos:2012mg}
\bibinfo{author}{\bibfnamefont{G.}~\bibnamefont{Petropoulos}},
  \bibinfo{author}{\bibfnamefont{A.}~\bibnamefont{Cheng}},
  \bibinfo{author}{\bibfnamefont{A.}~\bibnamefont{Hasenfratz}}
  \bibnamefont{and} \bibinfo{author}{\bibfnamefont{D.}~\bibnamefont{Schaich}},
  \bibinfo{journal}{PoS} \textbf{\bibinfo{volume}{Lattice 2012}},
  \bibinfo{pages}{051} (\bibinfo{year}{2012}), \eprint{1212.0053}.

\bibitem[{\citenamefont{Ishikawa et~al.}(2014)\citenamefont{Ishikawa, Iwasaki,
  Nakayama, and Yoshie}}]{Ishikawa:2013tua}
\bibinfo{author}{\bibfnamefont{K.~I.} \bibnamefont{Ishikawa}},
  \bibinfo{author}{\bibfnamefont{Y.}~\bibnamefont{Iwasaki}},
  \bibinfo{author}{\bibfnamefont{Y.}~\bibnamefont{Nakayama}} \bibnamefont{and}
  \bibinfo{author}{\bibfnamefont{T.}~\bibnamefont{Yoshie}},
  \bibinfo{journal}{Phys. Rev.} \textbf{\bibinfo{volume}{D89}},
  \bibinfo{pages}{114503} (\bibinfo{year}{2014}), \eprint{1310.5049}.

\bibitem[{\citenamefont{Ishikawa et~al.}(2015)\citenamefont{Ishikawa, Iwasaki,
  Nakayama, and Yoshie}}]{Ishikawa:2015iwa}
\bibinfo{author}{\bibfnamefont{K.~I.} \bibnamefont{Ishikawa}},
  \bibinfo{author}{\bibfnamefont{Y.}~\bibnamefont{Iwasaki}},
  \bibinfo{author}{\bibfnamefont{Y.}~\bibnamefont{Nakayama}} \bibnamefont{and}
  \bibinfo{author}{\bibfnamefont{T.}~\bibnamefont{Yoshie}},
  \bibinfo{journal}{Phys. Lett.} \textbf{\bibinfo{volume}{B748}},
  \bibinfo{pages}{289} (\bibinfo{year}{2015}), \eprint{1503.02359}.

\bibitem[{\citenamefont{da~Silva et~al.}(2015)\citenamefont{da~Silva, Pallante,
  and Robroek}}]{daSilva:2015vna}
\bibinfo{author}{\bibfnamefont{T.~N.} \bibnamefont{da~Silva}},
  \bibinfo{author}{\bibfnamefont{E.}~\bibnamefont{Pallante}} \bibnamefont{and}
  \bibinfo{author}{\bibfnamefont{L.}~\bibnamefont{Robroek}}
  (\bibinfo{year}{2015}), \eprint{1506.06396}.

\bibitem[{\citenamefont{Aoki et~al.}(2017{\natexlab{b}})\citenamefont{Aoki,
  Aoyama, Bennett, Kurachi, Maskawa, Miura, Nagai, Ohki, Rinaldi, Shibata
  et~al.}}]{Aoki:2015aqa}
\bibinfo{author}{\bibfnamefont{Y.}~\bibnamefont{Aoki}},
  \bibinfo{author}{\bibfnamefont{T.}~\bibnamefont{Aoyama}},
  \bibinfo{author}{\bibfnamefont{E.}~\bibnamefont{Bennett}},
  \bibinfo{author}{\bibfnamefont{M.}~\bibnamefont{Kurachi}},
  \bibinfo{author}{\bibfnamefont{T.}~\bibnamefont{Maskawa}},
  \bibinfo{author}{\bibfnamefont{K.}~\bibnamefont{Miura}},
  \bibinfo{author}{\bibfnamefont{K.-i.} \bibnamefont{Nagai}},
  \bibinfo{author}{\bibfnamefont{H.}~\bibnamefont{Ohki}},
  \bibinfo{author}{\bibfnamefont{E.}~\bibnamefont{Rinaldi}},
  \bibinfo{author}{\bibfnamefont{A.}~\bibnamefont{Shibata}},
  \bibinfo{author}{\bibfnamefont{K.}~\bibnamefont{Yamawaki}}
  \bibnamefont{and} \bibinfo{author}{\bibfnamefont{T.}~\bibnamefont{Yamazaki}}
  (\bibinfo{collaboration}{LatKMI Collaboration}),
  \bibinfo{journal}{Int. J. Mod. Phys.}
  \textbf{\bibinfo{volume}{A32}}, \bibinfo{pages}{1747005}
  (\bibinfo{year}{2017}{\natexlab{b}}), \eprint{1510.05863}.

\bibitem[{\citenamefont{Noaki et~al.}(2015)\citenamefont{Noaki, Cossu,
  Ishikawa, Iwasaki, and Yoshie}}]{Noaki:2015xpx}
\bibinfo{author}{\bibfnamefont{J.}~\bibnamefont{Noaki}},
  \bibinfo{author}{\bibfnamefont{G.}~\bibnamefont{Cossu}},
  \bibinfo{author}{\bibfnamefont{K.-I.} \bibnamefont{Ishikawa}},
  \bibinfo{author}{\bibfnamefont{Y.}~\bibnamefont{Iwasaki}} \bibnamefont{and}
  \bibinfo{author}{\bibfnamefont{T.}~\bibnamefont{Yoshie}},
  \bibinfo{journal}{PoS} \textbf{\bibinfo{volume}{LATTICE 2015}},
  \bibinfo{pages}{312} (\bibinfo{year}{2015}), \eprint{1511.06474}.

\bibitem[{\citenamefont{Aoki et~al.}(2016{\natexlab{c}})\citenamefont{Aoki,
  Aoyama, Bennett, Kurachi, Maskawa, Miura, Nagai, Ohki, Rinaldi, Shibata
  et~al.}}]{Aoki:2016yrm}
\bibinfo{author}{\bibfnamefont{Y.}~\bibnamefont{Aoki}},
  \bibinfo{author}{\bibfnamefont{T.}~\bibnamefont{Aoyama}},
  \bibinfo{author}{\bibfnamefont{E.}~\bibnamefont{Bennett}},
  \bibinfo{author}{\bibfnamefont{M.}~\bibnamefont{Kurachi}},
  \bibinfo{author}{\bibfnamefont{T.}~\bibnamefont{Maskawa}},
  \bibinfo{author}{\bibfnamefont{K.}~\bibnamefont{Miura}},
  \bibinfo{author}{\bibfnamefont{K.-i.} \bibnamefont{Nagai}},
  \bibinfo{author}{\bibfnamefont{H.}~\bibnamefont{Ohki}},
  \bibinfo{author}{\bibfnamefont{E.}~\bibnamefont{Rinaldi}},
  \bibinfo{author}{\bibfnamefont{A.}~\bibnamefont{Shibata}},
  \bibinfo{author}{\bibfnamefont{K.}~\bibnamefont{Yamawaki}}
  \bibnamefont{and} \bibinfo{author}{\bibfnamefont{T.}~\bibnamefont{Yamazaki}}
  (\bibinfo{collaboration}{LatKMI Collaboration}),
  \bibinfo{journal}{PoS} \textbf{\bibinfo{volume}{LATTICE
  2015}}, \bibinfo{pages}{214} (\bibinfo{year}{2016}{\natexlab{c}}),
  \eprint{1601.04687}.

\bibitem[{\citenamefont{Kawarabayashi and Suzuki}(1966)}]{Kawarabayashi:1966kd}
\bibinfo{author}{\bibfnamefont{K.}~\bibnamefont{Kawarabayashi}}
  \bibnamefont{and} \bibinfo{author}{\bibfnamefont{M.}~\bibnamefont{Suzuki}},
  \bibinfo{journal}{Phys. Rev. Lett.} \textbf{\bibinfo{volume}{16}},
  \bibinfo{pages}{255} (\bibinfo{year}{1966}).

\bibitem[{\citenamefont{Riazuddin and Fayyazuddin}(1966)}]{Riazuddin:1966sw}
\bibinfo{author}{\bibnamefont{Riazuddin}} \bibnamefont{and}
  \bibinfo{author}{\bibnamefont{Fayyazuddin}}, \bibinfo{journal}{Phys. Rev.}
  \textbf{\bibinfo{volume}{147}}, \bibinfo{pages}{1071} (\bibinfo{year}{1966}).

\bibitem[{\citenamefont{Hasenfratz and Knechtli}(2001)}]{Hasenfratz:2001hp}
\bibinfo{author}{\bibfnamefont{A.}~\bibnamefont{Hasenfratz}} \bibnamefont{and}
  \bibinfo{author}{\bibfnamefont{F.}~\bibnamefont{Knechtli}},
  \bibinfo{journal}{Phys. Rev.} \textbf{\bibinfo{volume}{D64}},
  \bibinfo{pages}{034504} (\bibinfo{year}{2001}), \eprint{hep-lat/0103029}.

\bibitem[{\citenamefont{Hasenfratz et~al.}(2007)\citenamefont{Hasenfratz,
  Hoffmann, and Schaefer}}]{Hasenfratz:2007rf}
\bibinfo{author}{\bibfnamefont{A.}~\bibnamefont{Hasenfratz}},
  \bibinfo{author}{\bibfnamefont{R.}~\bibnamefont{Hoffmann}} \bibnamefont{and}
  \bibinfo{author}{\bibfnamefont{S.}~\bibnamefont{Schaefer}},
  \bibinfo{journal}{JHEP} \textbf{\bibinfo{volume}{0705}}, \bibinfo{pages}{029}
  (\bibinfo{year}{2007}), \eprint{hep-lat/0702028}.

\bibitem[{\citenamefont{Deuzeman et~al.}(2013)\citenamefont{Deuzeman, Lombardo,
  Nunes~da Silva, and Pallante}}]{Deuzeman:2012ee}
\bibinfo{author}{\bibfnamefont{A.}~\bibnamefont{Deuzeman}},
  \bibinfo{author}{\bibfnamefont{M.~P.} \bibnamefont{Lombardo}},
  \bibinfo{author}{\bibfnamefont{T.}~\bibnamefont{Nunes~da Silva}}
  \bibnamefont{and} \bibinfo{author}{\bibfnamefont{E.}~\bibnamefont{Pallante}},
  \bibinfo{journal}{Phys. Lett.} \textbf{\bibinfo{volume}{B720}},
  \bibinfo{pages}{358} (\bibinfo{year}{2013}), \eprint{1209.5720}.

\bibitem[{\citenamefont{Fodor et~al.}(2012{\natexlab{a}})\citenamefont{Fodor,
  Holland, Kuti, Nogradi, Schroeder, and Wong}}]{Fodor:2012et}
\bibinfo{author}{\bibfnamefont{Z.}~\bibnamefont{Fodor}},
  \bibinfo{author}{\bibfnamefont{K.}~\bibnamefont{Holland}},
  \bibinfo{author}{\bibfnamefont{J.}~\bibnamefont{Kuti}},
  \bibinfo{author}{\bibfnamefont{D.}~\bibnamefont{Nogradi}},
  \bibinfo{author}{\bibfnamefont{C.}~\bibnamefont{Schroeder}}
  \bibnamefont{and} \bibinfo{author}{\bibfnamefont{C.~H.} \bibnamefont{Wong}},
  \bibinfo{journal}{PoS} \textbf{\bibinfo{volume}{Lattice 2012}},
  \bibinfo{pages}{279} (\bibinfo{year}{2012}{\natexlab{a}}),
  \eprint{1211.4238}.

\bibitem[{\citenamefont{Brown et~al.}(1992)\citenamefont{Brown, Chen, Christ,
  Dong, Mawhinney, Schaffer, and Vaccarino}}]{Brown:1992fz}
\bibinfo{author}{\bibfnamefont{F.~R.} \bibnamefont{Brown}},
  \bibinfo{author}{\bibfnamefont{H.}~\bibnamefont{Chen}},
  \bibinfo{author}{\bibfnamefont{N.~H.} \bibnamefont{Christ}},
  \bibinfo{author}{\bibfnamefont{Z.}~\bibnamefont{Dong}},
  \bibinfo{author}{\bibfnamefont{R.~D.} \bibnamefont{Mawhinney}},
  \bibinfo{author}{\bibfnamefont{W.}~\bibnamefont{Schaffer}} \bibnamefont{and}
  \bibinfo{author}{\bibfnamefont{A.}~\bibnamefont{Vaccarino}},
  \bibinfo{journal}{Phys. Rev.} \textbf{\bibinfo{volume}{D46}},
  \bibinfo{pages}{5655} (\bibinfo{year}{1992}), \eprint{hep-lat/9206001}.

\bibitem[{\citenamefont{Jin and Mawhinney}(2011)}]{Jin:2012dw}
\bibinfo{author}{\bibfnamefont{X.-Y.} \bibnamefont{Jin}} \bibnamefont{and}
  \bibinfo{author}{\bibfnamefont{R.~D.} \bibnamefont{Mawhinney}},
  \bibinfo{journal}{PoS} \textbf{\bibinfo{volume}{Lattice 2011}},
  \bibinfo{pages}{066} (\bibinfo{year}{2011}), \eprint{1203.5855}.

\bibitem[{\citenamefont{Fodor et~al.}(2012{\natexlab{b}})\citenamefont{Fodor,
  Holland, Kuti, Nogradi, and Wong}}]{Fodor:2012td}
\bibinfo{author}{\bibfnamefont{Z.}~\bibnamefont{Fodor}},
  \bibinfo{author}{\bibfnamefont{K.}~\bibnamefont{Holland}},
  \bibinfo{author}{\bibfnamefont{J.}~\bibnamefont{Kuti}},
  \bibinfo{author}{\bibfnamefont{D.}~\bibnamefont{Nogradi}} \bibnamefont{and}
  \bibinfo{author}{\bibfnamefont{C.~H.} \bibnamefont{Wong}},
  \bibinfo{journal}{JHEP} \textbf{\bibinfo{volume}{1211}}, \bibinfo{pages}{007}
  (\bibinfo{year}{2012}{\natexlab{b}}), \eprint{1208.1051}.

\bibitem[{\citenamefont{Ryttov and Shrock}(2011)}]{Ryttov:2010iz}
\bibinfo{author}{\bibfnamefont{T.~A.} \bibnamefont{Ryttov}} \bibnamefont{and}
  \bibinfo{author}{\bibfnamefont{R.}~\bibnamefont{Shrock}},
  \bibinfo{journal}{Phys. Rev.} \textbf{\bibinfo{volume}{D83}},
  \bibinfo{pages}{056011} (\bibinfo{year}{2011}), \eprint{1011.4542}.

\bibitem[{\citenamefont{Pica and Sannino}(2011)}]{Pica:2010xq}
\bibinfo{author}{\bibfnamefont{C.}~\bibnamefont{Pica}} \bibnamefont{and}
  \bibinfo{author}{\bibfnamefont{F.}~\bibnamefont{Sannino}},
  \bibinfo{journal}{Phys. Rev.} \textbf{\bibinfo{volume}{D83}},
  \bibinfo{pages}{035013} (\bibinfo{year}{2011}), \eprint{1011.5917}.

\bibitem[{\citenamefont{Osborn}(2014)}]{Osborn:2014kda}
\bibinfo{author}{\bibfnamefont{J.}~\bibnamefont{Osborn}},
  \bibinfo{journal}{PoS} \textbf{\bibinfo{volume}{LATTICE2014}},
  \bibinfo{pages}{028} (\bibinfo{year}{2014}),
  \urlprefix\url{http://usqcd-software.github.io/FUEL.html}.

\bibitem[{\citenamefont{Takaishi and de~Forcrand}(2006)}]{Takaishi:2005tz}
\bibinfo{author}{\bibfnamefont{T.}~\bibnamefont{Takaishi}} \bibnamefont{and}
  \bibinfo{author}{\bibfnamefont{P.}~\bibnamefont{de~Forcrand}},
  \bibinfo{journal}{Phys. Rev.} \textbf{\bibinfo{volume}{E73}},
  \bibinfo{pages}{036706} (\bibinfo{year}{2006}), \eprint{hep-lat/0505020}.

\bibitem[{\citenamefont{Hasenbusch and Jansen}(2003)}]{Hasenbusch:2002ai}
\bibinfo{author}{\bibfnamefont{M.}~\bibnamefont{Hasenbusch}} \bibnamefont{and}
  \bibinfo{author}{\bibfnamefont{K.}~\bibnamefont{Jansen}},
  \bibinfo{journal}{Nucl. Phys.} \textbf{\bibinfo{volume}{B659}},
  \bibinfo{pages}{299} (\bibinfo{year}{2003}), \eprint{hep-lat/0211042}.

\bibitem[{\citenamefont{Urbach et~al.}(2006)\citenamefont{Urbach, Jansen,
  Shindler, and Wenger}}]{Urbach:2005ji}
\bibinfo{author}{\bibfnamefont{C.}~\bibnamefont{Urbach}},
  \bibinfo{author}{\bibfnamefont{K.}~\bibnamefont{Jansen}},
  \bibinfo{author}{\bibfnamefont{A.}~\bibnamefont{Shindler}} \bibnamefont{and}
  \bibinfo{author}{\bibfnamefont{U.}~\bibnamefont{Wenger}},
  \bibinfo{journal}{Comput. Phys. Commun.} \textbf{\bibinfo{volume}{174}},
  \bibinfo{pages}{87} (\bibinfo{year}{2006}), \eprint{hep-lat/0506011}.

\bibitem[{\citenamefont{Omelyan et~al.}(2002)\citenamefont{Omelyan, Mryglod,
  and Folk}}]{Omelyan:2002fg}
\bibinfo{author}{\bibfnamefont{I.~P.} \bibnamefont{Omelyan}},
  \bibinfo{author}{\bibfnamefont{I.~M.} \bibnamefont{Mryglod}}
  \bibnamefont{and} \bibinfo{author}{\bibfnamefont{R.}~\bibnamefont{Folk}},
  \bibinfo{journal}{Phys. Rev.} \textbf{\bibinfo{volume}{E66}},
  \bibinfo{pages}{026701} (\bibinfo{year}{2002}), \eprint{cond-mat/0111055}.

\bibitem[{\citenamefont{Yin and Mawhinney}(2011)}]{Yin:2011np}
\bibinfo{author}{\bibfnamefont{H.}~\bibnamefont{Yin}} \bibnamefont{and}
  \bibinfo{author}{\bibfnamefont{R.~D.} \bibnamefont{Mawhinney}},
  \bibinfo{journal}{PoS} \textbf{\bibinfo{volume}{LATTICE2011}},
  \bibinfo{pages}{051} (\bibinfo{year}{2011}), \eprint{1111.5059}.

\bibitem[{\citenamefont{Narayanan and Neuberger}(2006)}]{Narayanan:2006rf}
\bibinfo{author}{\bibfnamefont{R.}~\bibnamefont{Narayanan}} \bibnamefont{and}
  \bibinfo{author}{\bibfnamefont{H.}~\bibnamefont{Neuberger}},
  \bibinfo{journal}{JHEP} \textbf{\bibinfo{volume}{0603}}, \bibinfo{pages}{064}
  (\bibinfo{year}{2006}), \eprint{hep-th/0601210}.

\bibitem[{\citenamefont{Bowick and Brezin}(1991)}]{Bowick:1991ky}
\bibinfo{author}{\bibfnamefont{M.~J.} \bibnamefont{Bowick}} \bibnamefont{and}
  \bibinfo{author}{\bibfnamefont{E.}~\bibnamefont{Brezin}},
  \bibinfo{journal}{Phys. Lett.} \textbf{\bibinfo{volume}{B268}},
  \bibinfo{pages}{21} (\bibinfo{year}{1991}).

\bibitem[{\citenamefont{Akemann et~al.}(1997)\citenamefont{Akemann, Damgaard,
  Magnea, and Nishigaki}}]{Akemann:1996vr}
\bibinfo{author}{\bibfnamefont{G.}~\bibnamefont{Akemann}},
  \bibinfo{author}{\bibfnamefont{P.~H.} \bibnamefont{Damgaard}},
  \bibinfo{author}{\bibfnamefont{U.}~\bibnamefont{Magnea}} \bibnamefont{and}
  \bibinfo{author}{\bibfnamefont{S.}~\bibnamefont{Nishigaki}},
  \bibinfo{journal}{Nucl. Phys.} \textbf{\bibinfo{volume}{B487}},
  \bibinfo{pages}{721} (\bibinfo{year}{1997}), \eprint{hep-th/9609174}.

\bibitem[{\citenamefont{Damgaard et~al.}(2000)\citenamefont{Damgaard, Heller,
  Niclasen, and Rummukainen}}]{Damgaard:2000cx}
\bibinfo{author}{\bibfnamefont{P.~H.} \bibnamefont{Damgaard}},
  \bibinfo{author}{\bibfnamefont{U.~M.} \bibnamefont{Heller}},
  \bibinfo{author}{\bibfnamefont{R.}~\bibnamefont{Niclasen}} \bibnamefont{and}
  \bibinfo{author}{\bibfnamefont{K.}~\bibnamefont{Rummukainen}},
  \bibinfo{journal}{Nucl. Phys.} \textbf{\bibinfo{volume}{B583}},
  \bibinfo{pages}{347} (\bibinfo{year}{2000}), \eprint{hep-lat/0003021}.

\bibitem[{\citenamefont{L{\"u}scher}(2010)}]{Luscher:2010iy}
\bibinfo{author}{\bibfnamefont{M.}~\bibnamefont{L{\"u}scher}},
  \bibinfo{journal}{JHEP} \textbf{\bibinfo{volume}{1008}}, \bibinfo{pages}{071}
  (\bibinfo{year}{2010}) \bibinfo{note}{[Erratum \bibinfo{journal}{JHEP} \textbf{\bibinfo{volume}{1403}}, \bibinfo{pages}{092} (\bibinfo{year}{2014})}],
  \eprint{1006.4518}.

\bibitem[{\citenamefont{Fodor et~al.}(2014)\citenamefont{Fodor, Holland, Kuti,
  Mondal, Nogradi, and Wong}}]{Fodor:2014cpa}
\bibinfo{author}{\bibfnamefont{Z.}~\bibnamefont{Fodor}},
  \bibinfo{author}{\bibfnamefont{K.}~\bibnamefont{Holland}},
  \bibinfo{author}{\bibfnamefont{J.}~\bibnamefont{Kuti}},
  \bibinfo{author}{\bibfnamefont{S.}~\bibnamefont{Mondal}},
  \bibinfo{author}{\bibfnamefont{D.}~\bibnamefont{Nogradi}} \bibnamefont{and}
  \bibinfo{author}{\bibfnamefont{C.~H.} \bibnamefont{Wong}},
  \bibinfo{journal}{JHEP} \textbf{\bibinfo{volume}{1409}}, \bibinfo{pages}{018}
  (\bibinfo{year}{2014}), \eprint{1406.0827}.

\bibitem[{\citenamefont{Cheng et~al.}(2014)\citenamefont{Cheng, Hasenfratz,
  Liu, Petropoulos, and Schaich}}]{Cheng:2014jba}
\bibinfo{author}{\bibfnamefont{A.}~\bibnamefont{Cheng}},
  \bibinfo{author}{\bibfnamefont{A.}~\bibnamefont{Hasenfratz}},
  \bibinfo{author}{\bibfnamefont{Y.}~\bibnamefont{Liu}},
  \bibinfo{author}{\bibfnamefont{G.}~\bibnamefont{Petropoulos}}
  \bibnamefont{and} \bibinfo{author}{\bibfnamefont{D.}~\bibnamefont{Schaich}},
  \bibinfo{journal}{JHEP} \textbf{\bibinfo{volume}{1405}}, \bibinfo{pages}{137}
  (\bibinfo{year}{2014}), \eprint{1404.0984}.

\bibitem[{\citenamefont{Bar and Golterman}(2014)}]{Bar:2013ora}
\bibinfo{author}{\bibfnamefont{O.}~\bibnamefont{Bar}} \bibnamefont{and}
  \bibinfo{author}{\bibfnamefont{M.}~\bibnamefont{Golterman}},
  \bibinfo{journal}{Phys. Rev.} \textbf{\bibinfo{volume}{D89}},
  \bibinfo{pages}{034505} (\bibinfo{year}{2014}), \bibinfo{note}{[Erratum:
  Phys. Rev. D 89, 099905 (2014)]}, \eprint{1312.4999}.

\bibitem[{\citenamefont{DeGrand}(2017)}]{DeGrand:2017gbi}
\bibinfo{author}{\bibfnamefont{T.}~\bibnamefont{DeGrand}},
  \bibinfo{journal}{Phys. Rev.} \textbf{\bibinfo{volume}{D95}},
  \bibinfo{pages}{114512} (\bibinfo{year}{2017}), \eprint{1701.00793}.

\bibitem[{\citenamefont{Kilcup and Sharpe}(1987)}]{Kilcup:1986dg}
\bibinfo{author}{\bibfnamefont{G.~W.} \bibnamefont{Kilcup}} \bibnamefont{and}
  \bibinfo{author}{\bibfnamefont{S.~R.} \bibnamefont{Sharpe}},
  \bibinfo{journal}{Nucl. Phys.} \textbf{\bibinfo{volume}{B283}},
  \bibinfo{pages}{493} (\bibinfo{year}{1987}).

\bibitem[{\citenamefont{Aoki et~al.}(2017{\natexlab{c}})}]{Aoki:2016frl}
\bibinfo{collaboration}{Flavor Lattice Averaging Group},
  \bibinfo{journal}{Eur. Phys. J.} \textbf{\bibinfo{volume}{C77}},
  \bibinfo{pages}{112} (\bibinfo{year}{2017}{\natexlab{c}}),
  \eprint{1607.00299}.

\bibitem[{\citenamefont{Gupta et~al.}(1991)\citenamefont{Gupta, Guralnik,
  Kilcup, and Sharpe}}]{Gupta:1990mr}
\bibinfo{author}{\bibfnamefont{R.}~\bibnamefont{Gupta}},
  \bibinfo{author}{\bibfnamefont{G.}~\bibnamefont{Guralnik}},
  \bibinfo{author}{\bibfnamefont{G.~W.} \bibnamefont{Kilcup}}
  \bibnamefont{and} \bibinfo{author}{\bibfnamefont{S.~R.}
  \bibnamefont{Sharpe}}, \bibinfo{journal}{Phys. Rev.}
  \textbf{\bibinfo{volume}{D43}}, \bibinfo{pages}{2003} (\bibinfo{year}{1991}).

\bibitem[{\citenamefont{Wolff}(2004)}]{Wolff:2003sm}
\bibinfo{author}{\bibfnamefont{U.}~\bibnamefont{Wolff}}
  (\bibinfo{collaboration}{Alpha Collaboration}), \bibinfo{journal}{Comput.
  Phys. Commun.} \textbf{\bibinfo{volume}{156}}, \bibinfo{pages}{143}
  (\bibinfo{year}{2004}) \bibinfo{note}{[Erratum: \bibinfo{journal}{Comput. Phys. Commun.} \textbf{\bibinfo{volume}{176}},
  \bibinfo{pages}{383} (\bibinfo{year}{2007})]}, \eprint{hep-lat/0306017}.

\bibitem[{\citenamefont{Golterman}(1986)}]{Golterman:1985dz}
\bibinfo{author}{\bibfnamefont{M.~F.} \bibnamefont{Golterman}},
  \bibinfo{journal}{Nucl. Phys.} \textbf{\bibinfo{volume}{B273}},
  \bibinfo{pages}{663} (\bibinfo{year}{1986}).

\bibitem[{\citenamefont{Golterman and Smit}(1985)}]{Golterman:1984dn}
\bibinfo{author}{\bibfnamefont{M.~F.~L.} \bibnamefont{Golterman}}
  \bibnamefont{and} \bibinfo{author}{\bibfnamefont{J.}~\bibnamefont{Smit}},
  \bibinfo{journal}{Nucl. Phys.} \textbf{\bibinfo{volume}{B255}},
  \bibinfo{pages}{328} (\bibinfo{year}{1985}).

\bibitem[{\citenamefont{Foley et~al.}(2005)\citenamefont{Foley, Juge, O'Cais,
  Peardon, Ryan, and Skullerud}}]{Foley:2005ac}
\bibinfo{author}{\bibfnamefont{J.}~\bibnamefont{Foley}},
  \bibinfo{author}{\bibfnamefont{K.~J.} \bibnamefont{Juge}},
  \bibinfo{author}{\bibfnamefont{A.}~\bibnamefont{O'Cais}},
  \bibinfo{author}{\bibfnamefont{M.}~\bibnamefont{Peardon}},
  \bibinfo{author}{\bibfnamefont{S.~M.} \bibnamefont{Ryan}} \bibnamefont{and}
  \bibinfo{author}{\bibfnamefont{J.-I.} \bibnamefont{Skullerud}},
  \bibinfo{journal}{Comput. Phys. Commun.} \textbf{\bibinfo{volume}{172}},
  \bibinfo{pages}{145} (\bibinfo{year}{2005}), \eprint{hep-lat/0505023}.

\bibitem[{\citenamefont{Lepage}(2017)}]{Lepage:lsqfit}
\bibinfo{author}{\bibfnamefont{G.~P.} \bibnamefont{Lepage}},
  \emph{\bibinfo{title}{\texttt{lsqfit} v9.1.2}} (\bibinfo{year}{2017}),
  doi:10.5281/zenodo.60221.

\bibitem[{\citenamefont{Bitar et~al.}(1990)\citenamefont{Bitar, Edwards,
  Heller, Kennedy, Liu, DeGrand, Gottlieb, Krasnitz, Kogut, Renken
  et~al.}}]{Bitar:1990cb}
\bibinfo{author}{\bibfnamefont{K.~M.} \bibnamefont{Bitar}},
  \bibinfo{author}{\bibfnamefont{R.}~\bibnamefont{Edwards}},
  \bibinfo{author}{\bibfnamefont{U.~M.} \bibnamefont{Heller}},
  \bibinfo{author}{\bibfnamefont{A.~D.} \bibnamefont{Kennedy}},
  \bibinfo{author}{\bibfnamefont{W.-q.} \bibnamefont{Liu}},
  \bibinfo{author}{\bibfnamefont{T.~A.} \bibnamefont{DeGrand}},
  \bibinfo{author}{\bibfnamefont{S.~A.} \bibnamefont{Gottlieb}},
  \bibinfo{author}{\bibfnamefont{A.}~\bibnamefont{Krasnitz}},
  \bibinfo{author}{\bibfnamefont{J.~B.} \bibnamefont{Kogut}},
  \bibinfo{author}{\bibfnamefont{R.~L.} \bibnamefont{Renken}},
  \bibinfo{author}{\bibfnamefont{M.~C.} \bibnamefont{Ogilvie}},
  \bibinfo{author}{\bibfnamefont{R.}~\bibnamefont{Pietro}},
  \bibinfo{author}{\bibfnamefont{D.~K.} \bibnamefont{Sinclair}},
  \bibinfo{author}{\bibfnamefont{K.~C.} \bibnamefont{Wang}},
  \bibinfo{author}{\bibfnamefont{R.~L.} \bibnamefont{Sugar}},
  \bibinfo{author}{\bibfnamefont{M.}~\bibnamefont{Teper}}
  \bibnamefont{and} \bibinfo{author}{\bibfnamefont{D.}~\bibnamefont{Toussaint}},
  \bibinfo{journal}{Phys. Rev.}
  \textbf{\bibinfo{volume}{D42}}, \bibinfo{pages}{3794} (\bibinfo{year}{1990}).

\bibitem[{\citenamefont{Ayyar et~al.}(2018)\citenamefont{Ayyar, DeGrand,
  Golterman, Hackett, Jay, Neil, Shamir, and Svetitsky}}]{Ayyar:2017qdf}
\bibinfo{author}{\bibfnamefont{V.}~\bibnamefont{Ayyar}},
  \bibinfo{author}{\bibfnamefont{T.}~\bibnamefont{DeGrand}},
  \bibinfo{author}{\bibfnamefont{M.}~\bibnamefont{Golterman}},
  \bibinfo{author}{\bibfnamefont{D.~C.} \bibnamefont{Hackett}},
  \bibinfo{author}{\bibfnamefont{W.~I.} \bibnamefont{Jay}},
  \bibinfo{author}{\bibfnamefont{E.~T.} \bibnamefont{Neil}},
  \bibinfo{author}{\bibfnamefont{Y.}~\bibnamefont{Shamir}} \bibnamefont{and}
  \bibinfo{author}{\bibfnamefont{B.}~\bibnamefont{Svetitsky}},
  \bibinfo{journal}{Phys. Rev.} \textbf{\bibinfo{volume}{D97}},
  \bibinfo{pages}{074505} (\bibinfo{year}{2018}), \eprint{1710.00806}.

\bibitem[{\citenamefont{Kunihiro et~al.}(2004)\citenamefont{Kunihiro, Muroya,
  Nakamura, Nonaka, Sekiguchi, and Wada}}]{Kunihiro:2003yj}
\bibinfo{author}{\bibfnamefont{T.}~\bibnamefont{Kunihiro}},
  \bibinfo{author}{\bibfnamefont{S.}~\bibnamefont{Muroya}},
  \bibinfo{author}{\bibfnamefont{A.}~\bibnamefont{Nakamura}},
  \bibinfo{author}{\bibfnamefont{C.}~\bibnamefont{Nonaka}},
  \bibinfo{author}{\bibfnamefont{M.}~\bibnamefont{Sekiguchi}}
  \bibnamefont{and} \bibinfo{author}{\bibfnamefont{H.}~\bibnamefont{Wada}}
  (\bibinfo{collaboration}{Scalar Collaboration}), \bibinfo{journal}{Phys.
  Rev.} \textbf{\bibinfo{volume}{D70}}, \bibinfo{pages}{034504}
  (\bibinfo{year}{2004}), \eprint{hep-ph/0310312}.

\bibitem[{\citenamefont{Howarth and Giedt}(2017)}]{Howarth:2015caa}
\bibinfo{author}{\bibfnamefont{D.}~\bibnamefont{Howarth}} \bibnamefont{and}
  \bibinfo{author}{\bibfnamefont{J.}~\bibnamefont{Giedt}},
  \bibinfo{journal}{Int. J. Mod. Phys.} \textbf{\bibinfo{volume}{C28}},
  \bibinfo{pages}{1750124} (\bibinfo{year}{2017}), \eprint{1508.05658}.

\bibitem[{\citenamefont{Briceno et~al.}(2017)\citenamefont{Briceno, Dudek,
  Edwards, and Wilson}}]{Briceno:2016mjc}
\bibinfo{author}{\bibfnamefont{R.~A.} \bibnamefont{Briceno}},
  \bibinfo{author}{\bibfnamefont{J.~J.} \bibnamefont{Dudek}},
  \bibinfo{author}{\bibfnamefont{R.~G.} \bibnamefont{Edwards}}
  \bibnamefont{and} \bibinfo{author}{\bibfnamefont{D.~J.}
  \bibnamefont{Wilson}}, \bibinfo{journal}{Phys. Rev. Lett.}
  \textbf{\bibinfo{volume}{118}}, \bibinfo{pages}{022002}
  (\bibinfo{year}{2017}), \eprint{1607.05900}.

\bibitem[{\citenamefont{Del~Debbio and Zwicky}(2010)}]{DelDebbio:2010ze}
\bibinfo{author}{\bibfnamefont{L.}~\bibnamefont{Del~Debbio}} \bibnamefont{and}
  \bibinfo{author}{\bibfnamefont{R.}~\bibnamefont{Zwicky}},
  \bibinfo{journal}{Phys. Rev.} \textbf{\bibinfo{volume}{D82}},
  \bibinfo{pages}{014502} (\bibinfo{year}{2010}), \eprint{1005.2371}.

\bibitem[{\citenamefont{Wilson et~al.}(2015)\citenamefont{Wilson, Briceno,
  Dudek, Edwards, and Thomas}}]{Wilson:2015dqa}
\bibinfo{author}{\bibfnamefont{D.~J.} \bibnamefont{Wilson}},
  \bibinfo{author}{\bibfnamefont{R.~A.} \bibnamefont{Briceno}},
  \bibinfo{author}{\bibfnamefont{J.~J.} \bibnamefont{Dudek}},
  \bibinfo{author}{\bibfnamefont{R.~G.} \bibnamefont{Edwards}}
  \bibnamefont{and} \bibinfo{author}{\bibfnamefont{C.~E.}
  \bibnamefont{Thomas}}, \bibinfo{journal}{Phys. Rev.}
  \textbf{\bibinfo{volume}{D92}}, \bibinfo{pages}{094502}
  (\bibinfo{year}{2015}), \eprint{1507.02599}.

\bibitem[{\citenamefont{Briceno et~al.}(2018)\citenamefont{Briceno, Dudek,
  Edwards, and Wilson}}]{Briceno:2017qmb}
\bibinfo{author}{\bibfnamefont{R.~A.} \bibnamefont{Briceno}},
  \bibinfo{author}{\bibfnamefont{J.~J.} \bibnamefont{Dudek}},
  \bibinfo{author}{\bibfnamefont{R.~G.} \bibnamefont{Edwards}}
  \bibnamefont{and} \bibinfo{author}{\bibfnamefont{D.~J.}
  \bibnamefont{Wilson}}, \bibinfo{journal}{Phys. Rev.}
  \textbf{\bibinfo{volume}{D97}}, \bibinfo{pages}{054513}
  (\bibinfo{year}{2018}), \eprint{1708.06667}.

\bibitem[{\citenamefont{Appelquist et~al.}(2010)\citenamefont{Appelquist,
  Avakian, Babich, Brower, Cheng, Clark, Cohen, Fleming, Kiskis, Neil
  et~al.}}]{Appelquist:2009ka}
\bibinfo{author}{\bibfnamefont{T.}~\bibnamefont{Appelquist}},
  \bibinfo{author}{\bibfnamefont{A.}~\bibnamefont{Avakian}},
  \bibinfo{author}{\bibfnamefont{R.}~\bibnamefont{Babich}},
  \bibinfo{author}{\bibfnamefont{R.~C.} \bibnamefont{Brower}},
  \bibinfo{author}{\bibfnamefont{M.}~\bibnamefont{Cheng}},
  \bibinfo{author}{\bibfnamefont{M.~A.} \bibnamefont{Clark}},
  \bibinfo{author}{\bibfnamefont{S.~D.} \bibnamefont{Cohen}},
  \bibinfo{author}{\bibfnamefont{G.~T.} \bibnamefont{Fleming}},
  \bibinfo{author}{\bibfnamefont{J.}~\bibnamefont{Kiskis}},
  \bibinfo{author}{\bibfnamefont{E.~T.} \bibnamefont{Neil}},
  \bibinfo{author}{\bibfnamefont{J.~C.}~\bibnamefont{Osborn}},
  \bibinfo{author}{\bibfnamefont{C.}~\bibnamefont{Rebbi}},
  \bibinfo{author}{\bibfnamefont{D.}~\bibnamefont{Schaich}}
  \bibnamefont{and} \bibinfo{author}{\bibfnamefont{P.}~\bibnamefont{Vranas}}
  (\bibinfo{collaboration}{LSD Collaboration}),
  \bibinfo{journal}{Phys. Rev. Lett.} \textbf{\bibinfo{volume}{104}},
  \bibinfo{pages}{071601} (\bibinfo{year}{2010}), \eprint{0910.2224}.

\bibitem[{\citenamefont{Neil et~al.}(2009)\citenamefont{Neil, Avakian, Babich,
  Brower, Cheng, Clark, Cohen, Fleming, Kiskis, Osborn et~al.}}]{Neil:2010sc}
\bibinfo{author}{\bibfnamefont{E.~T.} \bibnamefont{Neil}},
  \bibinfo{author}{\bibfnamefont{A.}~\bibnamefont{Avakian}},
  \bibinfo{author}{\bibfnamefont{R.}~\bibnamefont{Babich}},
  \bibinfo{author}{\bibfnamefont{R.~C.} \bibnamefont{Brower}},
  \bibinfo{author}{\bibfnamefont{M.}~\bibnamefont{Cheng}},
  \bibinfo{author}{\bibfnamefont{M.~A.} \bibnamefont{Clark}},
  \bibinfo{author}{\bibfnamefont{S.~D.} \bibnamefont{Cohen}},
  \bibinfo{author}{\bibfnamefont{G.~T.} \bibnamefont{Fleming}},
  \bibinfo{author}{\bibfnamefont{J.}~\bibnamefont{Kiskis}},
  \bibinfo{author}{\bibfnamefont{J.~C.} \bibnamefont{Osborn}},
  \bibinfo{author}{\bibfnamefont{C.}~\bibnamefont{Rebbi}},
  \bibinfo{author}{\bibfnamefont{D.}~\bibnamefont{Schaich}}
  \bibnamefont{and} \bibinfo{author}{\bibfnamefont{P.}~\bibnamefont{Vranas}}
  (\bibinfo{collaboration}{LSD Collaboration}),
  \bibinfo{journal}{PoS} \textbf{\bibinfo{volume}{CD09}},
  \bibinfo{pages}{088} (\bibinfo{year}{2009}), \eprint{1002.3777}.

\bibitem[{\citenamefont{Fleming and Neil}(2014)}]{Fleming:2013tra}
\bibinfo{author}{\bibfnamefont{G.~T.} \bibnamefont{Fleming}} \bibnamefont{and}
  \bibinfo{author}{\bibfnamefont{E.~T.} \bibnamefont{Neil}}
  (\bibinfo{collaboration}{LSD Collaboration}), in
  \emph{\bibinfo{booktitle}{{Strong Coupling Gauge Theories in the LHC
  Perspective (SCGT12)}}} (\bibinfo{year}{2014}), \bibinfo{pages}{58--64},
  \eprint{1312.5298}.

\bibitem[{\citenamefont{Kurachi and Shrock}(2006)}]{Kurachi:2006ej}
\bibinfo{author}{\bibfnamefont{M.}~\bibnamefont{Kurachi}} \bibnamefont{and}
  \bibinfo{author}{\bibfnamefont{R.}~\bibnamefont{Shrock}},
  \bibinfo{journal}{JHEP} \textbf{\bibinfo{volume}{0612}}, \bibinfo{pages}{034}
  (\bibinfo{year}{2006}), \eprint{hep-ph/0605290}.

\bibitem[{\citenamefont{Appelquist et~al.}(2011)\citenamefont{Appelquist,
  Babich, Brower, Cheng, Clark, Cohen, Fleming, Kiskis, Lin, Neil
  et~al.}}]{Appelquist:2010xv}
\bibinfo{author}{\bibfnamefont{T.}~\bibnamefont{Appelquist}},
  \bibinfo{author}{\bibfnamefont{R.}~\bibnamefont{Babich}},
  \bibinfo{author}{\bibfnamefont{R.~C.} \bibnamefont{Brower}},
  \bibinfo{author}{\bibfnamefont{M.}~\bibnamefont{Cheng}},
  \bibinfo{author}{\bibfnamefont{M.~A.} \bibnamefont{Clark}},
  \bibinfo{author}{\bibfnamefont{S.~D.} \bibnamefont{Cohen}},
  \bibinfo{author}{\bibfnamefont{G.~T.} \bibnamefont{Fleming}},
  \bibinfo{author}{\bibfnamefont{J.}~\bibnamefont{Kiskis}},
  \bibinfo{author}{\bibfnamefont{M.}~\bibnamefont{Lin}},
  \bibinfo{author}{\bibfnamefont{E.~T.} \bibnamefont{Neil}},
  \bibinfo{author}{\bibfnamefont{J.~C.} \bibnamefont{Osborn}},
  \bibinfo{author}{\bibfnamefont{C.}~\bibnamefont{Rebbi}},
  \bibinfo{author}{\bibfnamefont{D.}~\bibnamefont{Schaich}}
  \bibnamefont{and} \bibinfo{author}{\bibfnamefont{P.}~\bibnamefont{Vranas}}
  (\bibinfo{collaboration}{LSD Collaboration}),
  \bibinfo{journal}{Phys. Rev. Lett.} \textbf{\bibinfo{volume}{106}},
  \bibinfo{pages}{231601} (\bibinfo{year}{2011}), \eprint{1009.5967}.

\bibitem[{\citenamefont{Soto et~al.}(2013)\citenamefont{Soto, Talavera, and
  Tarrus}}]{Soto:2011ap}
\bibinfo{author}{\bibfnamefont{J.}~\bibnamefont{Soto}},
  \bibinfo{author}{\bibfnamefont{P.}~\bibnamefont{Talavera}} \bibnamefont{and}
  \bibinfo{author}{\bibfnamefont{J.}~\bibnamefont{Tarrus}},
  \bibinfo{journal}{Nucl. Phys.} \textbf{\bibinfo{volume}{B866}},
  \bibinfo{pages}{270} (\bibinfo{year}{2013}), \eprint{1110.6156}.

\bibitem[{\citenamefont{Matsuzaki and Yamawaki}(2014)}]{Matsuzaki:2013eva}
\bibinfo{author}{\bibfnamefont{S.}~\bibnamefont{Matsuzaki}} \bibnamefont{and}
  \bibinfo{author}{\bibfnamefont{K.}~\bibnamefont{Yamawaki}},
  \bibinfo{journal}{Phys. Rev. Lett.} \textbf{\bibinfo{volume}{113}},
  \bibinfo{pages}{082002} (\bibinfo{year}{2014}), \eprint{1311.3784}.

\bibitem[{\citenamefont{Golterman and Shamir}(2016)}]{Golterman:2016lsd}
\bibinfo{author}{\bibfnamefont{M.}~\bibnamefont{Golterman}} \bibnamefont{and}
  \bibinfo{author}{\bibfnamefont{Y.}~\bibnamefont{Shamir}},
  \bibinfo{journal}{Phys. Rev.} \textbf{\bibinfo{volume}{D94}},
  \bibinfo{pages}{054502} (\bibinfo{year}{2016}), \eprint{1603.04575}.

\bibitem[{\citenamefont{Hansen et~al.}(2017)\citenamefont{Hansen, Langaeble,
  and Sannino}}]{Hansen:2016fri}
\bibinfo{author}{\bibfnamefont{M.}~\bibnamefont{Hansen}},
  \bibinfo{author}{\bibfnamefont{K.}~\bibnamefont{Langaeble}}
  \bibnamefont{and} \bibinfo{author}{\bibfnamefont{F.}~\bibnamefont{Sannino}},
  \bibinfo{journal}{Phys. Rev.} \textbf{\bibinfo{volume}{D95}},
  \bibinfo{pages}{036005} (\bibinfo{year}{2017}), \eprint{1610.02904}.

\bibitem[{\citenamefont{Appelquist et~al.}(2017)\citenamefont{Appelquist,
  Ingoldby, and Piai}}]{Appelquist:2017wcg}
\bibinfo{author}{\bibfnamefont{T.}~\bibnamefont{Appelquist}},
  \bibinfo{author}{\bibfnamefont{J.}~\bibnamefont{Ingoldby}} \bibnamefont{and}
  \bibinfo{author}{\bibfnamefont{M.}~\bibnamefont{Piai}},
  \bibinfo{journal}{JHEP} \textbf{\bibinfo{volume}{1707}}, \bibinfo{pages}{035}
  (\bibinfo{year}{2017}), \eprint{1702.04410}.

\bibitem[{\citenamefont{Appelquist
  et~al.}(2018{\natexlab{a}})\citenamefont{Appelquist, Ingoldby, and
  Piai}}]{Appelquist:2017vyy}
\bibinfo{author}{\bibfnamefont{T.}~\bibnamefont{Appelquist}},
  \bibinfo{author}{\bibfnamefont{J.}~\bibnamefont{Ingoldby}} \bibnamefont{and}
  \bibinfo{author}{\bibfnamefont{M.}~\bibnamefont{Piai}},
  \bibinfo{journal}{JHEP} \textbf{\bibinfo{volume}{1803}}, \bibinfo{pages}{039}
  (\bibinfo{year}{2018}{\natexlab{a}}), \eprint{1711.00067}.

\bibitem[{\citenamefont{Meurice}(2017{\natexlab{a}})}]{Meurice:2017jry}
\bibinfo{author}{\bibfnamefont{Y.}~\bibnamefont{Meurice}}, in
  \emph{\bibinfo{booktitle}{{Meeting of the APS Division of Particles and
  Fields (DPF 2017) Batavia, Illinois, 31 July--4 August 2017}}}
  (\bibinfo{year}{2017}{\natexlab{a}}), \eprint{1709.10017}.

\bibitem[{\citenamefont{Meurice}(2017{\natexlab{b}})}]{Meurice:2017zng}
\bibinfo{author}{\bibfnamefont{Y.}~\bibnamefont{Meurice}},
  \bibinfo{journal}{Phys. Rev.} \textbf{\bibinfo{volume}{D96}},
  \bibinfo{pages}{114507} (\bibinfo{year}{2017}{\natexlab{b}}),
  \eprint{1709.09264}.

\bibitem[{\citenamefont{Gasbarro}(2018)}]{Gasbarro:2017ccf}
\bibinfo{author}{\bibfnamefont{A.}~\bibnamefont{Gasbarro}},
  \bibinfo{journal}{EPJ Web Conf.} \textbf{\bibinfo{volume}{175}},
  \bibinfo{pages}{08024} (\bibinfo{year}{2018}), \eprint{1710.08545}.

\bibitem[{\citenamefont{Golterman and Shamir}(2018)}]{Golterman:2018mfm}
\bibinfo{author}{\bibfnamefont{M.}~\bibnamefont{Golterman}} \bibnamefont{and}
  \bibinfo{author}{\bibfnamefont{Y.}~\bibnamefont{Shamir}}
  (\bibinfo{year}{2018}), \eprint{1805.00198}.

\bibitem[{\citenamefont{De~Floor et~al.}(2018)\citenamefont{De~Floor,
  Gustafson, and Meurice}}]{DeFloor:2018xrp}
\bibinfo{author}{\bibfnamefont{D.}~\bibnamefont{De~Floor}},
  \bibinfo{author}{\bibfnamefont{E.}~\bibnamefont{Gustafson}},
  \bibnamefont{and} \bibinfo{author}{\bibfnamefont{Y.}~\bibnamefont{Meurice}}
  (\bibinfo{year}{2018}), \eprint{1807.05047}.

\bibitem[{\citenamefont{Appelquist
  et~al.}(2018{\natexlab{b}})\citenamefont{Appelquist, Brower, Fleming,
  Gasbarro, Hasenfratz, Ingoldby, Kiskis, Osborn, Rebbi, Rinaldi
  et~al.}}]{Appelquist:2018tyt}
\bibinfo{author}{\bibfnamefont{T.}~\bibnamefont{Appelquist}},
  \bibinfo{author}{\bibfnamefont{R.~C.} \bibnamefont{Brower}},
  \bibinfo{author}{\bibfnamefont{G.~T.} \bibnamefont{Fleming}},
  \bibinfo{author}{\bibfnamefont{A.}~\bibnamefont{Gasbarro}},
  \bibinfo{author}{\bibfnamefont{A.}~\bibnamefont{Hasenfratz}},
  \bibinfo{author}{\bibfnamefont{J.}~\bibnamefont{Ingoldby}},
  \bibinfo{author}{\bibfnamefont{J.}~\bibnamefont{Kiskis}},
  \bibinfo{author}{\bibfnamefont{J.~C.} \bibnamefont{Osborn}},
  \bibinfo{author}{\bibfnamefont{C.}~\bibnamefont{Rebbi}},
  \bibinfo{author}{\bibfnamefont{E.}~\bibnamefont{Rinaldi}},
  \bibinfo{author}{\bibfnamefont{D.}~\bibnamefont{Schaich}},
  \bibinfo{author}{\bibfnamefont{P.}~\bibnamefont{Vranas}},
  \bibinfo{author}{\bibfnamefont{E.}~\bibnamefont{Weinberg}}
  \bibnamefont{and} \bibinfo{author}{\bibfnamefont{O.}~\bibnamefont{Witzel}}
  (\bibinfo{collaboration}{LSD Collaboration}),
  \bibinfo{journal}{Phys. Rev.} \textbf{\bibinfo{volume}{D98}},
  \bibinfo{pages}{114510} (\bibinfo{year}{2018}{\natexlab{b}}),
  \eprint{1809.02624}.

\bibitem[{\citenamefont{Erdmenger et~al.}(2015)\citenamefont{Erdmenger, Evans,
  and Scott}}]{Erdmenger:2014fxa}
\bibinfo{author}{\bibfnamefont{J.}~\bibnamefont{Erdmenger}},
  \bibinfo{author}{\bibfnamefont{N.}~\bibnamefont{Evans}} \bibnamefont{and}
  \bibinfo{author}{\bibfnamefont{M.}~\bibnamefont{Scott}},
  \bibinfo{journal}{Phys. Rev.} \textbf{\bibinfo{volume}{D91}},
  \bibinfo{pages}{085004} (\bibinfo{year}{2015}), \eprint{1412.3165}.

\bibitem[{\citenamefont{Aubin et~al.}(2004)\citenamefont{Aubin, Bernard, DeTar,
  Osborn, Gottlieb, Gregory, Toussaint, Heller, Hetrick, and
  Sugar}}]{Aubin:2004wf}
\bibinfo{author}{\bibfnamefont{C.}~\bibnamefont{Aubin}},
  \bibinfo{author}{\bibfnamefont{C.}~\bibnamefont{Bernard}},
  \bibinfo{author}{\bibfnamefont{C.}~\bibnamefont{DeTar}},
  \bibinfo{author}{\bibfnamefont{J.}~\bibnamefont{Osborn}},
  \bibinfo{author}{\bibfnamefont{S.}~\bibnamefont{Gottlieb}},
  \bibinfo{author}{\bibfnamefont{E.~B.} \bibnamefont{Gregory}},
  \bibinfo{author}{\bibfnamefont{D.}~\bibnamefont{Toussaint}},
  \bibinfo{author}{\bibfnamefont{U.~M.} \bibnamefont{Heller}},
  \bibinfo{author}{\bibfnamefont{J.~E.} \bibnamefont{Hetrick}}
  \bibnamefont{and} \bibinfo{author}{\bibfnamefont{R.}~\bibnamefont{Sugar}},
  \bibinfo{journal}{Phys. Rev.} \textbf{\bibinfo{volume}{D70}},
  \bibinfo{pages}{094505} (\bibinfo{year}{2004}), \eprint{hep-lat/0402030}.

\bibitem[{\citenamefont{Lee and Sharpe}(1999)}]{Lee:1999zxa}
\bibinfo{author}{\bibfnamefont{W.-J.} \bibnamefont{Lee}} \bibnamefont{and}
  \bibinfo{author}{\bibfnamefont{S.~R.} \bibnamefont{Sharpe}},
  \bibinfo{journal}{Phys. Rev.} \textbf{\bibinfo{volume}{D60}},
  \bibinfo{pages}{114503} (\bibinfo{year}{1999}), \eprint{hep-lat/9905023}.

\end{thebibliography}
\end{document}